\documentclass[12pt,a4paper]{article}
\pdfoutput=1
\usepackage[utf8]{inputenc}
\usepackage[T1]{fontenc}
\usepackage{style}
\usepackage[english]{babel}
\usepackage{url}
\usepackage{footnote}

\usepackage{amsmath}
\usepackage{graphicx} 
\usepackage{verbatim}

\newcommand{\Mpc}{\text{Mpc}}

\renewcommand{\a}{\alpha}

\newcommand{\be}{\begin{equation}}
\newcommand{\ee}{\end{equation}}
\newcommand{\beqa}{\begin{eqnarray}}
\newcommand{\eeqa}{\end{eqnarray}}
\newcommand{\bsm}{\begin{smallmatrix}}
\newcommand{\esm}{\end{smallmatrix}}

\newcommand\kmax{k_{\rm max}}
\newcommand\hMpc{h\text{Mpc}^{-1}}
\newcommand\Mpch{h^{-1}\text{Mpc}}

\renewcommand\a{\alpha}

\renewcommand\k{{\bf k}}

\newcommand{\e}{\eta}

\def\e{{\rm e}}

\newcommand{\bseq}{\begin{subequations}}
\newcommand{\eseq}{\end{subequations}}

\renewcommand{\ln}{\mathop{\rm ln}\nolimits}

\renewcommand{\k}{{\bf k}}

\newcommand{\z}{{\bf z}}

\newcommand{\lin}{\mathrm{lin}}
\newcommand{\tree}{\mathrm{tree}}
\newcommand{\lp}{\mathrm{1\text{-}loop}}

\newcommand{\maxx}{{\rm max}}

\title{
Optimizing large-scale structure data analysis
with the theoretical error likelihood
}

\author[a,b]{Anton Chudaykin\footnote{\texttt{chudy@ms2.inr.ac.ru}}}
\author[c,b]{Mikhail M. Ivanov\footnote{\texttt{mi1271@nyu.edu}}}
\author[d]{Marko Simonovi\'c\footnote{\texttt{marko.simonovic@cern.ch}}}
\affiliation[a]{Department of Physics \& Astronomy, McMaster University,\\ 
\normalsize \it  1280 Main Street West, Hamilton, ON L8S 4M1, Canada}
\affiliation[b]{Institute for Nuclear Research of the
Russian Academy of Sciences, \\ 
\normalsize \it  60th October Anniversary Prospect, 7a, 117312
Moscow, Russia}
\affiliation[c]{Center for Cosmology and Particle Physics, Department of Physics,
New York University,\\
New York, NY 10003, USA}
\affiliation[d]{Theoretical Physics Department, CERN,\\
1 Esplanade des Particules, Geneva 23, CH-1211, Switzerland}

\abstract{
An important aspect of large-scale structure data 
analysis is the presence of non-negligible theoretical uncertainties,
which become increasingly important on small scales. 
We show how to incorporate these uncertainties
in realistic power spectrum likelihoods by an appropriate change of the fitting model 
and the covarance matrix.
The inclusion of the theoretical error has several advantages over the standard practice
of using the sharp momentum cut $k_{\rm max}$. 
First, the theoretical error covariance gradually suppresses the information 
from the short scales as the 
employed theoretical model becomes less reliable. 
This allows one to avoid laborious measurements of $k_{\rm max}$, 
which is an essential part of the standard methods.
Second, the theoretical error likelihood gives unbiased constrains with reliable error bars that are not artificially shrunk due to over-fitting. 
In realistic settings, 
the theoretical
error likelihood yields
essentially the same
parameter constraints
as
the standard analysis 
with an appropriately selected $k_{\rm max}$,
thereby effectively
optimizing the
choice 
of $k_{\rm max}$.
We demonstrate these points using the large-volume N-body 
data for 
the clustering of matter and galaxies in real and redshift space.
In passing, we validate the effective field theory description of the redshift space distortions
and show that the use of the 
one-parameter 
phenomenological 
Gaussian damping
model for fingers-of-God causes
significant biases in parameter recovery.
}

\begin{document}

\begin{flushright}
	INR-TH-2020-040, CERN-TH-2020-154
\end{flushright}

\vspace{-1.5cm}

\maketitle
\flushbottom


\section{Introduction}

Galaxy clustering on large scales becomes ever more important in modern cosmology.
The measurements of baryon acoustic oscillations (BAO)
and the power spectrum shape in the current data allow one to determine cosmological parameters with precision
that rivals the cosmic microwave background analysis~\cite{Ivanov:2019pdj,Colas:2019ret,Troster:2019ean,Ivanov:2019hqk,Philcox:2020vvt,Philcox:2020xbv,Chudaykin:2020ghx}. 
Even more progress is expected in the 
era of the upcoming high-precision surveys like
Euclid~\cite{Laureijs:2011gra,Amendola:2016saw} 
and DESI~\cite{Aghamousa:2016zmz}, see e.g.~\cite{Chudaykin:2019ock,Audren:2012vy,Yankelevich:2018uaz,Ivanov:2020ril}. 

One of the crucial ingredients for measurement of cosmological parameters from large-scale structure (LSS) 
is an accurate covariance matrix for a given summary statistic. 
The question of how to
estimate
covariance matrices 
and how they impact cosmological constraints
has 
stimulated a broad line of research over the past decades. 
This includes perturbative calculations~\cite{Scoccimarro:1999kp,Wadekar:2019rdu,Yamamoto:2005dz,Grieb:2015bia,Blake:2018tou,Li:2018scc},
measurements from mock catalogs (e.g.~\cite{Kitaura:2015uqa,Zhao:2020bib,Lin:2020nef}) and 
studying the systematic biases that can arise due to the uncertainties in the covariance matrix~\cite{Hartlap:2006kj,Percival:2013sga,Sellentin:2015waz,Philcox:2020zyp}. 
While these significant efforts focus mainly on statistical errors, it is important to note that there is another aspect 
of the covariance matrix treatment that has attracted attention only recently.
This is the theoretical error (TE) covariance, which is as important as the statistical covariance for any 
realistic large-scale structure analysis~\cite{Audren:2012vy,Baldauf:2016sjb,Chudaykin:2019ock,Steele:2020tak}. 
The TE covariance originates from imperfect knowledge of the theory model that is used to fit the data.
The LSS observables are sensitive to various nonlinear effects whose accurate description 
is quite challenging, both in analytical approaches to LSS clustering and in N-body or hydrodynamical 
simulations. The importance of these nonlinear effects grows on small scales and therefore impacts 
a large number of Fourier modes that are important to constrain cosmological parameters. 

Theoretical error becomes a leading source of uncertainty when the statistical errors become sufficiently small
(either due to a large volume of a survey or the analysis being pushed to smaller scales), at which point it has to be included in the analysis to avoid biasing the output cosmology.
The standard approach to deal with this situation is to assume that the fitted theory model is perfect up to 
a certain scale (e.g.~$\kmax$ in the power spectrum case) and perform the analysis with this data cut.
This approach has a number of disadvantages. First, it is very time-consuming, since the only way to determine the optimal $\kmax$ is to run many Markov-Chain Monte-Carlo analyses 
of the realistic mock data samples, see e.g.~\cite{DAmico:2019fhj,Ivanov:2019pdj,Nishimichi:2020tvu,Rossi:2020wxx,Alam:2020jvh}. 
Moreover, if a different model or a set of priors are used, one has to re-validate the $\kmax$ choice
because the parameter variances and degeneracy orientations change in this case~\cite{DAmico:2019fhj,DAmico:2020kxu}.
The second drawback of the $\kmax$ analysis is that 
biases for
different parameters become sizeable at very different scales. 
This makes the choice of $\kmax$ ambiguous. The usual approach is to choose the cuts such that 
biases on all {\em cosmological}
parameters are significantly smaller than
the statistical error.
On the one hand, this choice is the most conservative as it ignores possible improvements on 
cosmological parameters that are unbiased for larger $\kmax$. On the other hand, it still does not guarantee 
the absence of bias in {\em nuisance} parameters, which can be problematic 
when doing a joint analysis of different observables (for example, power spectrum and bispectrum) or combining 
different data sets. Finally, using a sharp data cut neglects two important properties of the power spectrum: 
(a) the broadband power spectrum is smooth, and (b) the shape of the common features, such as BAO wiggles, 
can be reliably calculated for any wavenumber~\cite{Senatore:2014via,Baldauf:2015xfa,Vlah:2015zda,Blas:2016sfa,Senatore:2017pbn,Ivanov:2018gjr,Chen:2020fxs}.\footnote{This statement can be generalized to the primordial
oscillating features
in the power spectrum and bispectrum~\cite{Vasudevan:2019ewf,Chen:2020ckc}.
} 
These two facts reveal yet another important limitation of the standard $\kmax$ analysis:
it neglects all information 
from scales beyond $\kmax$, a part of which can be recovered assuming some reasonable smoothens of 
the power spectrum and using the shape information from the BAO wiggles.

In this paper we show how all these issues can be resolved including the theoretical error covariance in the analysis. 
As a first step, we 
re-derive the theoretical 
error covariance
emphasizing the similarity 
of this procedure with 
the exact
marginalization over 
a nuisance parameter.
This example naturally suggests
that the theoretical error
likelihood should include 
both 
the mean and the covariance
of the theoretical uncertainty, 
with identical shapes. 
Then, we propose a new
way of extracting 
these quantities directly
from the mock data,
and argue that this method
gives a more reliable estimate
of theoretical error
than the original approach based
exclusively 
on perturbation theory.

We apply the TE formalism to the N-body simulation data with volume $\sim 100~(\text{Gpc}/h)^3$, similar to the cumulative 
volume of upcoming spectroscopic galaxy surveys. 
We explicitly illustrate how the TE approach allows one to obtain accurate and optimal constrains
that are independent of the choice of $\kmax$ and that are not affected by over-fitting. 
Moreover, the TE guarantees that both the principal components and 
their 1d projections onto the particular 
parameter planes are unbiased, along with the variances of these parameters.
Finally, we show that using the TE covariance indeed improves constrains on cosmological parameters
by including extra information from the BAO wiggles and exploiting the smoothness of the broadband power 
spectrum.\footnote{This approach has been recently validated in a slightly different context of 
the re-analysis of the BAO data from the BOSS survey in Ref.~\cite{Philcox:2020vvt}.}
It is important to stress that all these results can be obtained with a single MCMC analysis, in contrast with the standard $\kmax$ approach.

We scrutinize the effect of TE covariance on the power spectrum analyses in four difference setups: 
dark matter and galaxies in real and redshift space. 
We analyze the large-volume N-body simulation data using the power spectra calculated in the framework of 
the effective field theory of large-scale structure (see~\cite{Ivanov:2019pdj,DAmico:2019fhj} and references therein)
and explicitly demonstrate that the true input cosmology is extracted in an unbiased manner in all of these different examples. 
Our analyses also validate the implementation of the effective field theory for various tracers
with the \texttt{CLASS-PT} code~\cite{Chudaykin:2020aoj}.
In this context, our work can be viewed as a companion paper of Ref.~\cite{Chudaykin:2020aoj}, as it proves
that the accuracy of \texttt{CLASS-PT} meets the requirements of future high-precision surveys.

The paper is organized as follows. 
We give a theoretical background on the theoretical error in Section~\ref{sec:theory}.
Section~\ref{sec:method} specifies our methodology, the theoretical model, and N-body simulations. 
Then, we present the TE analysis and confront it with the standard $\kmax$ approach 
for dark matter in real space in Section~\ref{sec:dm_rs}, dark matter in redshift space in Section~\ref{sec:dm_rsd},
galaxies in real space in Section~\ref{sec:gal_rs}, and galaxies in redshift space in Section~\ref{sec:gal_rsd}.
We conclude in Section~\ref{sec:concl}.
Some additional material is presented in the 
appendices. Some details on the choice
of the theoretical error are given in Appendix~\ref{app:TEenvelope}.
Some details on the fingers-of-God modeling
for dark matter in redshift space
are given in Appendix~\ref{app:fog}.
Finally, Appendix~\ref{app:fid}
presents tests of the stability of our
constraints w.r.t. the choice of 
fiducial cosmology used to calibrate 
the theoretical error.

\section{Theoretical error likelihood}
\label{sec:theory}

In this section we review the theoretical error formalism. 
Let us first repeat the derivation of Ref.~\cite{Baldauf:2016sjb}
focusing on the power spectrum likelihood,
\be
\label{eq:L0}
-2\ln \mathcal{L} = C^{-1}_{ij}(P^{\rm th}_i-P^{\rm d}_i)(P^{\rm th}_j-P^{\rm d}_j)\,,
\ee
where 
the sum over $(i,j)$ is 
implicitly assumed, 
$C_{ij}$ is the data covariance matrix, $P^{\rm th}$ is the true theoretical model
and $P^{\rm d}$ is the datavector. 
In practice, the true theoretical model is not known and we use an 
approximate model $P^{\rm approx}_i$, e.g. the one-loop perturbation theory, which 
becomes less and less accurate on small scales. Theoretical error is the difference between the (unknown) true theory and this approximation,
\be
P^{\rm TE}_i \equiv P^{\rm th}_i - P^{\rm approx}_i\,.
\ee
Let us assume that $P^{\rm TE}_i$ is drawn from a Gaussian distribution with mean $\bar P^{\rm TE}_i$
and the covariance $C^{({\rm E})}$, 
\be
\mathcal{P}(P^{\rm TE})\propto 
\exp\left\{-\frac{1}{2}(C^{(E)})_{ij}^{-1} (P^{\rm TE}_i-\bar P^{\rm TE}_i)(P^{\rm TE}_j-\bar P^{\rm TE}_j)\right\} \,.
\ee
Marginalizing our original likelihood \eqref{eq:L0}
over the theoretical error we get the following marginalized likelihood:
\be
\label{eq:L1}
\begin{split}
 -2\ln \mathcal{L}_{\rm marg} & = (C^{({\rm tot})})^{-1}_{ij}(P^{\rm approx}_i+\bar P^{\rm TE}_i-P^{\rm d}_i)(P^{\rm approx}_j+\bar P^{\rm TE}_j-P^{\rm d}_j)\,,\\
 C^{({\rm tot}) } & = C + C^{({\rm E})}\,.
\end{split}
\ee

It is instructive to consider a few illustrative examples. 
First, let us assume that the theoretical error covariance 
is diagonal, $C^{({\rm E})}_{ij}\propto E^2_i\delta_{ij}$, the mean $ \bar P^{\rm TE}_i  = 0$, and 
\be
E_i= 
\begin{cases}
    \infty,& \text{if } k \geq k_{\rm max} \\
    0,              & \text{otherwise}
\end{cases}\,.
\ee
With this choice the bins with $k \geq k_{\rm max}$ are thrown away from the likelihood 
and the bins with $k < k_{\rm max}$ are assumed to have no theoretical error.
This limit corresponds to the standard analysis with fixed $k_{\rm max}$.

\paragraph{Theoretical error from analytic nuisance parameter marginalization.}
Let us now take a look at the opposite extreme, where the theoretical error is correlated in all $k$ bins.
A simple example is given by
\be
\label{eq:pte}
P^{\rm TE}_i = \alpha \left(\frac{k_i}{k_{\rm NL}}\right)^4 P_{\rm lin}(k_i)\equiv \alpha \mathcal{E}_i\,.
\ee
whose coefficient is expected to be Gaussian-distributed as
\be 
\label{eq:aprior}
\alpha \sim \mathcal{N}(\bar \alpha , \sigma^2_{\alpha})\,.
\ee
Such form of the TE as well as the prior on $\alpha$ can be
motivated either from perturbation theory (as higher derivative counterterms) or 
from N-body simulations. Let us now marginalize over $\alpha$
just like one normally marginalizes over nuisance parameters in 
realistic cosmological analyses.
The marginalized Gaussian likelihood can be easily obtained from \eqref{eq:L0},
\be 
\label{eq:L2}
\begin{split}
-2\ln \mathcal{L}^{\a}_{\rm marg} = 
(& C^{({\rm tot})})^{-1}_{ij}(P^{\rm approx}_i+\bar\alpha \mathcal{E}_i-P^{\rm d}_i)
(P^{\rm approx}_j+\bar\alpha \mathcal{E}_j-P^{\rm d}_j)\,,\\
 C^{({\rm tot}) }_{ij} &= C_{ij} + \sigma^2_{\alpha}\mathcal{E}_i \mathcal{E}_j\,,
\end{split}
\ee
where we have used the Sherman–Morrison identity.
We can see that in this case, since the $k$-dependence of the TE is entirely fixed,
the new covariance is fully correlated even if the data covariance $C_{ij}$
is diagonal. We stress that so far our calculation has been exact, i.e.~up to marginalization over $\alpha$,
the likelihood \eqref{eq:L2}
contains the same information as the original likelihood \eqref{eq:L0}. 

\paragraph{Theoretical error model.}
In reality, the exact $k$-dependence of the theoretical error is not known (otherwise it would be included 
in the theory model), but we want to marginalize over it just like we did with $\alpha$.
In other words, we want to marginalize over all possible curves within some bounds given by the 
expected size of the theoretical error and with sufficient degree of smoothness.
This can be achieved choosing an appropriate TE covariance matrix.
Following~\cite{Baldauf:2016sjb}, we use the ansatz
\be 
\label{eq:CE}
C^{({\rm E})}_{ij}=E_i E_j e^{-\frac{(k_i - k_j)^2}{2\Delta k^2}}\,,
\ee 
where we introduced finite coherence scale $\Delta k$, whereas $E_i$ is some $k$-dependent smooth envelope
for the theoretical error. Note that we neglect the cosmology-dependence 
of $C^{({\rm E})}$ in the same way as it is customarily done for the data covariance~\cite{Tegmark:1997rp} (see \cite{Tegmark:1996qt} for the justification of this practice for the CMB).
The coherence scale ensures that the neighboring bins at 
distance smaller than $\Delta k$ are almost fully correlated, but the 
allowed theoretical uncertainties can freely vary only on separations larger than $\Delta k$.   
Choosing the appropriate limits, we can recover the two previous examples. 
In the limit $\Delta k \to \infty$ all $k$ bins are correlated and 
the theoretical error corresponds to adding 
the shape $E(k)$ to the theory model and marginalizing over its amplitude.
In the opposite limit $\Delta k \to 0$ all bins can fluctuate independently and 
the theoretical error becomes diagonal as in our first example. 
Note that this limit is unphysical because it allows theoretical uncertainty to have arbitrarily 
fast oscillations and it makes the result dependent on the binning.

We will proceed with the following choice for the coherence scale
\be
\label{eq:dk}
\Delta  k =0.1~h/\text{Mpc}\,.
\ee
On the one hand, it is small enough to allow for typical variations of nonlinear corrections 
within the usual range of scales used in the data analysis. On the other hand,
it is big enough to ensure that the broadband theoretical error is correlated on 
scales corresponding to the frequency of the BAO wiggles, $k_{\rm BAO}\sim 0.01~h$/Mpc. 
This can help extract the information from the BAO even if the envelope $E_i$
is large.

Indeed, the theoretical error with this coherence length will effectively discard the broadband information,
but will retain the 
oscillatory features like the BAO wiggles~\cite{Philcox:2020vvt}.
Therefore, while increasing the total covariance, the theoretical error can possibly {\em improve}
the constraints on cosmological parameters by taking into account the 
information on the smoothness of $E_i$ and exploiting any additional features.

Once the coherence length is fixed, the envelope and the mean of the theoretical error have 
to be chosen as well. This is the most difficult task, since it strongly depends on the exact
model being used and the observable being analyzed. In the context of perturbation theory a reasonable guess is $\bar{P}_i = 0$ and the envelope can be estimated as the typical size 
of higher loop corrections not included in the model~\cite{Baldauf:2016sjb}. 
This theory-inspired estimate 
does not use any input from simulations or data, but, strictly speaking,
it is expected to be 
accurate only on 
the
order-of-magnitude basis.
Indeed, we will see that
the perturbation theory-inspired theoretical error can overestimate the actual
uncertainty in real space by a factor of few.
The perturbation
theory 
estimate
is even less reliable
in redshift space,
and can significantly underestimate the true 
theoretical uncertainty, particularly for higher order multipoles. We illustrate this issue 
in detail in Appendix~\ref{app:TEenvelope}.

In order to avoid making some overly optimistic or pessimistic choices in our analyses, we choose an alternative, simulation-driven strategy to estimate the mean and the envelope of 
the theoretical error. 
The main idea is to estimate the TE covariance from the difference between the data and the best-fit model inferred from an analysis 
based on large scales. 
We assume that the typical size of the envelope 
is equal to the TE mean,
\be
\label{eq:EP} 
E_i=\bar{P}^{({\rm TE})}_i\,,
\ee
in which case the theoretical error is fully characterized by a single shape $\bar{P}^{({\rm TE})}(k)$.
Eq.~\eqref{eq:EP} is motivated by the example of the counterterm marginalization~\eqref{eq:L2}, where the mean and the variance of the theoretical error have identical shapes.

\paragraph{Practical realization.}
In practice, we suggest the following
concrete
algorithm to estimate the theoretical error and 
construct the likelihood:

\begin{enumerate}
    \item Choose 
    some fiducial cosmological model.
    Given the real of mock N-body data, we compute the theory prediction for 
cosmological parameters fixed to some fiducial values. 
This step is similar, in spirit,
to choosing a fiducial model
for the statistical covariance matrix.
\item 
Select some fiducial 
data cut $\kmax^{\rm fid.}$.
This data cut
should be reasonably small, 
such that the theoretical error is negligible for this data cut.
We use the following procedure.
Having fixed some fiducial cosmology in the first step, we
minimize the usual 
power spectrum likelihoods 
w.r.t.~nuisance parameters
for a set of $\kmax$ and plot the best-fit reduced $\chi^2$ statistic as a function of $\kmax$. The corresponding profiles are flat up to a certain
scale $k_{\rm 2-loop}$, at which they start exhibiting 
significant scale-dependence. 
This is the scale at which the two-loop corrections become 
important. 
Any choice of $\kmax^{\rm fid.}$ that is lower than $k_{\rm 2-loop}$ is suitable 
for our purposes.
In practice, we use 
$\kmax^{\rm fid.}=k_{\rm 2-loop}-\Delta k'$, where 
$\Delta k'=0.04~\hMpc$.
We have tested our procedure by varying $\kmax^{\rm fid.}$
and found that it does not have any significant effect on the results.
The details about the choice of the data cuts $\kmax^{\rm fid.}$ can be found in Appendix~\ref{app:TEenvelope}.
\item 
Obtain a fiducial 
theoretical spectrum at
$\kmax^{\rm fid.}$.
To that end we fit 
the power spectrum data varying
{\em only} the nuisance parameters at our fiducial $\kmax^{\rm fid.}$. This analysis is very fast, since it requires computing a single best-fit point involving only the nuisance 
parameters. As a result
we obtain the best-fit 
theoretical
curve
$P^{\rm best-fit}(k)$.
\item Take the best-fit theory curve and compute
the theoretical curve envelop:
\be 
{P}^{({\rm TE})}_i=P^{\rm d}_i - P^{\rm best-fit}(k_i)\,.
\ee
This curve may have some stochastic scatter induced by the data vector $P^{\rm d}_i$.
To remove it, one can fit ${P}^{({\rm TE})}_i$ with a smooth polynomial. We 
will call this smooth 
curve $\bar{P}^{({\rm TE})}_i$.
\item 
Construct the TE likelihood
using $\bar{P}^{({\rm TE})}_i$
as follows:
\be 
-2\ln\mathcal{L}(P(\vec\theta))
=(C+C^{\rm (E)})^{-1}_{ij}
(P(\vec\theta)+\bar{P}^{({\rm TE})}_i-P^{\rm d}_i)
(P(\vec\theta)+\bar{P}^{({\rm TE})}_i-P^{\rm d}_j)\,,
\ee 
where $\vec\theta$
is the vector of cosmological parameters that we want to fit and
\be 
C^{\rm (E)}_{ij}=
\bar{P}^{({\rm TE})}_i
\bar{P}^{({\rm TE})}_j e^{-\frac{(k_i - k_j)^2}{2\Delta k^2}}\,,
\quad \text{with}\quad
k=0.1~h/\text{Mpc}.
\ee 
\end{enumerate}

An important comment is in order.
The key ingredient of our 
algorithm 
is the theoretical error envelope
that depends on the three particular choices:
the fiducial data cut 
$\kmax^{\rm fid.}$, the fiducial cosmological 
model, and the mock (or real)
data points.
In principle,
one has to verify that the final
result is stable w.r.t. 
these three choices.
This can be done 
by iterating the fiducial
cosmology to match 
the best-fit output spectra,\footnote{
In the context of the statistical 
error, this is a common practice 
in photometric and spectrosopic 
surveys, see~\cite{Wadekar:2020rdu}
and references therein.
}
and by varying 
$\kmax^{\rm fid.}$
and simulated data points.
However, all these three choices are correlated 
and 
effectively they
have the same result --- the change
of the theoretical error envelope.
Hence, essentially one only needs
to verify 
that the final results 
do not vary much 
as the
theoretical error envelope
is changed within some 
reasonable range.
We have performed these
consistency checks 
focusing on the variation
of the fiducial cosmology and $k_{\rm max}^{\rm fid.}$.
This test has shown that 
constraints on some cosmological parameters are 
affected by the choice of theoretical error envelope
in the unrealistic cases of pure dark matter clustering, 
while 
the realistic 
redshift-space galaxy power spectrum results 
are fully 
consistent and stable w.r.t. the variation of the theoretical error envelope.

Finally, it is worth pointing out  that the most robust 
strategy is to calibrate the theoretical error from simulated mock catalogs of a given survey 
and use it in all analyses of the real data.
Unlike the statistical uncertainty, the theoretical error covariance does not scale with volume or shot noise and hence needs to be calibrated only once for a particular tracer and redshift bin.
Besides, in this case one 
can include the scatter produced by
variations of N-body simulation
parameters in the theoretical error budget, which is required 
in order to ensure 
the stability of the 
final cosmological constraints.

\paragraph{The choice of 
theoretical cross-covariance 
for
redshift-space multipoles.}
So far all our formulae were written for the simplest case of the real space power spectrum analysis.
The treatment of the theoretical error becomes slightly more complicated in redshift space.
In the plane-parallel (flat sky) approximation, the redshift space power spectrum is the function of two variables, the wavenumber $k$ and the cosine 
between the wavevector $\k$ and the unit line-of-sight direction $\hat{\z}$, 
denoted by $\mu$,
\be 
\mu \equiv {(\k\cdot \hat{\bf z})}/{k}\,.
\ee
The redshift space power spectrum can be conveniently cast in the form of Legendre multipoles $L_\ell$,
\be
P (k,\mu) = \sum_{\ell = 0}P_\ell (k){L}_\ell(\mu) \,,\quad 
P_\ell \equiv \frac{2\ell+1}{2}\int_{-1}^{1}d\mu \, {L}_\ell(\mu) P (k,\mu) \,.
\ee 
where $\ell=0,2,4$ are
monopole, quadrupole, and hexadecapole moments, respectively.


We need to implement the condition that the theoretical error is a smooth function
both in $k$ and $\mu$. 
If the theoretical error curve is a single 
fixed function of $k$ and $\mu$, and only its amplitude is unknown, an explicit marginalization
over this amplitude (similar to Eq.~\eqref{eq:L2})
suggests that the theoretical error covariance is fully correlated across $\mu$ and $k$ bins.
At the level of the Legendre multipoles, this means that the theoretical error multipoles are $100\%$ correlated. 
However, since the precise shape of the theoretical error is not known \textit{by definition}, 
we will impose some finite correlation between the TE multipoles
using the model similar to that describing the correlation of $k$-bins,
\be 
\label{eq:CEl}
C^{({\rm E})~(\ell \ell')}_{ij}=E^{(\ell)}_i E^{(\ell')}_j e^{-\frac{(k_i - k_j)^2}{2\Delta k^2}}e^{-\frac{(\ell - \ell')^2}{2\Delta \ell^2}}\,.
\ee 
where $i,j$ now run only over the $k$-bins. In what follows, we will use the minimal non-trivial multipole 
coherence length $\Delta \ell =2$. 

The appearance of $e^{-\frac{(\ell - \ell')^2}{2\Delta \ell^2}}$ in Eq.~\eqref{eq:CEl}
can be understood from the following argument. 
If the theoretical error in $(k,\mu)$ space were characterized by a single parameter, 
i.e.~in analogy with Eq.~\eqref{eq:pte} it were given by
\be
\label{eq:ptemu1}
P^{({\rm TE})}(k,\mu)=\alpha \left(\frac{k}{k_{\rm NL}} \right)^4 \mu^4 P_{\rm lin}(k)(b_1+f\mu^2)^2\,,
\ee
then the theoretical error covariance between the multipoles, obtained after marginalization over $\alpha$, would be $100\%$ correlated. The shape \eqref{eq:ptemu1} indeed appears at the next-to-leading order (NLO) in the Taylor expansion
of the redshift-space mapping~\cite{Ivanov:2019pdj}. However, if the theoretical error were characterized 
by an independent parameter for every single multipole ($L_{\ell}$ is the Legendre polynomial of order $\ell$), i.e. 
\be 
\label{eq:ptemu2}
P^{({\rm TE})}(k,\mu)=\alpha_0 \mathcal{E}_0(k_i){L}_0(\mu)+\alpha_2 \mathcal{E}_2(k_i){L}_2(\mu)+...\,,
\ee
then marginalizations over the nuisance parameters $\alpha_\ell$ would produce the effective theoretical error 
covariance that is diagonal in multipole numbers. Thus, having a finite coherence length in the multipole
space represents a sensible compromise between these two extreme situations.

\section{Methodology, simulations and covariances}
\label{sec:method}

In this section we discuss technical details of our analysis:
the simulation data, theoretical model and the covariance matrces.

\subsection{Theoretical model}
\label{subsec:theor}

We use one-loop cosmological perturbation theory to 
compute power spectra for dark mater and galaxies in real and redshift spaces. The details of our theoretical model 
can be found in Ref.~\cite{Chudaykin:2020aoj};
it includes IR resummation to capture the non-linear evolution of the BAO wiggles \cite{Senatore:2014via,Baldauf:2015xfa,Vlah:2015zda,Blas:2015qsi,Blas:2016sfa,Senatore:2017pbn,Ivanov:2018gjr} and UV counterterms that are required 
to account for the effects of short-scale dynamics, whose description is impossible within perturbation theory itself~\cite{Baumann:2010tm,Carrasco:2012cv,Perko:2016puo}.

\subsection{Data}

Throughout the paper we will use a suite of LasDamas Oriana simulations \cite{2009AAS...21342506M}, 
which consists of 40 boxes with $L=2.4~\text{Gpc}/h$ on a side, totaling 
the volume of $553~(\text{Gpc}/h)^3$. 
The cosmological parameters used to generate mock catalogs 
are\footnote{We also use the CMB monopole temperature $T_0=2.725$ K, which is required to set various normalizations in \texttt{CLASS}~\cite{Ivanov:2020mfr}.
}
$h=0.7$, $\Omega_m=0.25$, 
$\Omega_b=0.04$, $\sigma_8=0.8~(A_s=2.22\cdot 10^{-9})$, $n_s = 1$, and $\sum m_\nu = 0$.
The details of LasDamas simulation can be found at.\footnote{\href{http://lss.phy.vanderbilt.edu/lasdamas/overview.html}{
\textcolor{blue}{http://lss.phy.vanderbilt.edu/lasdamas/overview.html}}\,
}
To reduce the stochastic scatter, we fit the mean of power spectra extracted from 
40 independent simulation boxes.
However, the statistical error corresponding to the total simulation volume is so small that the 
two loop corrections and inaccuracies of N-body modeling can supersede cosmic variance already on very large scales. 
In order to be more realistic,
we will use the covariance that corresponds to $V=100~(\text{Gpc}/h)^3$ and not to the actual LasDamas volume.
This reduced volume is comparable to the total volume of future spectroscopic surveys like DESI~\cite{Aghamousa:2016zmz} or Euclid~\cite{Chudaykin:2019ock}, which allows one to interpret our results in the context of these surveys.

We will analyze the statistics of dark matter particles and galaxies in real and redshift space.
The redshift snapshots available to us are $z=0$ and $z=0.974$ for dark matter, and $z=0.342$ for galaxies.
The galaxy distribution is generated using the HOD model that matches the luminous red galaxy (LRG) sample of the BOSS survey~\cite{Alam:2016hwk},
with the shot noise level 
\be
\label{ng}
\bar{n}^{-1} = 1.05906\times 10^4 ~[\text{Mpc}/h]^3\,.
\ee
The theoretical non-linear spectra are generated with the \texttt{CLASS-PT} code~\cite{Chudaykin:2020aoj} that computes
1-loop perturbation theory integrals using the FFTLog method \cite{Simonovic:2017mhp}. 
We run our MCMC chains with the 
\texttt{Montepython} sampler~\cite{Audren:2012wb,Brinckmann:2018cvx}, 
analyze these chains and produce triangle plots with the \texttt{GetDist} package~\cite{Lewis:2019xzd}.
In all analyses of this paper, we fit the following set of cosmological parameters:
\be 
\{\omega_{cdm},h,A^{1/2},n_s\}\,,\quad A\equiv {A_s}/{A_{s,{\rm fid.}}} \;,
\ee 
and also present the derived parameters $\Omega_m$ and $\sigma_8$. We fix $\omega_b$
to the fiducial value used in simulations. This is done in order to simulate the $\omega_b$
prior that can be taken from BBN or Planck in a realistic analysis~\cite{Ivanov:2019pdj}.

\subsection{Statistical covaraince}
\label{subsec:covmat}

We use the covariance matrix in the Gaussian approximation to 
describe sample variance~\cite{Scoccimarro:1999kp}. This approximation is very accurate 
for the power spectrum at least for scales $\kmax\leq 0.4~\hMpc$~\cite{Wadekar:2020rdu}.
The Gaussian covariance for the real space power spectrum takes the following form 
for the matter-matter, galaxy-galaxy power spectra ($P_{\rm mm}$ and $P_{\rm gg}$) and galaxy-matter cross-spectrum
($P_{\rm gm}$), respectively,
\be
\label{Creal}
\begin{split}
& C[P_{\rm mm}(k),P_{\rm mm}(k')]= \frac{2}{N_k}P^2_{\rm mm}(z,k)\delta_{kk'}\,,\\
& C[P_{\rm gg}(k),P_{\rm gg}(k')]= \frac{2}{N_k}\left(P_{\rm gg}(z,k)+\frac{1}{\bar n(z)}\right)^2\delta_{kk'}\,,\\
& C[P_{\rm gm}(k),P_{\rm gm}(k')]= \frac{2}{N_k}\left(P_{\rm gg}(z,k)+\frac{1}{2\bar n(z)}\right)P_{\rm mm}(z,k)~\delta_{kk'}\,,
\end{split}
\ee
where 
$N_k =V {4\pi k^2dk}/(2\pi)^3$ is the number of modes in a $k$-bin of size (\mbox{$dk=0.0025~h$/Mpc} in all our analyses) and $\delta_{kk'}$ is the Kronecker delta. 
In practice, we use the measurements $P_{\rm mm}$ and $P_{\rm gm}$ in the covariance matrix estimates~\eqref{Creal}.

The Gaussian covariance between the redshift space galaxy multipoles is given by~\cite{Chudaykin:2019ock},
\be
\label{Crsd}
\begin{split}
	C_{k_i k_j}^{(\ell_1 \ell_2)}=\frac{2}{N_k}(2\ell_1+1)\left(2\ell_2+1\right)
	\int_{0}^1d\mu \,L_{\ell_1}(\mu)L_{\ell_2}(\mu)\left[P(k_i,\mu)+\frac{1}{\bar n}\right]^2
	\delta_{ij}.
\end{split} 
\ee
Expressions \eqref{Creal}, \eqref{Crsd} are valid for both dark mater and galaxy statistics. It is worth noting that the galaxy power spectrum should be considered along with shot noise term \eqref{ng} which accounts for the discrete nature of galaxies. In practice, we evaluate \eqref{Crsd} in the linear (Kaiser) approximation \cite{Kaiser:1987qv}. 

All analyses for sharp data cuts are performed with the usual 
statistical covariances. To run the analyses with the theoretical error, 
we supplement the statistical covariances with the theoretical ones, \eqref{eq:CE} in real space,
\eqref{eq:CEl} in redshift space.

\section{Dark matter in real space}
\label{sec:dm_rs}

We start by analyzing the dark matter power spectrum in real space. To that end we use the one-loop
IR-resummed template of \cite{Blas:2016sfa} with the addition of 
the effective dark matter sound speed counterterm $\gamma$~\cite{Baumann:2010tm,Carrasco:2012cv}, 
which will be the only nuisance parameter in our analysis,
\be 
\label{eq:dmreal}
P=P_\tree+P_\lp^{\rm SPT}-2\gamma k^2 P_{\tree}(k)\,,
\ee
where $P^{\rm SPT}_{\rm 1-loop}$
is the one-loop 
correction computed 
in standard perturbation theory (SPT)~\cite{Bernardeau:2001qr}.

\begin{figure}[h!]
	\begin{center}
		\includegraphics[width=0.49\textwidth]{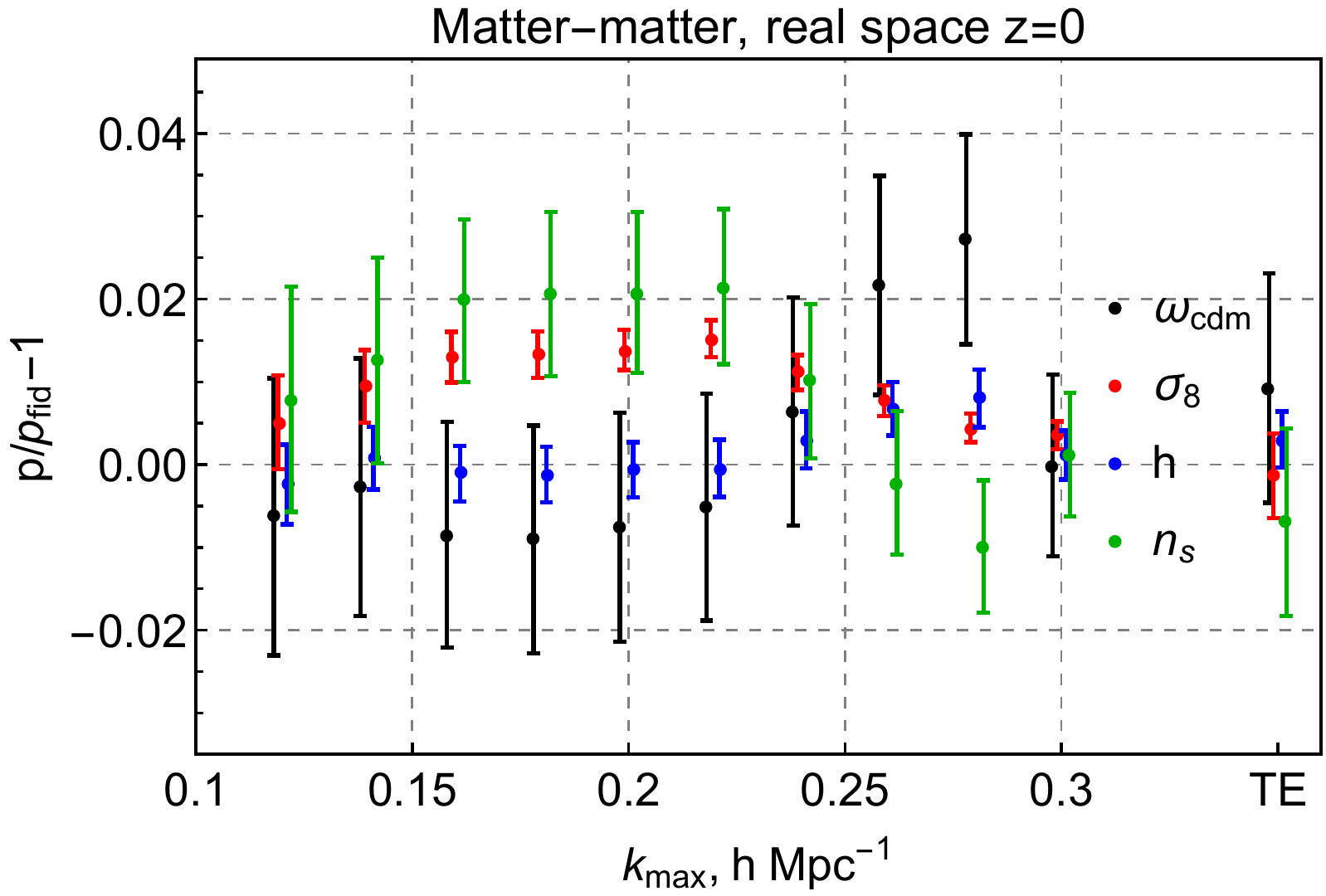}
		\includegraphics[width=0.49\textwidth]{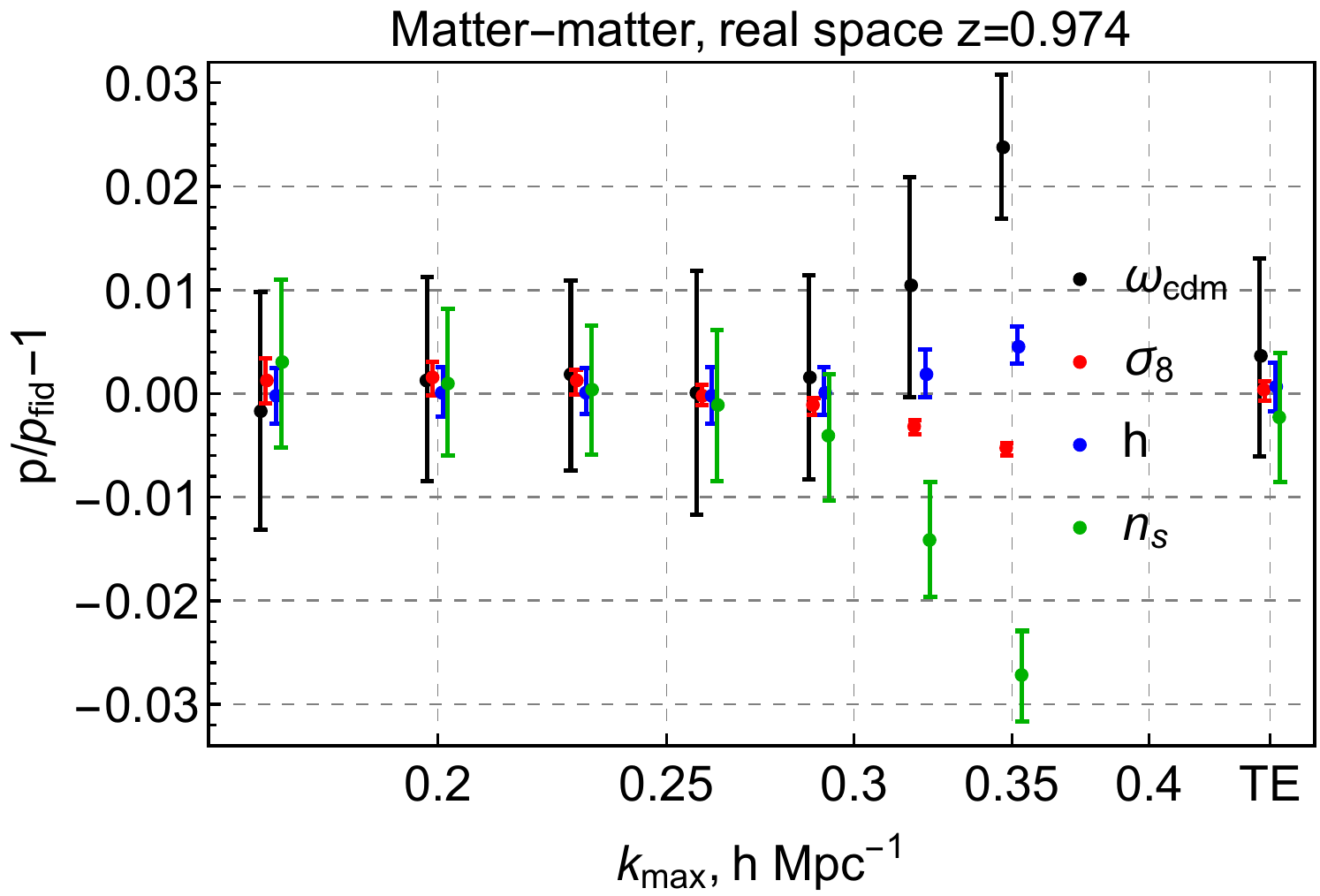}
		\end{center}
	\caption{\label{fig:rsdmz0ps}
		Marginalized 1d limits for cosmological parameters from the matter power spectrum at $z=0$ (left panel) and $z=0.974$ (right panel) as a function of $\kmax$. The rightmost group of points in each panel corresponds to the TE analysis.
		All parameters are normalized to their fiducial values.
	}
\end{figure}

As a first step, we fit the datavectors with sharp cuts $k_{\rm max}$ and pure statistical covariance. 
As a second step, we add the theoretical error to the analysis. 
The results in this case do not depend on $k_{\rm max}$ provided that it is reasonably high.
In practice, we cut the datavector at $k= k_{NL} \approx 0.3\,h$/Mpc and $0.5~h/$Mpc for $z=0$ and $z=0.974$, respectively.
The results for the marginalized 1d variances of cosmological parameters 
as functions of $k_{\text{max}}$ are shown in Fig.~\ref{fig:rsdmz0ps}. Note that we have run a complete MCMC analysis for every $\kmax$ in this plot.
The rightmost points correspond to the theoretical error analysis. Let us begin with the $z=0$ case. 
One can see from the left panel of Fig.~\ref{fig:rsdmz0ps} 
that the measured optimal parameters start deviating 
from the fiducial ones at $\kmax=0.12~h/$Mpc. 
But starting from $k_{\rm max}=0.24\,h$/Mpc, the estimates start moving in the opposite direction and accidentally 
become fully compatible with the fiducial cosmology at $k_{\max}=0.3\,h$/Mpc.
Clearly, this behavior is caused by over-fitting and the constraints at $k_{\max}=0.3\,h$/Mpc
cannot be trusted even though they enclose the fiducial values within $1\sigma$.
Importantly, the error bars obtained at $k_{\max}=0.3\,h$/Mpc are much smaller than the ones we got before 
the fit started deviating from the truth at $k_{\max}=0.12\,h$/Mpc.
This illustrates the danger of the $\kmax$ approach, which can be significantly affected by over-fitting.
In what follows, we choose $k_{\text{max}}=0.12$ to be a baseline cut at $z=0$ as 
the estimated parameters match the fiducial cosmology within $1\sigma$ there, but 
$\sigma_8$ becomes biased 
by more than $2\sigma$ for larger $\kmax$.

Now we focus on the  $z=0.974$ case, whose results are shown in the right panel of Fig.~\ref{fig:rsdmz0ps}.
We see that $\sigma_8$ undergoes an excursion beyond $1\sigma$
when $\kmax$ is varied between $0.14~\hMpc$ and $0.3~\hMpc$, but accidentally crosses the fiducial value at 
$\kmax=0.28~\hMpc$. 
The error bars at $\kmax=0.14~\hMpc$ and $\kmax=0.28~\hMpc$ are significantly different,
but both measurements enclose the fiducial cosmology within $1\sigma$, hence the results depend on whether one wants to be more conservative or aggressive.
In practice, the second option is often more popular, thus we choose $k_{\text{max}}=0.28$~$h$/Mpc
as a baseline data cut at $z=0.974$.

\begin{figure}[ht!]
	\begin{center}
		\includegraphics[width=1\textwidth]{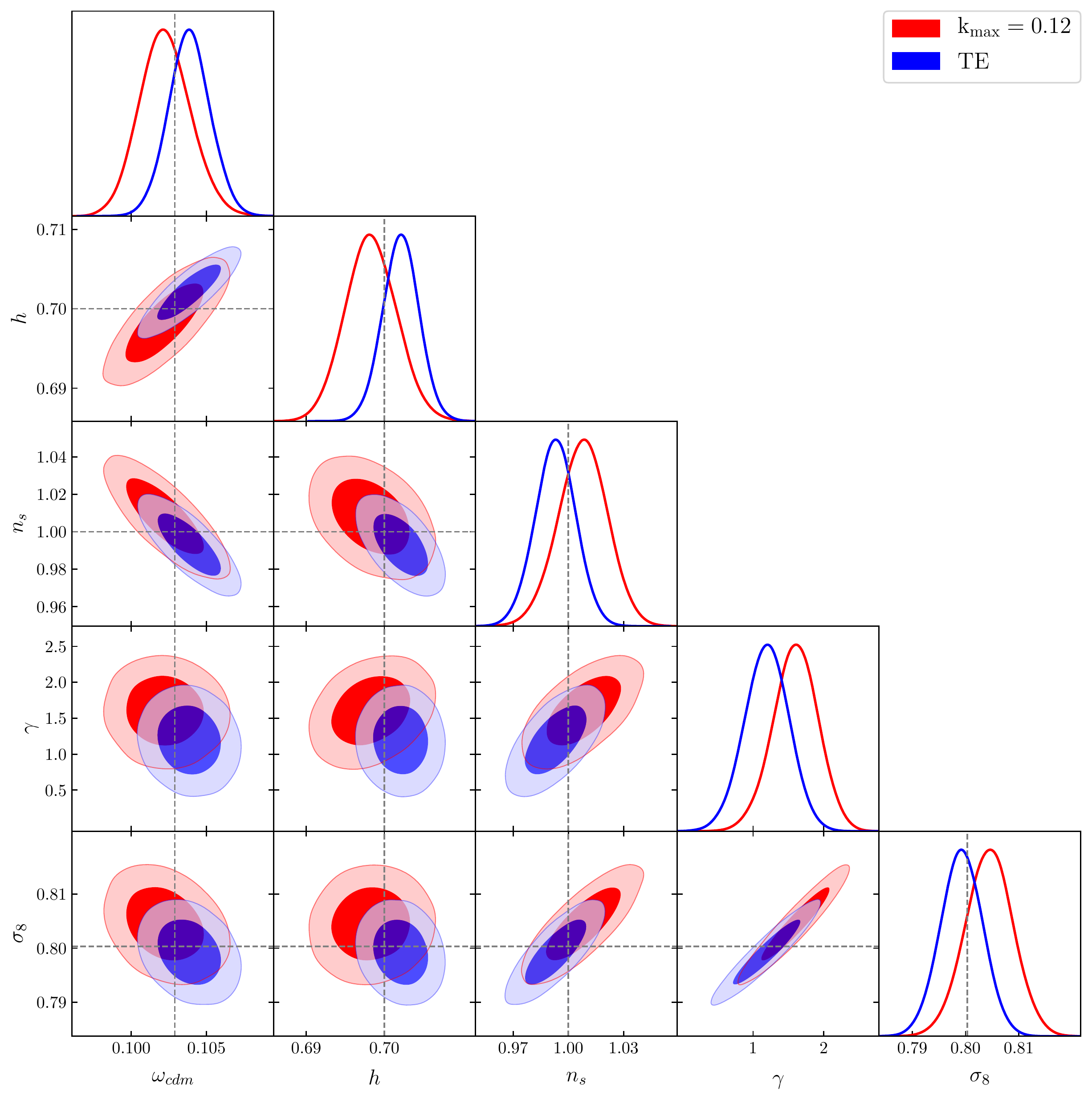}\end{center}
	\caption{\label{fig:rsdmz0}
		Triangle plot for the cosmological and nuisance parameters 
		measured from the real space dark matter power spectrum
		of the Las Damas simulations at $z=0$. 
	}
\end{figure}

Let us compare the baseline $\kmax$ results with the theoretical error analysis, focusing first on the $z=0$ case. 
The values of the marginalized 1d variances are displayed in Table \ref{tab:cos_dm_real}.
The marginalized 2d triangle plot for the $k_{\text{max}}=0.12$~$h$/Mpc and the theoretical error (TE) analysis are displayed in Fig.~\ref{fig:rsdmz0}.
The theoretical error envelope for TE analysis was computed
using a best-fit model at fiducial $\kmax^{\rm fid.}=0.10\,h$/Mpc (see Appendix~\ref{app:TEenvelope} for more detail).
\begin{table}[h!]
	\begin{center}
		\begin{tabular}{|c|c|c|c|c|c|}  \hline 
			\small{par.} & fid. & \multicolumn{2}{|c|}{$z=0$} & \multicolumn{2}{|c|}{$z=0.974$}\\  \hline
			&  & $k_{\text{max}}=0.12$ & TE & $k_{\text{max}}=0.28$ & TE \\  \hline\hline
			$\omega_{cdm}$ & $0.1029$  & $0.1023^{+1.6\cdot 10^{-3}}_{-1.8\cdot 10^{-3}}$   &
			$0.1039^{+1.4\cdot 10^{-3}}_{-1.5\cdot 10^{-3}}$ &
			$0.1033^{+1.0\cdot 10^{-3}}_{-1.1\cdot 10^{-3}}$  &
			$0.1033^{+9.4\cdot 10^{-4}}_{-1.0\cdot 10^{-3}}$ \\ \hline
			$h$ & $0.7$& $0.6983^{+3.3\cdot 10^{-3}}_{-3.4\cdot 10^{-3}}$ &
			$0.7021^{+2.4\cdot 10^{-3}}_{-2.4\cdot 10^{-3}}$ &
			$0.7004^{+1.6\cdot 10^{-3}}_{-1.8\cdot 10^{-3}}$&
			$0.7005^{+1.6\cdot 10^{-3}}_{-1.7\cdot 10^{-3}}$\\ \hline
			$n_s$ & $1$ & $1.008^{+0.014}_{-0.013}$&
			$0.9930^{+0.011}_{-0.011}$ &
			$0.9955^{+6.4\cdot 10^{-3}}_{-6.1\cdot 10^{-3}}$ &
			$0.9977^{+6.3\cdot 10^{-3}}_{-6.1\cdot 10^{-3}}$\\ \hline
			$A$ & $1$ & $1.014^{+0.020}_{-0.020}$ &
			$0.9897^{+0.016}_{-0.016}$ &
			$0.9972^{+9.7\cdot 10^{-3}}_{-9.4\cdot 10^{-3}}$ &
			$0.9977^{+9.4\cdot 10^{-3}}_{-9.0\cdot 10^{-3}}$\\ \hline
			$\Omega_m$ & $0.25$ & $0.2499^{+2.2\cdot10^{-3}}_{-2.3\cdot10^{-3}}$ &
			$0.2504^{+1.6\cdot10^{-3}}_{-1.7\cdot10^{-3}}$ &
			$0.2504^{+1.2\cdot10^{-3}}_{-1.2\cdot10^{-3}}$ &
			$0.2504^{+1.1\cdot10^{-3}}_{-1.2\cdot10^{-3}}$\\ \hline
			$\sigma_8$  & $0.8003$ & $ 0.8044^{+4.6\cdot 10^{-3}}_{-4.5\cdot 10^{-3}}$ &
			$0.7992^{+4.1\cdot 10^{-3}}_{-4.1\cdot 10^{-3}}$ &
			$0.7996^{+6.8\cdot 10^{-4}}_{-6.8\cdot 10^{-4}}$ &
			$0.8005^{+7.8\cdot 10^{-4}}_{-7.7\cdot 10^{-3}}$  \\  \hline\hline
			$\gamma$ & -- & $1.60^{+0.33}_{-0.31}$ &
			$1.19^{+0.33}_{-0.32}$ &
			$0.498^{+0.025}_{-0.024}$ &
			$0.535^{+0.029}_{-0.029}$\\ \hline
		\end{tabular}
		\caption{
			The marginalized 1d intervals for the cosmological parameters 
			estimated from the Las Damas real space dark matter power spectra at $z=0$ and $z=0.974$. 
			The table contains fitted parameters (first column), fiducial values used 
			in simulations (second column), and the results for two different redshifts: $z=0$ (third and fourth columns)
			and $z=0.974$ (fifth and sixth columns). 
			In either case we display the results of 
			the baseline $k_{\text{max}}$ analysis and the outcome of the theoretical error approach. $\gamma$ is quoted in units $[\Mpch]^2$.
		}
		\label{tab:cos_dm_real}
	\end{center}
\end{table} 
We see that the TE covariance helps us measure all cosmological parameters without any significant bias.
Moreover, there is a moderate improvement over the $\kmax$ analysis for all cosmological parameters of interest:
the errors on $\omega_{cdm}$, $h$ , $n_s$ and $\sigma_8$ reduce by
$20\%$, $30\%$, $20\%$, and $10\%$, respectively. 

It is important to test whether our results depend on the 
shape of the theoretical error envelope.
The details on this test can be found 
in Appendix~\ref{app:fid}.
Choosing a different 
fiducial cosmology to compute the theoretical error template, 
we have found that the constraints on $\omega_{cdm}$ and $h$
do no change, whereas the posteriors for $n_s$ and $\sigma_8$ shrink by $10\%$
and $25\%$, respectively.
While these 
changes are very marginal, they suggest
that the improvements that we have obtained for $n_s$ and $\sigma_8$
should be taken with a grain of salt
and in conjunction with our choice
of the theoretical error envelope.

Now we turn to the $z=0.974$ case, which is more relevant for upcoming surveys. 
The values of the marginalized 1d variances for the $k_{\text{max}}=0.28$~$h$/Mpc and the TE analysis are displayed in Table \ref{tab:cos_dm_real}.
The envelope for TE analysis was computed with $\kmax^{\rm fid.}=0.26\,h$/Mpc (see Appendix~\ref{app:TEenvelope} for more detail).
 We found that using the TE covariance matrix 
 leads to unbiased estimates of all cosmological parameters. 
 However, unlike the $z=0$ case, it does not noticeably improve the error bars compared to the baseline $k_{\text{max}}$ results. This happens mainly because the BAO wiggles are strongly suppressed for wavenumbers larger than $0.28$~$h$/Mpc~\cite{Blas:2016sfa}, thus going to larger wavenumbers with the TE does not yield as much new information as in the $z=0$ case since the baseline $\kmax$ is larger.
Moreover, our TE constraints on $\sigma_8$ are somewhat weaker (by $13\%$) than those from the baseline $k_{\text{max}}$ analysis, and the mean
value is not shifted from the true cosmology. 
\begin{figure}[ht!]
	\begin{center}
		\includegraphics[width=1\textwidth]{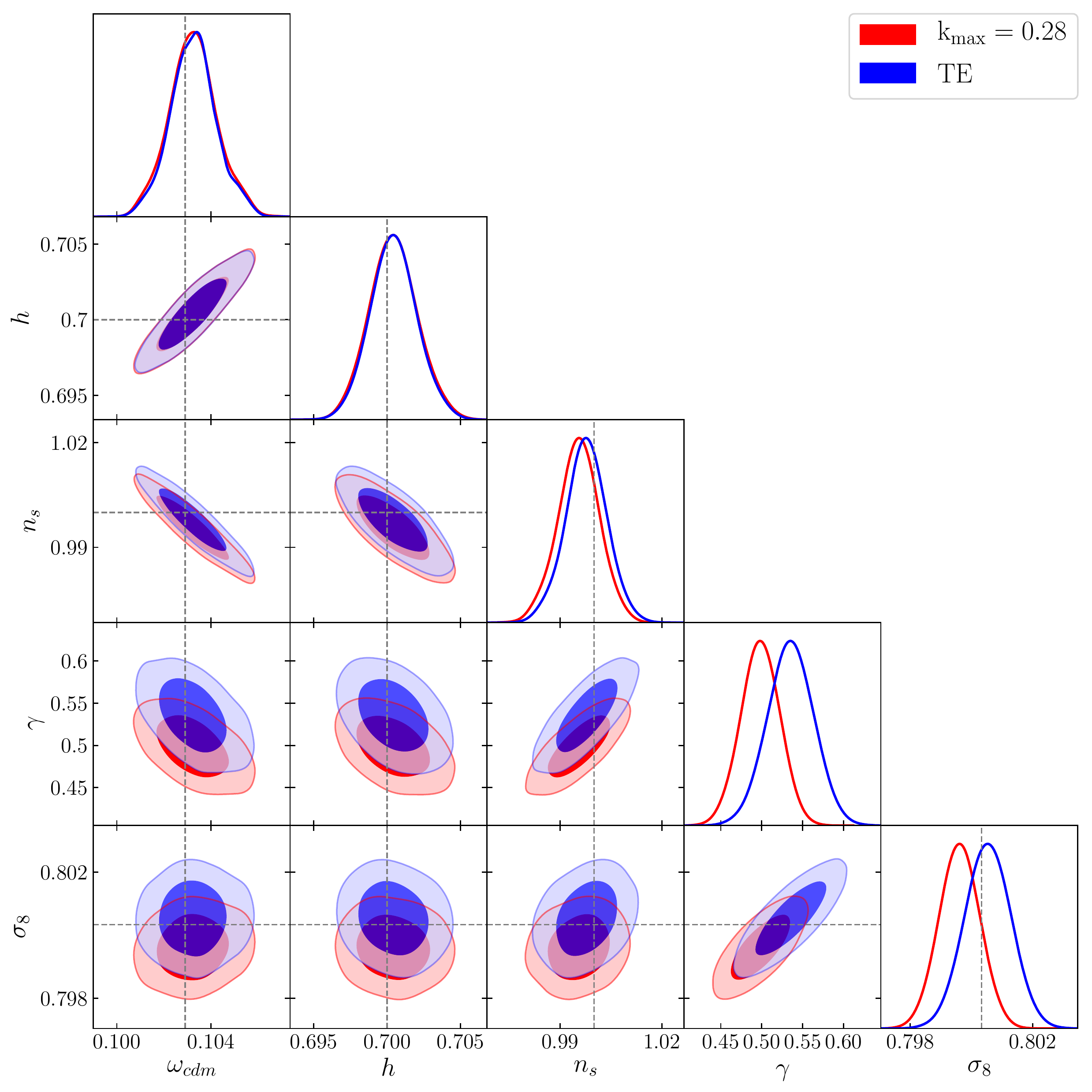}\end{center}
	\caption{\label{fig:rsdmz1}
		Triangle plot for the cosmological and nuisance parameters 
		measured from the real space dark matter power spectrum
		of the Las Damas simulations at $z=0.974$. 
	}
\end{figure}
From the marginalized 2d triangle plot shown in Fig. \ref{fig:rsdmz1}
one clearly sees the source of this worsening: the degeneracy between $\sigma_8$ 
and the counterterm $\gamma$ is broken less efficiently when the theoretical error is taken into account.
Recall that the measurements of the effective sound speed are very sensitive to the two-loop
corrections~\cite{Baldauf:2015aha}. 
If we ignore these corrections, the measurement of $\gamma$ will be biased and the 
error bars will be underestimated. 
Given the apparent degeneracy $\gamma-\sigma_8$, this propagates into the posterior of $\sigma_8$, 
whose width is underestimated too. 
In contrast, including the theoretical error allows one 
to marginalize over the two-loop contributions and get a more correct measurement of $\gamma$ and $\sigma_8$
with unbiased, albeit larger, error bars. This shows that the TE covariance is imperative in order to get 
accurate estimates of the parameter variances.

\section{Dark matter in redshift space}
\label{sec:dm_rsd}

In this section we study the clustering of dark matter in redshift space. 
In a real survey one always observes dark matter tracers (like galaxies) 
in redshift space. In this case, it is hard to disentangle between 
the nonlinear effects of redshift-space and the nonlinear galaxy bias. 
Thus, the case of redshift space dark matter will clearly show us how well perturbation
theory can describe nonlinearities induced specifically by redshift-space mapping. 
We will see that the theoretical error plays a crucial role in this analysis too,
and allows for noticeable improvement of the cosmological constraints.

Normally the analysis of the redshift space power 
spectrum 
is limited to the first three non-trivial moments $\ell=0,2,4$ because only these moments exist in linear theory.
We have found that 
the hexadecapole signal is dominated by a systematic leakage from lower moments due to discreteness effects. 
Given this reason, we perform our analysis only for the monopole and quadrupole moments.
We fit the IR-resummed one-loop power spectrum,
\be
\label{eq:RSDth}
P^{(z)}(k,\mu)=P_{\text{tree}}^{\rm IR-res}(k,\mu)+ P^{\text{SPT, IR-res}}_{\text{1-loop}}(k,\mu)+P_{\rm ctr}(k,\mu) \,,
\ee
where $P_{\rm ctr}$ denotes conterterms which fix the UV-dependence of the loops and parameterizes the ignorance about short scale-dynamics. This term also addresses the ``fingers-of-God'' effect~\cite{Jackson:2008yv}.
In what follows we describe this phenomenon perturbatively along the lines of the effective field theory.

The section has the following outline.
First, we validate the effective field theory treatment of fingers-of-God and compare it with popular phenomenological prescriptions in Sec.~\ref{subsec:fog}. Then, we present the measurements of cosmological parameters from the mock redshift-space power spectrum for the cases of  
the theoretical error approach and 
the sharp momentum cuts in Sec.~\ref{subsec:rsd_par}.

\subsection{Fingers-of-God modeling}
\label{subsec:fog}

The biggest challenge in the analytic description of redshift-space distortions is the fingers-of-God effect, induced by virialized motions of dark matter particles (or galaxies)
in halos. The virialized velocity field is fully non-linear and there is very little hope that it 
can be described analytically from first principles. However, the effect of fingers-of-God on long-wavelength 
fluctuations can be captured within effective field theory by a finite set of operators.
In particular, the lowest order corrections are given by~\cite{Senatore:2014vja,Lewandowski:2015ziq},
\be
\label{eq:RSDc0c2}
P_{\rm \nabla^2\delta}(k,\mu)=-2\left(c'_0 + c'_2f\mu^2 + c'_4f^2\mu^4 \right)k^2 P_{\text{lin}}(k)\,,
\ee
where $c_0$, $c_2$ and $c_4$ are called ``counterterms'' and $f\equiv d\ln D/d\ln a$ is the logarithmic growth factor. The values and time-dependence of counterterms are not known {\it a priori}, and hence we treat them as free parameters and marginalize over their amplitudes.

It is quite natural to keep three different free coefficients since they fix the UV-dependence 
of different loop diagrams and capture different physical effects. 
Namely, the monopole counterterm includes the contribution similar to the higher-derivative bias $\nabla^2\delta$ (along with the dark matter effective sound speed operator), which is absent for higher order multipole moments. 
In contrast, the quadrupole counterterm captures the dispersion of the short-scale velocity field. 
In what follows, we will justify that keeping independent coefficients in every counterterm is necessary for proper description of dark matter clustering on large scales.

The expansion \eqref{eq:RSDc0c2} assumes that the dimensionfull coefficients $c_i$ are small, 
\be
\label{eq:c0c2c}
c_i k^2 \ll 1\,,\quad i=0,2,4\,.
\ee
However, the peculiar velocities can be rather large, which can violate the condition of the applicability of perturbation theory \eqref{eq:c0c2c} even on large scales. 
For instance, the BOSS galaxies are expected to have the velocity dispersion \mbox{$\sim 6~\Mpc/h$},
which gives an estimate $c_2 \sim 50\,[\Mpc/h]^2$~\cite{Ivanov:2019pdj}. 
It implies that characteristic momentum scale of higher order short-scale velocity cumulants is significantly lower than the non-linear scale that controls gravitational nonlinearities in real space, $k_{\rm NL}\sim 0.5~\hMpc$ \cite{Baldauf:2016sjb}. 
Hence, the usual one-loop power spectrum model \eqref{eq:RSDth}, \eqref{eq:RSDc0c2} can become insufficient for an accurate description of the data even on large scales. We proceed by introducing an additional counterterm to capture the redshift space nonlinearities at next-to-leading order \cite{Ivanov:2019pdj},
\be
\label{eq:RSDb4}
P_{\rm \nabla^4_\z\delta}=-b_4(\mu k f)^4(b_1+f\mu^2)^2P_\lin(k)
\ee
where $b_4$ denotes
the next-to-leading counterterm and $b_1=1$ for the dark matter. 
The redshift-space mapping effectively generates an expansion in powers of $\nabla_z^2 \delta$. From this point of view, the extra counterterm can be viewed as $\nabla_z^4 \delta$ contribution in this expansion. We choose the next-to-leading order counterterm \eqref{eq:RSDb4} to be universal for all multipole moments, as dictated by the redshift-space mapping. We assume that the contributions from other physical effects (higher-derivative counterterms etc.) 
are sub-dominant since their nonlinear scale is similar to $k_{\rm NL}$ for the real space dark matter.

It is worth mentioning that 
in a realistic LSS data analysis
the redshift determination errors generate 
additional
corrections to the power spectrum.
For example, the redshift smearing of quasars from the eBOSS survey
is described by a 
$k^2\mu^2 P_{\rm lin}$-like counterterm
at leading order in
the derivative expansion
\cite{Neveux:2020voa}, 
but a more realistic model requires higher order corrections with at least one additional free parameter to capture deviations from Gaussianity~\cite{Smith:2020stf}.
An accurate modeling of redshift errors
beyond simplistic one-parameter models 
will also be important for the Euclud/DESI-type 
emission line galaxies, see Ref.~\cite{Chen:2020fxs} for an extended discussion.
This is another reason why it is important to have free 
coefficients in front of the 
various redshift-space counterterms: they also capture different systematic effects.

The full counterterm contribution is given by $P_{\rm ctr}=  P_{\rm \nabla^2\delta} + P_{\nabla^4_{\z} \delta} $. Doing the integrals with the Legendre polynomials, we get\footnote{We stress that this is a symbolic expression. 
Once IR resummation is included, it becomes impossible to write explicit close formulas for the Legendre integrlas. In practice, we use the full expressions Eq.~\eqref{eq:RSDc0c2} and Eq.~\eqref{eq:RSDb4} and  
evaluate all Legendre integrals numerically
with \texttt{CLASS-PT}~\cite{Chudaykin:2020aoj}.}
\be
\label{eq:RSDctr}
\begin{split}
& P_{{\rm ctr},~\ell=0}=  -2c_0 k^2 P_{\rm lin}(k) - b_4 f^4 
\left(\frac{f^2}{9}+\frac{2 f b_1}{7}+\frac{b_1^2}{5}\right) k^4 P_{\rm lin}(k) \,,\\
& P_{{\rm ctr},~\ell=2}=  -2c_2 \frac{2f}{3} k^2 P_{\rm lin}(k) - b_4 f^4
\left(
\frac{40 f^2}{99}+\frac{20 f b_1}{21}+\frac{4 b_1^2}{7}
\right)
 k^4 P_{\rm lin}(k) \,.
\end{split}
\ee
where $c_0$, $c_2$ and $b_4$ are free fitting parameters and $b_1$ is 
the linear bias term which we explicitly inserted in Eq.~\eqref{eq:RSDctr}
for the future reference when we study redshift-space galaxies. For dark matter $b_1=1$.
Note that, unlike galaxies, 
the lowest order stochastic contribution to the dark matter redshift-space power spectrum 
starts with the $k^4-$term~\cite{Lewandowski:2015ziq}, and hence, it is a higher order correction 
that can be ignored. The effects of these corrections are taken into account 
by the theoretical error.

The first goal of this section is to 
validate the inclusion of the next-to-leading order conterterm \eqref{eq:RSDb4} 
into the one-loop theoretical model \eqref{eq:RSDth}, \eqref{eq:RSDc0c2}. 
Along the way, we will test the accuracy of the fingers-of-God modeling with phenomenological 
fitting functions that are often used in the literature, see e.g.~\cite{Beutler:2016arn}.
These popular models approximate the fingers-of-God effect by a simple one-parameter Gaussian or Lorentzian damping, e.g.
\be
\label{eq:fog}
P^{\rm FOG}(k,\mu)=\e^{-(k\mu\Sigma f)^2}P_{\rm NL}(k,\mu)\,,
\ee
where $\Sigma$ can be interpreted as a  short-scale velocity dispersion.
We stress that the phenomenological models like \eqref{eq:fog} are not derived from first principles
and introduce uncontrollable errors in parameter inference. To show this, we analyze the 
the redshift-space dark matter power spectrum data using the model~\eqref{eq:fog}
instead of the full EFT description.

In the following analyses, for simplicity, we fix all cosmological parameters to their fiducial values except for $\sigma_8$, which is allowed to float freely.
This choice is done only for simplicity. 
All our conclusions hold true even if we vary all relevant cosmological
parameter.
We consider three models:

(a) the fitting function \eqref{eq:fog} applied to the one-loop SPT power spectrum, i.e.~$P_{\rm NL}(k,\mu)=P_{\rm SPT}(k,\mu)$,

(b) the vanilla one-loop EFT with free $c_0,c_2$, but $b_4=0$, 

(c) the one-loop EFT model with additional NLO correction with free $b_4$. 

Note that we implement the full IR resummation in case (a). 
In this setup the model (a) is a non-linear resummation of perturbative models (b) and~(c), which implies the following relationship between the counterterms,
\be
\label{eq:fogfix}
\begin{split}
& c_0 = \frac{1}{210} f^2 \Sigma ^2 \left(35 b_1^2+42 b_1 f+15 f^2\right)\,,\\
&  
c_2=\frac{1}{14} f \Sigma ^2 \left(7 b_1^2+12 b_1 f+5 f^2\right)\,,\\
& b_4= -\Sigma^4 /2\,.
\end{split}
\ee

Let us first discuss the $z=0$ snapshot.
The resulting posterior distribution for nuisance parameters and clustering amplitude for 
the three models at \mbox{$\kmax=0.1~\hMpc$}
are shown in Fig.~\ref{fig:rsdSigma_z0} (see Appendix~\ref{app:fog} for tables with 1d marginalized limits).
\begin{figure}[ht!]
	\begin{center}
		\includegraphics[width=1\textwidth]{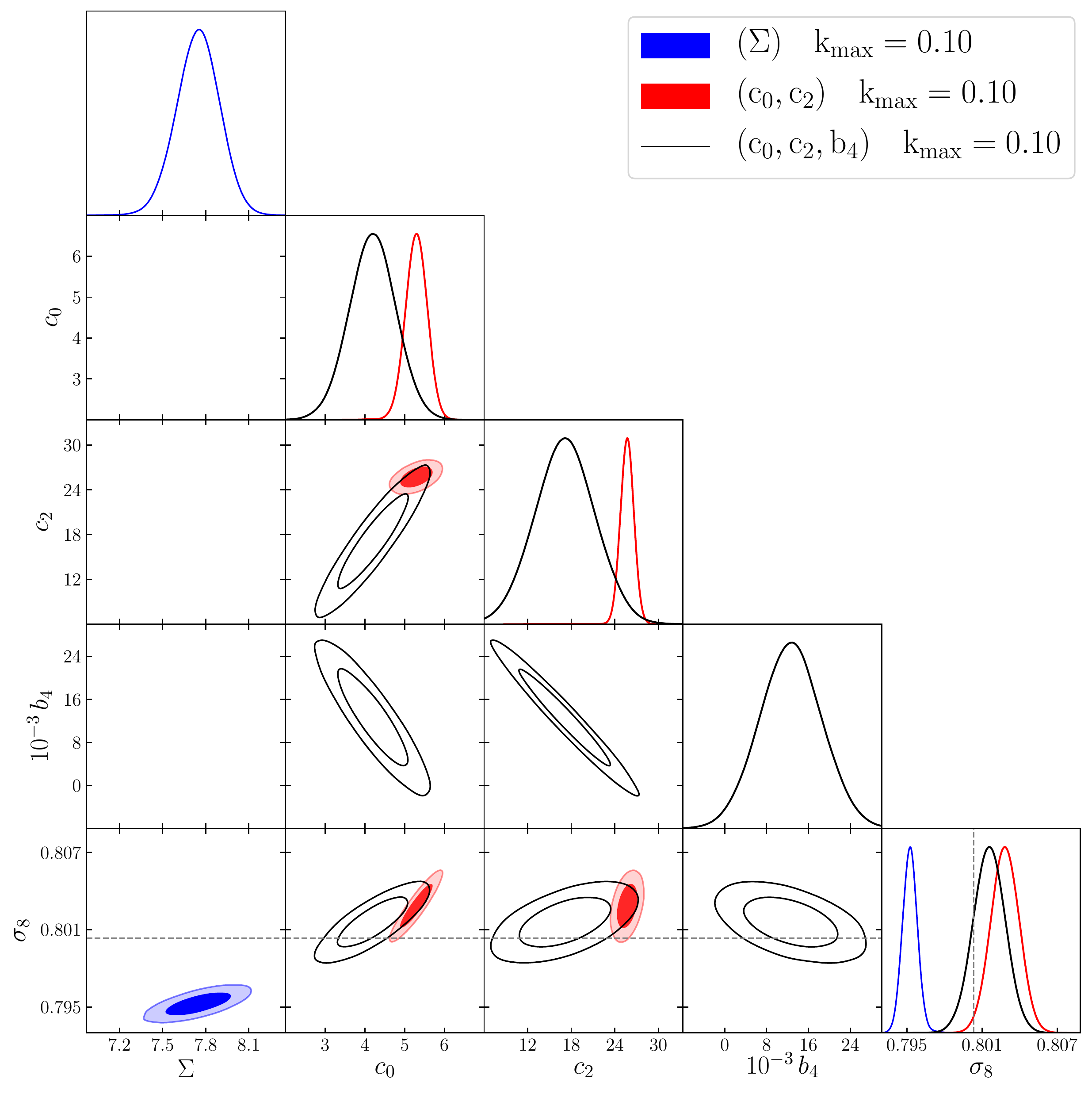}		
	\end{center}
	\caption{\label{fig:rsdSigma_z0}
		Triangle plot for conterterm normalizations and $\sigma_8$ measured from the redshift space dark matter power spectrum of the Las Damas simulations at $z=0$.
		$c_0,c_2,b_4$ are quoted in units $[\Mpch]^2,~[\Mpch]^2,~[\Mpch]^4$, respectively.
	}
\end{figure}
The first relevant observation is that 
the approximate model \eqref{eq:fog} biases $\sigma_8$ by more than $5\sigma$. 
However, varying $c_0$ and $c_2$ independently, one reduces the bias down to the $2\sigma$ level. 
This illustrates that the phenomenological model with a single parameter $\Sigma$ fails to reproduce the data 
because it does not have enough freedom to describe the monopole and quadrupole simultaneously. 
Thus, keeping free coefficients in different conterterms is a crucial part of any reliable analysis. 
The residual bias in $\sigma_8$ is removed by adding the next-to-leading order counterterm~\eqref{eq:RSDb4},
which justifies its presence in the data analysis. 
This counterterm, however, does not allow us to increase the $\kmax$ range, which 
may be a signal of perturbation theory breakdown.

\begin{figure}[ht!]
	\begin{center}
		\includegraphics[width=0.49\textwidth]{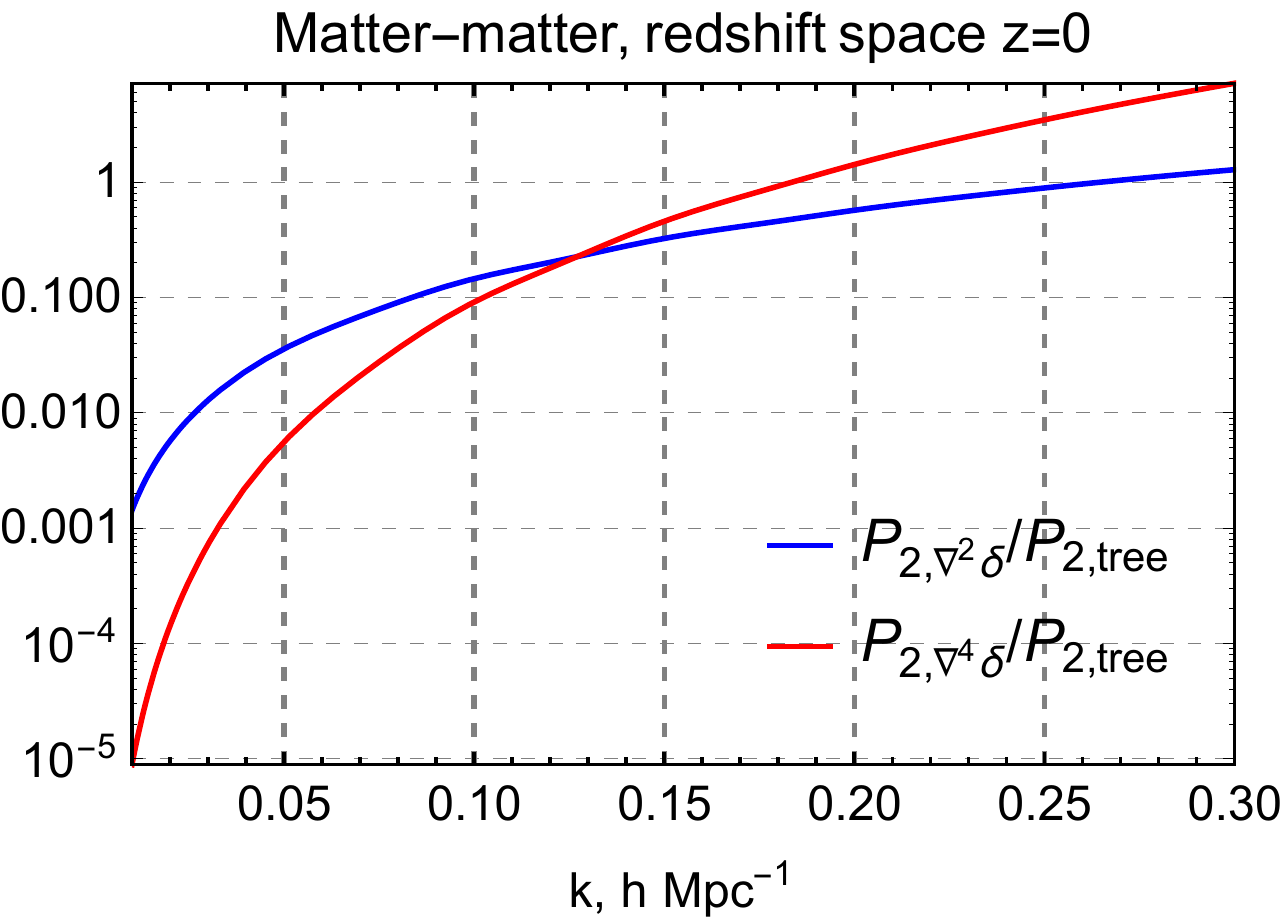}
		\includegraphics[width=0.49\textwidth]{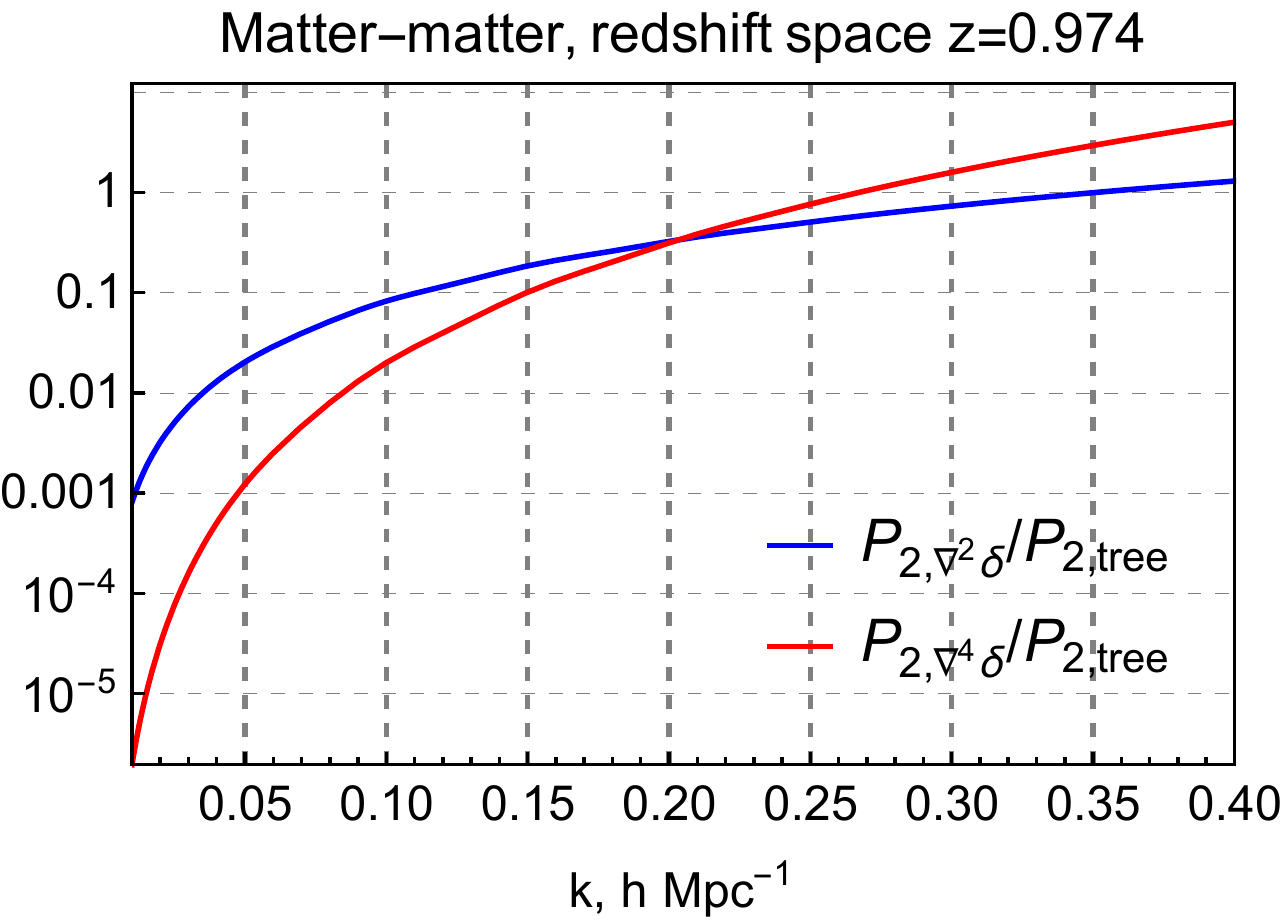}
	\end{center}
\vspace{-0.7cm}
	\caption{\label{fig:counters}
		The contribution of leading and next-to-leading order counterterm to quadrupole power spectrum for the best-fit parameters $c_2=17 \,[\Mpc/h]^2$, $b_4=1.3\cdot 10^4 \,[\Mpc/h]^4$ at $z=0$ (left panel) and $c_2=11 \,[\Mpc/h]^2$, $b_4=370 \,[\Mpc/h]^4$ at $z=0.974$~(right panel).
	}
\end{figure}

To demonstrate this, we perform the following exercise. 
We fix all cosmological parameters to their fiducial
values and extract the best-fit parameters $c_0$, $c_2$ and $b_4$ from the data at $\kmax=0.1~\hMpc$, 
where the fit is unbiased. 
Then we compare the contributions from leading and next-to-leading order counterterms of the quadrupole moment to the tree-level dark matter power spectrum evaluated for the best-fit parameters. 
We chose the quadrupole because this moment is most sensitive to the fingers-of-God.
The results are shown in the left panel
of Fig.~\ref{fig:counters}. One can see that the NLO contribution surpasses the LO counterterm at $k\approx 0.13~\hMpc$,
and the whole tree-level expression at $k\approx 0.18~\hMpc$.
This can be naturally interpreted as a breakdown of the perturbative description  
for the dark matter
in redshift space at $k\approx  0.15~\hMpc$ for $z=0$. 

\begin{figure}
	\begin{center}
		\includegraphics[width=1\textwidth]{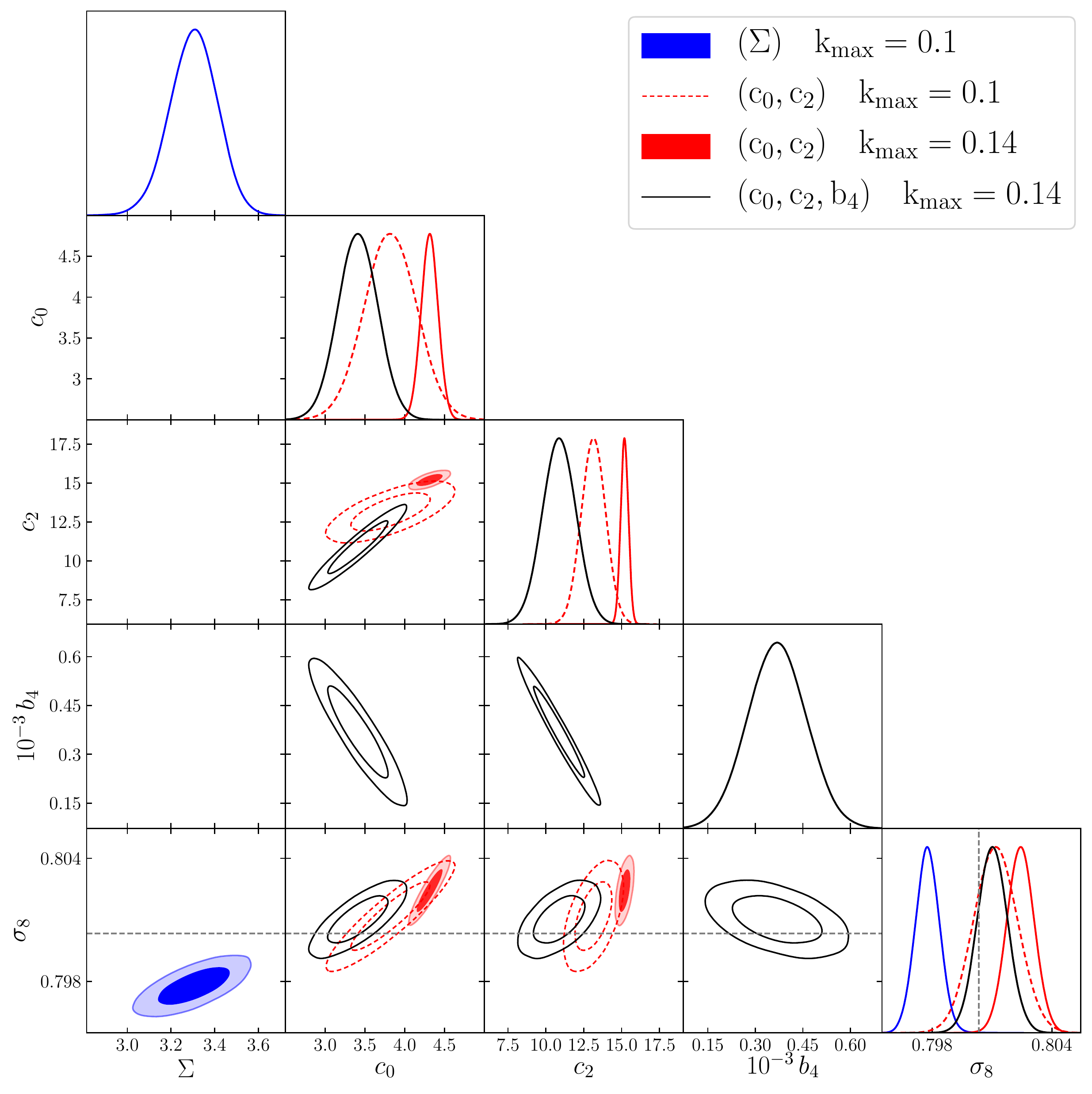}	
	\end{center}
	\caption{\label{fig:rsdSigma_z1}
Same as Fig.~\ref{fig:rsdSigma_z0}, but at redshift $z=0.974$.
	}
\end{figure}

Now let us consider the $z=0.974$ snapshot. 
The relevant posterior distributions are shown in Fig.~\ref{fig:rsdSigma_z1}.
One can observe the same trends as before: the fitting function~\eqref{eq:fog} (model (a)) yields a highly biased estimate of $\sigma_8$ already at $k_{\max}=0.1\,h$/Mpc. 
The naive EFT model (b) is accurate at $k_{\max}=0.1\,h$/Mpc, but fails to give a good fit beyond $k_{\max}=0.14\,h$/Mpc. 
The situation improves after the addition of the the next-to-leading order counterterm $b_4$.
We can clearly see that the presence of $b_4$ increases $\kmax$ and yields constraints better than the model (b)
at $\kmax=0.1~\hMpc$.
This illustrates the crucial importance of fourth order short-scale velocity cumulant~\eqref{eq:RSDb4} for robust parameter inference. 
In what follows, we always include \eqref{eq:RSDb4} in our baseline theoretical model \eqref{eq:RSDth}, \eqref{eq:RSDc0c2}.
Finally, we confirm the validity of perturbation theory by plotting the LO and NLO quadrupole counterterms, normalized to the linear theory prediction, in the right panel of Fig.~\ref{fig:counters}. We can see that the typical data cuts
of our analysis $\kmax=(0.1-0.15)~\hMpc$ are indeed 
lower than the nonlinear scale $k_{\rm NL, RSD}\approx 0.25~\hMpc$
at $z=0.974$.

One may wonder if a better model can be obtained by replacing $P_{\rm NL}$ in Eq.~\eqref{eq:fog}
computed in SPT by 
the EFT-like expression with one free counterterm,
\be 
\label{eq:fog2}
P^{\rm FOG}(k,\mu)=\e^{-(k\mu\Sigma f)^2}\left(P_{\rm SPT, 1-loop}(k,\mu)-2c'_0k^2P_{\rm lin}(k)\right)\,,
\ee 
which can be referred to as the
``EFT+FoG'' model.
In this case we have two free parameters 
and the resummed expression 
for the higher-derivative 
fingers-of-God
corrections. 
We have found that 
both at $\kmax=0.1~\hMpc$
and $\kmax=0.14~\hMpc$
this model is almost identical to 
the ``vanilla EFT'' case (b), see Fig.~\ref{fig:sigma2}.
This shows that the inclusion of the 
``FoG resummation" does not 
allow one to improve the fit over the basic EFT model.
After including the $b_4k^4\mu^4P_{\lin}$ term in the fit, we have found that the counterterm $b_4$ deviates from 
the prediction of the ``resummation'' 
formula $b^{\rm resum.}_4=-\Sigma^4/2$~\eqref{eq:fogfix}
at the $5\sigma$ level.
Note that even though $b^{\rm resum.}_4$ has the same order of magnitude as the actual $b_4$ measured from the data, its sign is wrong. 
This shows that the perturbative 
EFT expansion with the $k^4\mu^4P_{\rm lin}$ counterterm provides a better model than the resummed nonlinear damping function given in Eq.~\eqref{eq:fog}. This conclusion should also hold true for any non-linear damping function that predicts a positive coefficient in front of the $k^4\mu^4 P_{\rm lin}$-term, e.g.~the Lorentzian damping.

Let us summarize the results of this section so far. 
First, we have shown that the use of over-constrained fitting formulas for the fingers-of-God can generate
significant bias in the inferred cosmological parameters.
Second, we have justified the inclusion of the next-to-leading order $k^4\mu^4 P_{\rm lin}$-counterterm, by showing that
it allows us to extend the regime of applicability of the EFT description and improve the parameter constraints.

\begin{figure}[ht!]
	\begin{center}
		\includegraphics[width=0.6\textwidth]{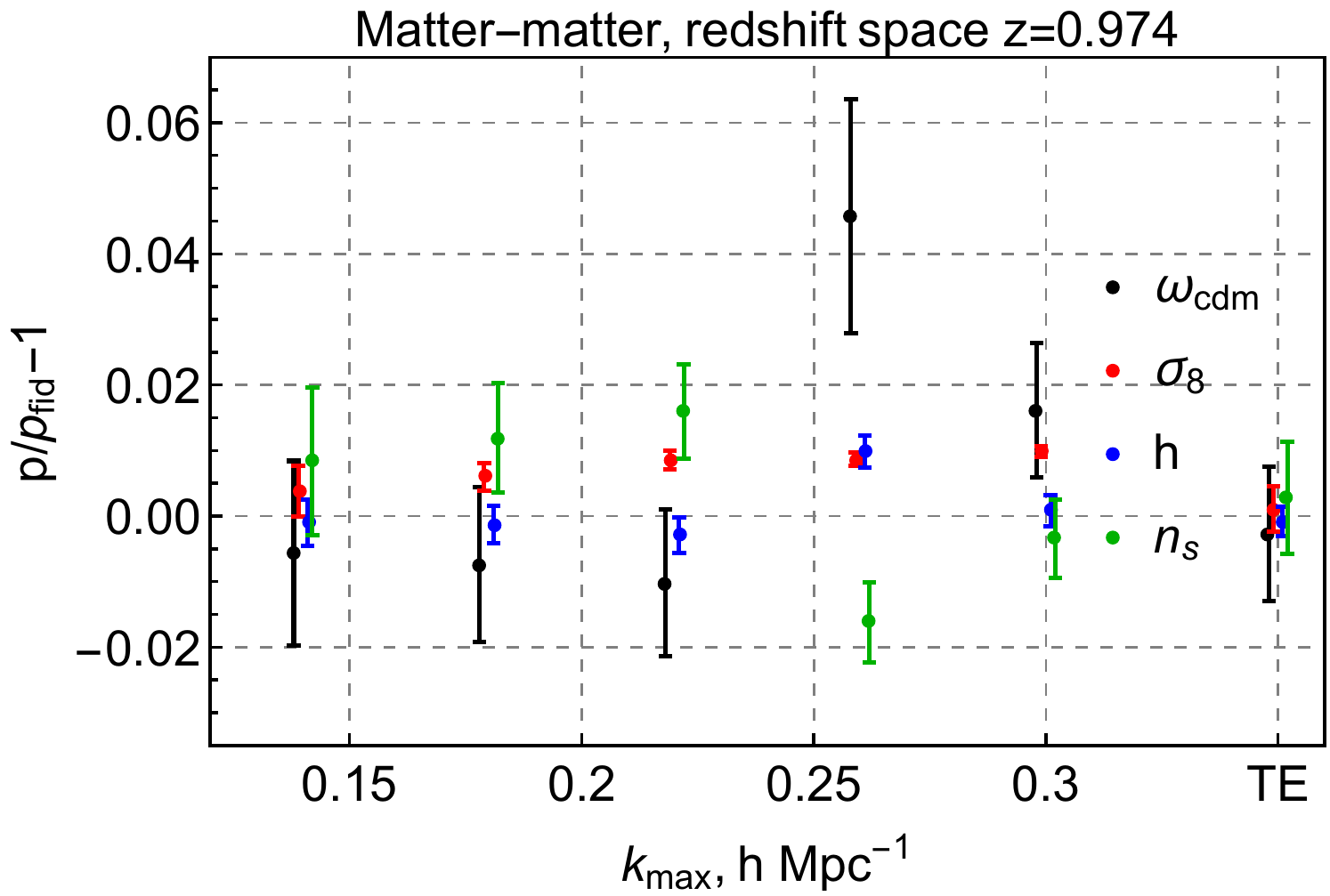}
	\end{center}
	\caption{\label{fig:dm_rsd_error}
		Same as Fig.~\ref{fig:rsdmz0ps} for the 
		redshift-space matter power spectrum multipoles at $z=0.974$.
	}
\end{figure} 
\begin{table}[h!]
	\begin{center}
		\begin{tabular}{|c|c|c|c|}  \hline 
			\small{par.} & fid. & \multicolumn{2}{|c|}{$z=0.974$} \\  \hline
			&  & $k_{\text{max}}=0.14\,h$/Mpc & TE  \\  \hline\hline
			$\omega_{cdm}$ & $0.1029$  & $0.1023^{+1.4\cdot 10^{-3}}_{-1.5\cdot 10^{-3}}$ &
			$0.1026^{+1.1\cdot 10^{-3}}_{-1.0\cdot 10^{-3}}$ \\ \hline
			$h$ & $0.7$& $0.6993^{+2.4\cdot 10^{-3}}_{-2.6\cdot 10^{-3}}$ &
			$0.6995^{+1.6\cdot 10^{-3}}_{-1.6\cdot 10^{-3}}$ \\ \hline
			$n_s$ & $1$ & $1.0083^{+0.012}_{-0.011}$&
			$1.0028^{+8.5\cdot 10^{-3}}_{-8.5\cdot 10^{-3}}$ \\ \hline
			$A$ & $1$ & $1.0096^{+0.016}_{-0.016}$&
			$1.0043^{+0.010}_{-0.011}$ \\ \hline
			$\Omega_m$ & $0.25$ & $0.2493^{+1.7\cdot10^{-3}}_{-1.8\cdot10^{-3}}$ &
			$0.2498^{+1.4\cdot10^{-3}}_{-1.3\cdot10^{-3}}$ \\ \hline
			$\sigma_8$  & $0.8003$ & $0.8033^{+3.1\cdot10^{-3}}_{-3.1\cdot10^{-3}}$ &
			$0.8012^{+2.8\cdot 10^{-3}}_{-2.8\cdot 10^{-3}}$ \\ \hline\hline
			$c_0$ & -- & $3.66^{+0.39}_{-0.37}$ &
			$3.29^{+0.40}_{-0.39}$ \\ \hline
			$c_2$ & -- & $11.26^{+1.32}_{-1.25}$ &
			$10.52^{+1.11}_{-1.10}$ \\ \hline
			$10^{-3}b_4$ & -- & $0.34^{+0.10}_{-0.11}$ &
			$0.41^{+0.08}_{-0.08}$ \\ \hline
		\end{tabular}
		\caption{
			The marginalized 1d intervals for the cosmological and nuisance parameters 
			estimated from the monopole and quadrupole moments of the Las Damas dark matter redshift space power spectrum
			at $z=0.974$. 
			We show the fitted parameters (first column), fiducial values used 
			in simulations (second column), and the resulting parameter constraints for the baseline $k_{\max}$ analysis (third column) and the theoretical error approach (fourth column).
				$c_0,c_2,b_4$ are quoted in units $[\Mpch]^2,~[\Mpch]^2,~[\Mpch]^4$, respectively.
		}
		\label{tab:cos_dm_rsd0}
	\end{center}
\end{table} 
\begin{figure}[ht!]
	\begin{center}
		\includegraphics[width=1\textwidth]{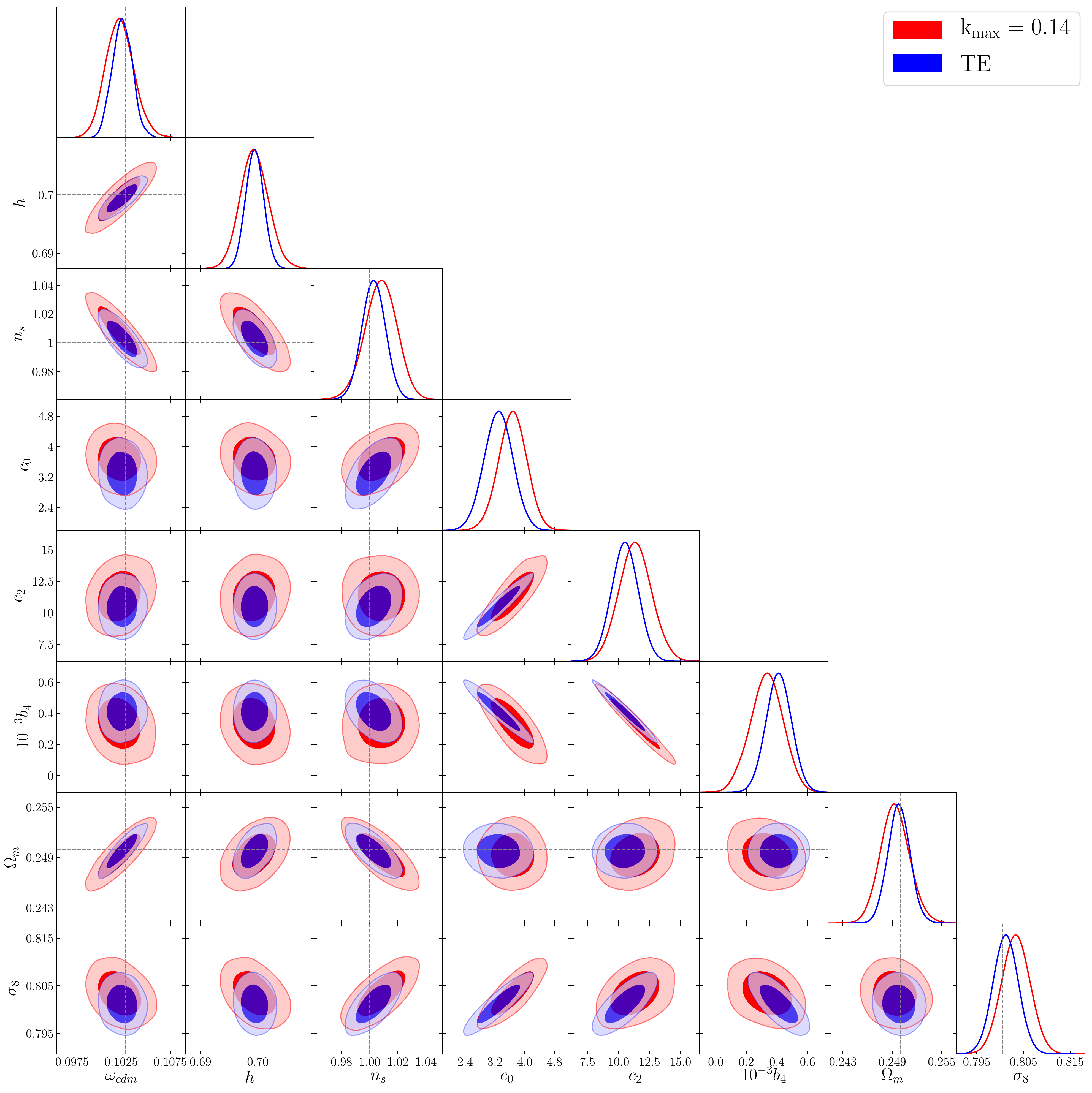}\end{center}
	\caption{\label{fig:rsddmz1}
		Triangle plot for the cosmological and nuisance parameters 
		measured from the redshift space dark matter power spectrum
		of the LasDamas simulations at $z=0.974$.
			$c_0,c_2,b_4$ are quoted in units $[\Mpch]^2,~[\Mpch]^2,~[\Mpch]^4$, respectively.
	}
\end{figure}

\subsection{Cosmological parameters}
\label{subsec:rsd_par}

We present now the complete cosmological analysis of the redshift-space dark matter power spectrum
in the $\kmax$ and TE analyses. 

In the previous section we have seen that the perturbative description breaks down at $z=0$ already on quite large scales $k\sim 0.1~\hMpc$, and hence this case may not be very illustrative.
We focus on the $z=0.974$ snapshot in what follows. 

The marginalized 1d constraints on cosmological parameters for different momentum cuts and the theoretical error are shown in Fig.~\ref{fig:dm_rsd_error}.
The theory prediction for TE analysis was computed at fiducial $\kmax^{\rm fid.}=0.12\,h$/Mpc.
We found that the momentum cutoff $k_{\max}=0.14\,h$/Mpc reproduces the true parameters within $1\sigma$ interval. However, the clustering amplitude becomes biased by more than $2\sigma$ starting from $k_{\max}=0.16\,h$/Mpc. Given this reason, we choose $k_{\max}=0.14\,h$/Mpc as our baseline data cut. The final constraints on cosmological and nuisance parameters for this $\kmax$ and for the theoretical error analysis are displayed in Tab.~\ref{tab:cos_dm_rsd0}
and in Fig.~\ref{fig:rsddmz1}. 
As in the previous section, we note three key features of the theoretical error covariance.
First, we can obtain the cosmological constraints with a single MCMC analysis.
This can be contrasted with the standard approach that requires running many analyses with different choices of $\kmax$. 
Second, we get unbiased estimates for cosmological parameters and reliable error bars. 
Third, these errorbars happened to be smaller than those obtained in the $\kmax$ analyses; 
the constraints on $\omega_{cdm}$, $h$, $n_s$ and $\sigma_8$ improve
by $30\%$, $40\%$, $25\%$, $10\%$, respectively.
This happens mainly due to additional BAO wiggles, which are included beyond $\kmax=0.14~\hMpc$.
Indeed, we can see that the 2d probability distribution in the $\omega_{cdm}-h$ plane, which 
reflects the BAO signal, is significantly narrower in the TE analysis.

Given some significant improvements 
that we have obtained, it is important
to check whether they remain 
if we alter the theoretical error
envelope. The details of this analysis
are presented in Appendix~\ref{app:fid}.
We have found that the parameter constraints
do depend 
on the choice of fiducial cosmology quite noticeably. This implies that in the regime where the constraints are dominated by the theoretical error
one has to be careful 
about its accurate modeling.

\section{Galaxies in real space}
\label{sec:gal_rs}

In this section we focus on the galaxy clustering in real space. 
Although this analysis is academic in nature,
it will allow us to assess the validity of the perturbative bias model 
and clearly see the implications
of the theoretical error for biased tracers.
We will use the effective field theory model characterized by the following set of nuisance parameters
(see~\cite{Chudaykin:2020aoj} for our notations),
\be
\{b_1,b_2,b_{\mathcal{G}_2},R^2_*, b_{\Gamma_3}, P_{\rm shot} \},
\ee
which includes the local quadratic bias $b_2$, the quadratic tidal bias $b_{\mathcal{G}_2}$, the higher-derivative counterterm\footnote{Since the degeneracy between $R_*^2$ and $c_s^2$ (in the notation of \cite{Chudaykin:2020aoj}) cannot be broken at the power spectrum level, we fix the dark matter counterterm $c_s^2=1~[\Mpch]^2$  to avoid clutter. } $R^2_*$ and the constant shot noise contribution $P_{\rm shot}$.
Even though the pair-counting shot noise prediction $\bar{n}^{-1}$ was subtracted from the power spectrum data,
we still need to keep a constant nuisance parameter in the fit because the value of shot noise is, in general, expected to deviate from $\bar{n}^{-1}$ due to exclusion effects \cite{Baldauf:2013hka} (or fiber collisions in realistic surveys~\cite{Hahn:2016kiy}). These deviations can be as large as $\sim 30\%$ of $\bar{n}^{-1}$ 
for the BOSS-like host halos~\cite{Schmittfull:2018yuk}.
Given this reason, we marginalize over $P_{\rm shot}$ within the following Gaussian prior:
\be
 P_{\rm shot} \sim (0, (0.3\cdot \bar{n}^{-1})^2 )\,.
\ee
Note that the value of the residual shot noise contribution can be negative. 
We do not impose any priors on the other biases and counterterms, except for the cubic tidal bias $b_{\Gamma_3}$, which we found to be very degenerate with $b_{\mathcal{G}_2}$.
We marginalize over $b_{\Gamma_3}$ assuming the physical prior centered at the prediction 
of the coevolution model~\cite{Desjacques:2016bnm,Abidi:2018eyd},
\[
b^{(Coev.)}_{\Gamma_3}=\frac{23}{42}(b_1-1)=0.66 \quad \text{for} \quad b_1=2.2\,, 
\]
and with the unit 
Gaussian variance,
\be
b_{\Gamma_3}\sim \mathcal{N}\left(b^{(Coev.)}_{\Gamma_3},1^2\right)\,. 
\ee
Finally, we have checked that the data does not show
any evidence for the scale-dependent stochastic contributions  $a_0k^2$ for $k\lesssim 0.3~\hMpc$, and hence we did not use it in our model.
This is consistent with the results of N-body simulations done in Ref.~\cite{Schmittfull:2018yuk}.

\begin{figure}[ht!]
	\begin{center}
		\includegraphics[width=1\textwidth]{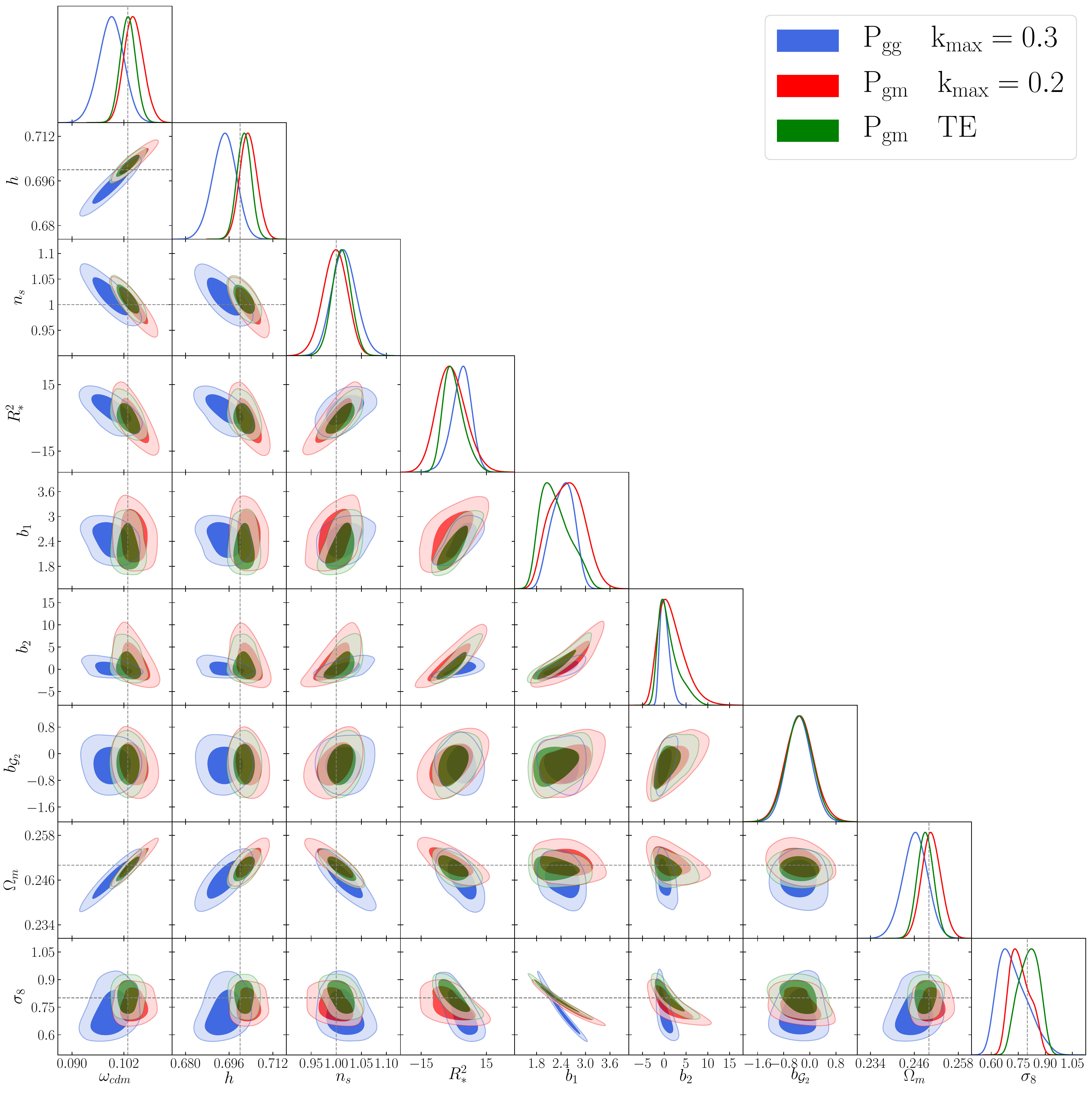}\end{center}
	\caption{\label{fig:rsdgal}
		Triangle plot for the cosmological and nuisance parameters 
		measured from the real space galaxy power spectrum
		of the Las Damas simulations at $z=0.342$.
	}
\end{figure}

We analyzed the galaxy-galaxy power spectrum and
the galaxy-matter cross-spectrum at $z=0.342$
using the $\kmax$ approach and found  
that the posterior distributions are unbiased for all $\kmax$ up to \mbox{$k_{\rm NL}\simeq 0.3~\hMpc$}.
We do not push to higher $k$'s because the perturbative expansion is clearly not valid there.
The resulting posterior distributions for nuisance and cosmological parameters are shown in Fig.~\ref{fig:rsdgal} (blue contrours), the 1d marginalized parameter limits are presented in Table~\ref{tab:gal_rs}.
\begin{table}[h!]
	\begin{center}
		\begin{tabular}{|c|c|c|c|c|}  \hline 
			\small{par.} & fid. & $\rm P_{gg}\quad k_{\max}=0.3\,h$/Mpc & $\rm P_{gm}\quad k_{\max}=0.2\,h$/Mpc & $\rm P_{gm}\quad TE$\\  \hline
			\hline
			$\omega_{cdm}$ & $0.1029$  & $0.099^{+3.0\cdot 10^{-3}}_{-2.9\cdot 10^{-3}}$   &
			$0.1043^{+2.1\cdot 10^{-3}}_{-2.4\cdot 10^{-3}}$ &
			$0.1030^{+1.8\cdot 10^{-3}}_{-1.9\cdot 10^{-3}}$ \\ \hline
			$h$ & $0.7$& $0.6942^{+4.6\cdot 10^{-3}}_{-4.3\cdot 10^{-3}}$ &
			$0.7031^{+3.1\cdot 10^{-3}}_{-3.3\cdot 10^{-3}}$ &
			$0.7015^{+2.7\cdot 10^{-3}}_{-2.7\cdot 10^{-3}}$ \\ \hline
			$n_s$ & $1$ & $1.0160^{+0.024}_{-0.026}$ &
			$0.9973^{+0.026}_{-0.024}$  &
			$1.0104^{+0.019}_{-0.019}$\\ \hline
			$A$ & $1$ & $0.8570^{+0.1085}_{-0.2509}$ &
			$0.8919^{+0.092}_{-0.184}$ &
			$1.0157^{+0.167}_{-0.157}$ \\ \hline
			$\Omega_m$ & $0.25$ & $0.2462^{+3.6\cdot10^{-3}}_{-3.3\cdot10^{-3}}$ &
			$0.2506^{+2.6\cdot10^{-3}}_{-2.8\cdot10^{-3}}$ &
			$0.2490^{+2.2\cdot10^{-3}}_{-2.3\cdot10^{-3}}$ \\ \hline
			$\sigma_8$  & $0.8003$ & $0.7208^{+0.057}_{-0.108}$  &
			$0.7600^{+0.043}_{-0.076}$  &
			$0.8080^{+0.072}_{-0.052}$ \\  \hline\hline
			$R_*^2$ & -- & $3.46^{+4.97}_{-3.93}$ &
			$-1.09^{+6.02}_{-7.21}$ &
			$-0.03^{+3.60}_{-5.54}$ \\ \hline
			$b_1$ & $-$ & $2.44^{+0.32}_{-0.26}$ &
			$2.54^{+0.41}_{-0.51}$ &
			$2.26^{+0.24}_{-0.48}$ \\ \hline
			$b_2$ & -- & $0.17^{+0.81}_{-1.34}$ &
			$1.74^{+2.03}_{-3.79}$ &
			$1.04^{+1.27}_{-2.96}$ \\ \hline
			$b_{\mathcal{G}_2}$ & -- & $-0.326^{+0.376}_{-0.397}$ &
			$-0.309^{+0.425}_{-0.457}$ &
			$-0.315^{+0.415}_{-0.423}$ \\ \hline
		\end{tabular}
		\caption{
			The marginalized 1d intervals for the cosmological parameters 
			estimated from the Las Damas real space galaxy power spectrum and galaxy-matter cross spectrum at $z=0.342$. 
			The shown are the fitted parameters (first column), fiducial values used 
			in simulations (second column), 
			the results for $P_{\rm gg}$ at $\kmax=0.3\hMpc$ (third column),
			for $P_{\rm gm}$ at $\kmax=0.2\hMpc$ (fourth column)
			for $P_{\rm gm}$ with theoretical error approach (fifth column).
		}
		\label{tab:gal_rs}
	\end{center}
\end{table}
\begin{figure}[h!]
	\begin{center}
		\includegraphics[width=0.7\textwidth]{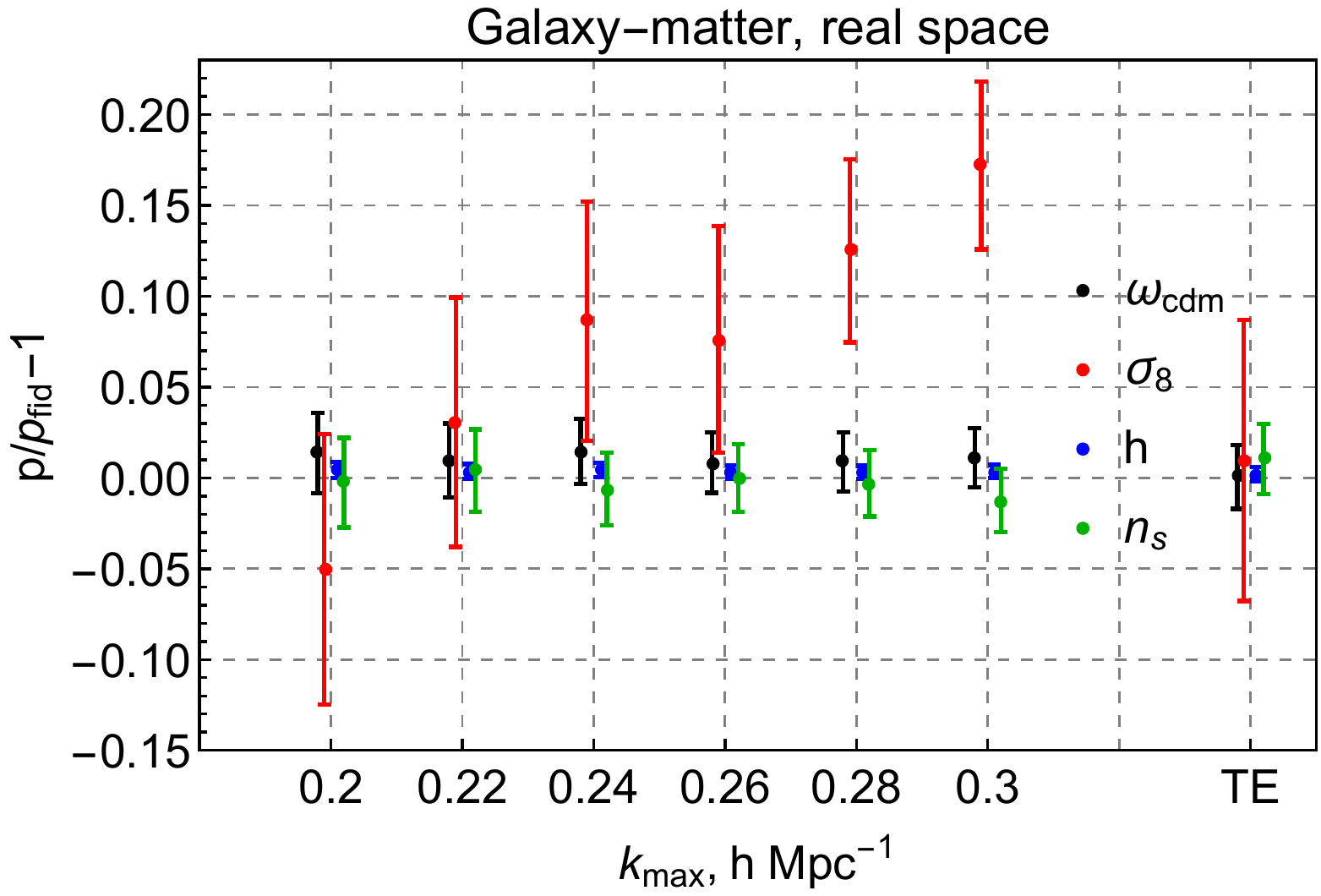}
	\end{center}
	\caption{\label{fig:gm_error}
		Marginalized 1d limits on cosmological parameters inferred from the real-space galaxy-matter
		cross spectrum as a function of $\kmax$. The rightmost points result from the TE analysis.
		All parameters are normalized to their fiducial values.
	}
\end{figure}

All in all, we do not see any evidence for the theoretical error up to very high $k_{\rm max}$ for 
the power spectrum of galaxies in real space.
We believe that it is caused by the following two reasons. 
First, the statistical covariance for galaxies contains a large shot noise contribution, which slows down the
reduction of the power spectrum errors on short scales compared to the dark matter case studied before.
Thus, unlike the dark matter case,
the theoretical error for galaxies is always smaller than the statistical one on the scales of interest. Note that the situation can be different
for another type of galaxies, whose shot noise is lower, e.g. the DESI bright 
galaxy sample~\cite{Aghamousa:2016zmz}.
The second reason is a large number of fitting parameters: the real space galaxy power spectrum data alone 
cannot efficiently break the degeneracies between these parameters. An example is the strong degeneracy $b_1-\sigma_8$, which also generates a highly not-Gaussian posterior distribution of $\sigma_8$ seen in Fig.~\ref{fig:rsdgal}.

Indeed, these two effects can be clearly assessed by analyzing the galaxy-matter cross spectrum $P_{\rm gm}$.
The shot noise contribution to the covariance matrix is reduced for $P_{\rm gm}$, see Eq.~\eqref{Creal} 
and there is no constant shot noise contribution
$P_{\rm shot}$ in the fit, which narrows the posterior distribution compared 
to the $P_{\rm gg}$ case for the same $\kmax$. The 1d marginalized limits on the cosmological 
parameters as a function of $\kmax$ are shown in Fig.~\ref{fig:gm_error}.
The theory prediction for TE analysis was computed at fiducial $\kmax^{\rm fid.}=0.18\,h$/Mpc.
	
We see that the 1d posterior distributions for $P_{\rm gm}$ are unbiased up to $\kmax\approx 0.22~\hMpc$.
However, a closer inspection of the MCMC results revealed a bias in the 2d posterior contours.
This suggests that using the marginalized 1d distributions to select $\kmax$ can be misleading:
one has to ensure that not only the 1d marginalized constraints 
are unbiased, but also the 2d contours enclose the fiducial cosmology e.g. within 68$\%$ CL. 
Indeed, for some data cuts the principal components of certain parameters can be biased, 
but their projections onto particular parameter planes can accidentally fall close to the fiducial values. 
Using such data cuts can lead to biases when different probes are combined, as in this case the resulting
posterior distributions are driven by the degeneracy breaking between different principal components (PCs)
of the combined data sets. This motivated us to choose $\kmax= 0.2~\hMpc$ as a baseline data cut for $P_{\rm gm}$.

Since the baseline $\kmax$ is quite low, we expect to gain some information with the theoretical error.
The posterior distributions for the $\kmax$ and the TE cases are shown in Fig.~\ref{fig:rsdgal}, whilst the parameter limits are presented in Table~\ref{tab:gal_rs}. 
Just like in the dark matter case, the TE covariance sharpens the constraints on the parameters related to the
power spectrum shape and the BAO: $\omega_{cdm}, h$, and $n_s$ improve by $20\%,20\%$ and $25\%$, respectively.
In contrast, the constraint on $\sigma_8$ is not significantly affected. Note also that  
the mean $\sigma_8$ shifts toward the true fiducial value. 
The constraint on $\sigma_8$ does not improve due to the notorious degeneracy $b_1-\sigma_8$.
Indeed, the linear-theory degeneracy $b_1-\sigma_8$ can be broken only by the one-loop corrections. This makes the result very sensitive 
to largest wavenumber bins used in the analysis, because for these bins the amplitude of the 
loop corrections is large.
However, when the one loop correction becomes large, the two-loop contribution becomes non-negligible as well.
The theoretical error covariance is exactly introduced to alleviate this problem and marginalize 
over all possible 2-loop shapes. 
It does not sharpen the constraints on $\sigma_8$, but yields some improvement for other parameters related to the shape and the BAO.

One possible way to shrink the posterior distribution in the galaxy power- and cross-spectra cases is to
fix some bias parameters to the predictions of some phenomenological models, i.e.~the coevolution model 
for the dark matter halos~\cite{Desjacques:2016bnm}. However, this can result in over-fitting and wrong estimation
of the parameter error bars. We believe that a more appropriate approach is to vary all relevant nuisance parameters 
in the fit within physical priors, as we do in this paper.
In a realistic survey one always observes galaxies in redshift space, which allows 
for breaking of the $b_1-\sigma_8$ degeneracy through the redshift-space distortions.
Let us now consider the latter case.

\section{Galaxies in redshift space}
\label{sec:gal_rsd}

In this section we scrutinize the galaxy clustering in redshift space in the context of the theoretical error covariance. We will analyze the monopole and quadrupole moments of the redshift space galaxy power 
spectrum at $z=0.342$.
The redshift of this sample is somewhat lower than the effective redshifts of future surveys, hence this represents the stringent test of our approach.

The EFT model with NLO fingers-of-God corrections is characterized by the following set 
of nuisance parameters (see Ref.~\cite{Chudaykin:2019ock} for our conventions),
\be 
\{b_1,b_2,b_{\mathcal{G}_2},b_{\Gamma_3},P_{\rm shot},c_0,c_2,b_4\}\,.
\ee 
We use the same priors on $b_{\Gamma_3}$ and $P_{\rm shot}$ as in the previous section,
and infinitely large flat priors for other parameters.
Note that on general grounds one can expect the presence of the additional stochastic contribution
$P_{\rm stoch}(k,\mu)=a_2 \mu^2 k^2$.
However, we have found that this contribution is completely degenerate with $b_4$ for the $P_{0}+P_{2}$ datavector,
which motivated us to fix $a_2=0$ as in Refs.~\cite{Nishimichi:2020tvu,Chudaykin:2020aoj}.

\begin{figure}[h!]
	\begin{center}
		\includegraphics[width=0.7\textwidth]{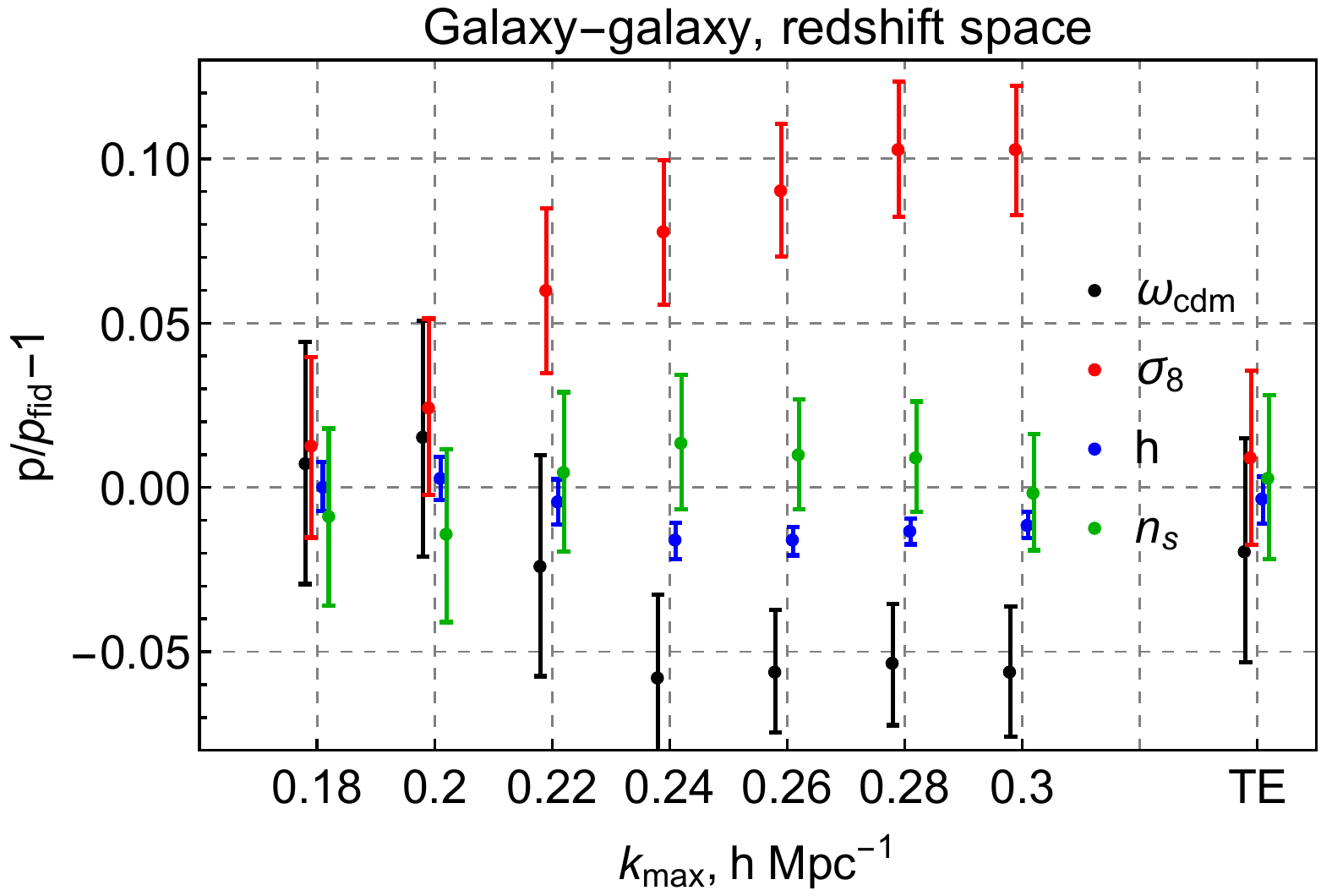}
	\end{center}
	\caption{\label{fig:gal_rsd_error}
Same as Fig.~\ref{fig:gm_error}, but for the redshift space galaxy multipoles.}
\end{figure}

As a first step,
we run a sequence of analyses varying the data cut $k_\maxx$.
The 1d marginalized constrains on the cosmological parameters as functions of $k_\maxx$ are presented in Fig.~\ref{fig:gal_rsd_error}.
The rightmost points correspond to the theoretical error analysis with the cutoff $k_\maxx=0.32\,h$/Mpc. 
One may observe that the standard analyses yields a biased estimate for $\sigma_8$ for 
 $k_\maxx>0.20\,h$/Mpc. Shortly after,
$\omega_{cdm}$ and $h$ start deviating from the true values at $k_\maxx\approx 0.23\,h$/Mpc.  

This motivates us to choose $k_\maxx=0.18\,h$/Mpc as our baseline analysis with the sharp momentum cutoff
as in this case all marginalized posteriors contain the true values well within $1\sigma$. 

\begin{table}[h!]
	\begin{center}
		\begin{tabular}{|c|c|c|c|}  \hline 
			\small{par.} & fid. & $k_\maxx=0.18\,h$/Mpc & TE \\  \hline
			\hline
			$\omega_{cdm}$ & $0.1029$  & $0.1037^{+3.5\cdot 10^{-3}}_{-4.1\cdot 10^{-3}}$ &
			$0.1009^{+3.5\cdot 10^{-3}}_{-3.5\cdot 10^{-3}}$ \\ \hline
			$h$ & $0.7$& $0.7002^{+5.1\cdot 10^{-3}}_{-5.3\cdot 10^{-3}}$ &
			$0.6973^{+5.0\cdot 10^{-3}}_{-5.0\cdot 10^{-3}}$ \\ \hline
			$n_s$ & $1$ & $0.9909^{+0.028}_{-0.026}$ &
			$1.0031^{+0.025}_{-0.025}$ \\ \hline
			$A$ & $1$ & $1.0169^{+0.073}_{-0.079}$& 
			$1.0571^{+0.069}_{-0.081}$ \\ \hline
			$\Omega_m$ & $0.25$ & $0.2513^{+4.6\cdot10^{-3}}_{-5.2\cdot10^{-3}}$ &
			$0.2479^{+4.3\cdot10^{-3}}_{-4.7\cdot10^{-3}}$ \\ \hline
			$\sigma_8$  & $0.8003$ & $ 0.8069^{+0.022}_{-0.022}$ &
			$0.8044^{+0.021}_{-0.021}$  \\  \hline\hline
			$c_0$ & -- & $0.77^{+17.93}_{-11.48}$ & 
			$9.40^{+12.45}_{-9.68}$ \\ \hline
			$c_2$ & -- & $36.22^{+28.01}_{-20.61}$ & 
			$29.61^{+23.11}_{-18.30}$ \\ \hline
			$10^{-3}b_4$ & -- & $1.40^{+0.21}_{-0.26}$ & 
			$1.73^{+0.25}_{-0.28}$ \\ \hline
			$b_1$ & $-$ & $2.17^{+0.07}_{-0.07}$ & 
			$2.16^{+0.07}_{-0.07}$ \\ \hline
			$b_2$ & -- & $-0.96^{+0.64}_{-0.87}$ & 
			$-0.92^{+0.57}_{-0.78}$ \\ \hline
			$b_{\mathcal{G}_2}$ & -- & $-0.350^{+0.364}_{-0.388}$ &
			$-0.330^{+0.351}_{-0.368}$ \\ \hline
		\end{tabular}
		\caption{
			The marginalized 1d intervals for the cosmological parameters 
			estimated from the Las Damas redshift space galaxy power spectra at $z=0.342$. 
			The table contains fitted parameters (first column), fiducial values used 
			in simulations (second column), and the results of the baseline $k_\maxx$ analysis (third column) and the outcome of the theoretical error approach (fourth columns). 
				$c_0,c_2,b_4$ are quoted in units $[\Mpch]^2,~[\Mpch]^2,~[\Mpch]^4$, respectively.
		}
		\label{tab:gal_rsd}
	\end{center}
\end{table} 
\begin{figure}[ht!]
	\begin{center}
		\includegraphics[width=1\textwidth]{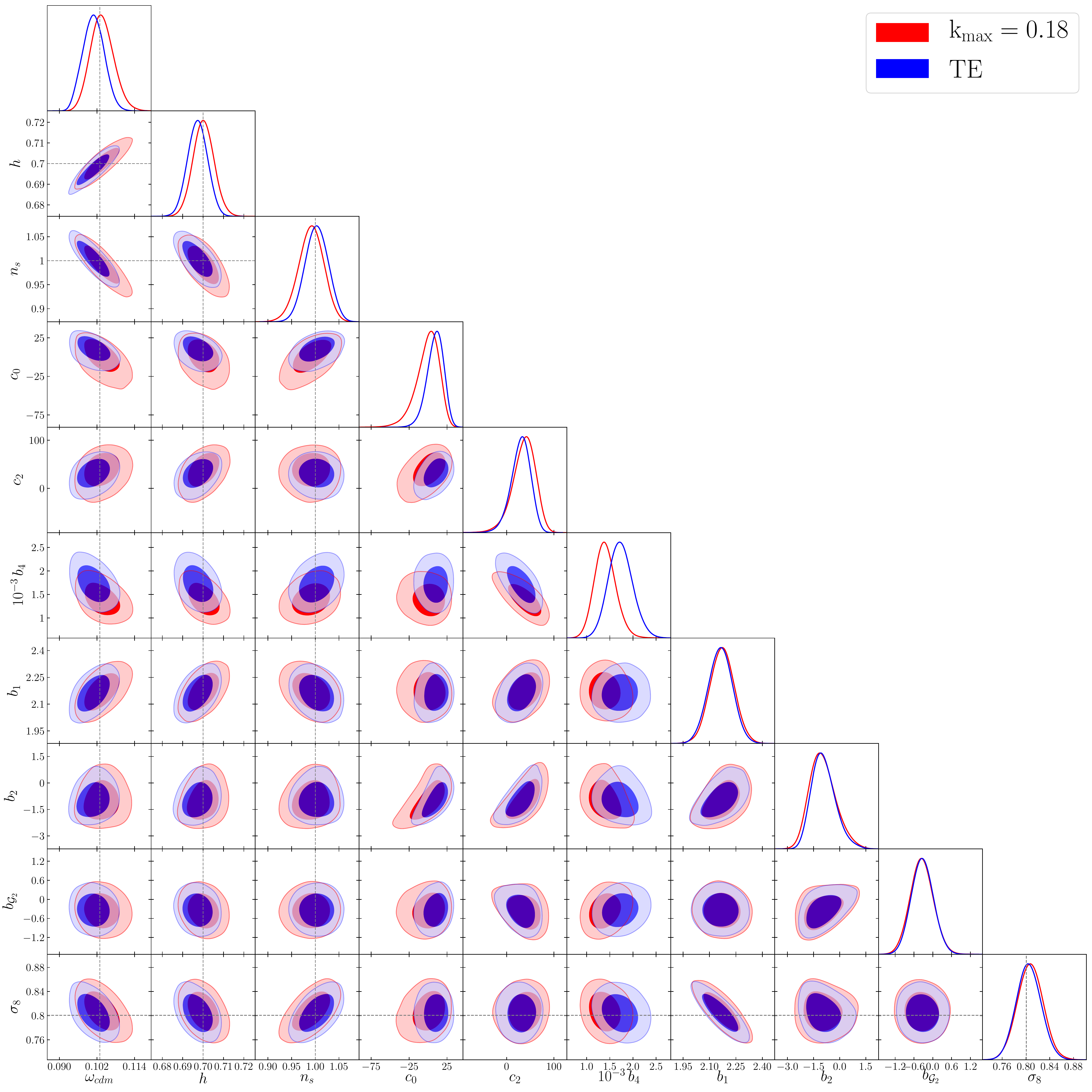}\end{center}
     	\caption{\label{fig:gal_rsd_cosmo}
		Triangle plot for the cosmological and nuisance parameters 
		measured from the redshift-space galaxy power spectrum
		of the LasDamas simulations at $z=0.342$.
			$c_0,c_2,b_4$ are quoted in units $[\Mpch]^2,~[\Mpch]^2,~[\Mpch]^4$, respectively.
	}
\end{figure}

The values of the marginalized 1d constraints for the baseline $k_\maxx$ and the TE analyses are shown in Tab. \ref{tab:gal_rsd}.
The 2d posteriors are presented in Fig.~\ref{fig:gal_rsd_cosmo}.
The theory prediction for TE analysis was computed at fiducial $\kmax^{\rm fid.}=0.16\,h$/Mpc.

Overall, we see that for the realistic example of redshift-space galaxies the theoretical error approach 
 improves the parameter constraints over the $\kmax$ case quite modestly ($\lesssim 10\%$), unlike the previous case of the galaxy-matter cross-spectrum.
This happens because 
the BAO wiggles at $k>0.18~\hMpc$, neglected in the $\kmax$ analysis, are strongly suppressed due 
to redshift-space displacements \cite{Ivanov:2018gjr} and the large shot noise covariance (c.f. Eq.~\eqref{Creal} and Eq.~\eqref{Crsd}).
There is, however, some non-negligible improvement for the nuisance parameters, whose values are quite sensitive 
to the large-$k$ modes. 
As far as the cosmological parameters are concerned, qualitatively, we can conclude
that the TE covariance, if properly chosen, automatically optimizes the choice of $\kmax$ and guarantees 
that parameter limits are unbiased, but does not noticeably improve them.
This implies that the application of the theoretical error approach most likely will not
sharpen the cosmological constraints from the current surveys like BOSS, where $\kmax$ has already been measured
for various analysis settings~\cite{Ivanov:2019pdj,DAmico:2019fhj}. 

Comparing the cases of 
redshift-space 
dark matter and redshift-space galaxies, we can understand
the reason why the gain
from the theoretical error
is so modest in the latter 
case. The main difference between
these two cases is the presence
of four additional shapes
in the non-linear bias model, 
which also
act like theoretical 
error, see Sec.~\ref{sec:theory}.
Hence, the addition of 
an 
extra theoretical error shape 
does not have a large impact 
on the likelihood because it 
already contains the theoretical error due to 
marginalization
over the non-linear bias parameters.

Finally, it is important to point out that we have checked that 
our results for galaxies in redshift space are invariant w.r.t. the choice
of the theoretical error template, see Appendix~\ref{app:fid}. This is an important consistency check which
validates the theoretical error approach for realistic analyses.

\section{Conclusions}
\label{sec:concl}

We have validated the theoretical error approach on large-volume N-body simulations data
for the clustering of dark matter and galaxies in real and redshift space.
First, we introduced a new mock-based approach to extract the theoretical error
covariance from the data. 
This approach allowed us to avoid uncertainties
in the theoretical estimates of higher-order nonlinearities. 
We have argued that the calibration
of the theoretical error from the N-body mock
simulations can be the optimal strategy
to build the theoretical error likelihood.

We have demonstrated that the use of the TE covariance in the power spectrum likelihoods 
allows us to recover
the input cosmological parameters used in the simulations of dark matter and galaxies in real and redshift spaces.
Crucially, using the TE approach we can avoid lengthy measurements of $\kmax$ 
and perform the full parameter inference with a single MCMC run. 
To put this result 
in context, let us 
consider the case of
the recent analysis of the BOSS data from Ref.~\cite{DAmico:2019fhj}.
This work 
calibrated their baseline $k_{\rm max}$ by analyzing
N-body simulation data
for 4 different choices of the data cut. 
Achieving good convergence
for nuisance and cosmological parameters 
for these chains
requires 184000 CPU hours on our cluster, which can
be saved if the theoretical 
error approach is implemented.
We stress that
all this CPU time is required
for only one particular choice 
of priors and fitted parameters. If a different set 
of priors or model parameters
are used, the 
data cut $k_{\rm max}$
needs to be re-measured, as 
explained in detail in Ref.~\cite{DAmico:2019fhj}.
Hence, the actual 
computational 
saving from the theoretical
error likelihood
can be even larger.

We have also performed a detailed 
comparison of the TE results with the $\kmax$ analyses.
For dark matter in real and redshift space, cosmological constraints actually improve with the
theoretical error covariance. This happens because it allows us to extract the information
encoded in the BAO wiggles at wavenumbers larger than $\kmax$, 
while marginalizing over the broadband uncertainties.

For galaxies, the presence of the large shot noise term in the covariance and proliferation of nuisance parameters 
inflate the error bars and force us to use more aggressive data cuts, where the BAO information in saturated.
Due to this reason, the TE approach does not significantly improve the parameter limits unlike in the dark matter case. 
For the galaxy auto-spectrum in real space, we have not found any evidence for 
theoretical error up 
to the nonlinear scale, and hence the TE approach is fully identical to the standard $\kmax$ analysis.
Finally, for galaxies in redshift space, the theoretical error is non-vanishing and its inclusion in the 
covariance matrix
effectively optimizes the choice of $\kmax$ at which 
the inferred parameters are unbiased. 
The cosmological parameter constrains improve quite marginally in this case, unlike 
the nuisance parameters, whose posterior volume shrinks appreciably.

As a by-product, we have validated the effective field theory implementation of
the \texttt{CLASS-PT} code \cite{Chudaykin:2020aoj} by showing that it can accurately reproduce 
the true fiducial parameters from various power spectrum datavectors used in this paper. 
In passing, we have shown that the use of over-constrained non-linear fingers-of-God models
leads to biases in parameter inferences. 
Besides, we have advocated the inclusion of the higher-order counterterm
in the description of nonlinear redshift-space distortions. 
Indeed, it allows us to extend
the range of scales where the modeling is accurate and eventually improve 
the cosmological constraints as compared to the standard one-loop effective field theory 
model without the higher-order corrections. This justifies 
the analyses of Refs.~\cite{Ivanov:2019pdj,Philcox:2020vvt,Nishimichi:2020tvu} that included the higher 
derivative fingers-of-God correction. 

It is worth pointing out that in this paper we have studied the mock galaxies that simulate the 
BOSS luminous red galaxy (LRG) sample~\cite{Alam:2016hwk}. This sample exhibits large fingers-of-God features,
which do not allow us to extend the perturbative treatment to sufficiently short scales.
The situation will be different for emission line galaxies (ELG), which are the main targets 
for the upcoming surveys like DESI~\cite{Aghamousa:2016zmz} or Euclid~\cite{Amendola:2016saw}.
The clustering properties of ELGs have been 
recently measured for the first time by the eBOSS survey~\cite{deMattia:2020fkb}. 
Crucially, these measurements already show that ELG sample is 
less affected by the fingers-of-God and hence one can expect 
some improvements in cosmological constraints over the LRG-based analysis,
due to larger effective $\kmax$.

Our analysis can be extended in various ways. First, the theoretical error approach can be applied 
to the bispectrum data~\cite{Baldauf:2016sjb}. 
Second, it would be curious to check to what extent it can be useful for the ELG sample.
Third, our formalism can be extended to the case of the projected statistics like weak lensing, 
where it could play an important role to minimize 
systematic biases due to imperfect theoretical modeling, see e.g.~\cite{Pandey:2020zyr}.
We leave these research directions for future work.

\section*{Acknowledgments}

We thank Roman Scoccimarro for sharing with us the power spectra from the LasDamas N-body simulation. 
We are grateful to 
Marcel Schmittfull and Matias Zaldarriaga for valuable discussions.
We thank Oliver Philcox
and Martin White
for their valuable comments on the draft. 
A.C.~and M.I.~are supported by the RFBR grant 20-02-00982. All
numerical calculations were performed with the HybriLIT heterogeneous computing platform (LIT, JINR) (\href{http://hlit.jinr.ru}{http://hlit.jinr.ru}).

\appendix 

\section{Theoretical error envelope}
\label{app:TEenvelope}

\begin{figure*}[ht!]
	\begin{center}
		\includegraphics[width=0.49\textwidth]{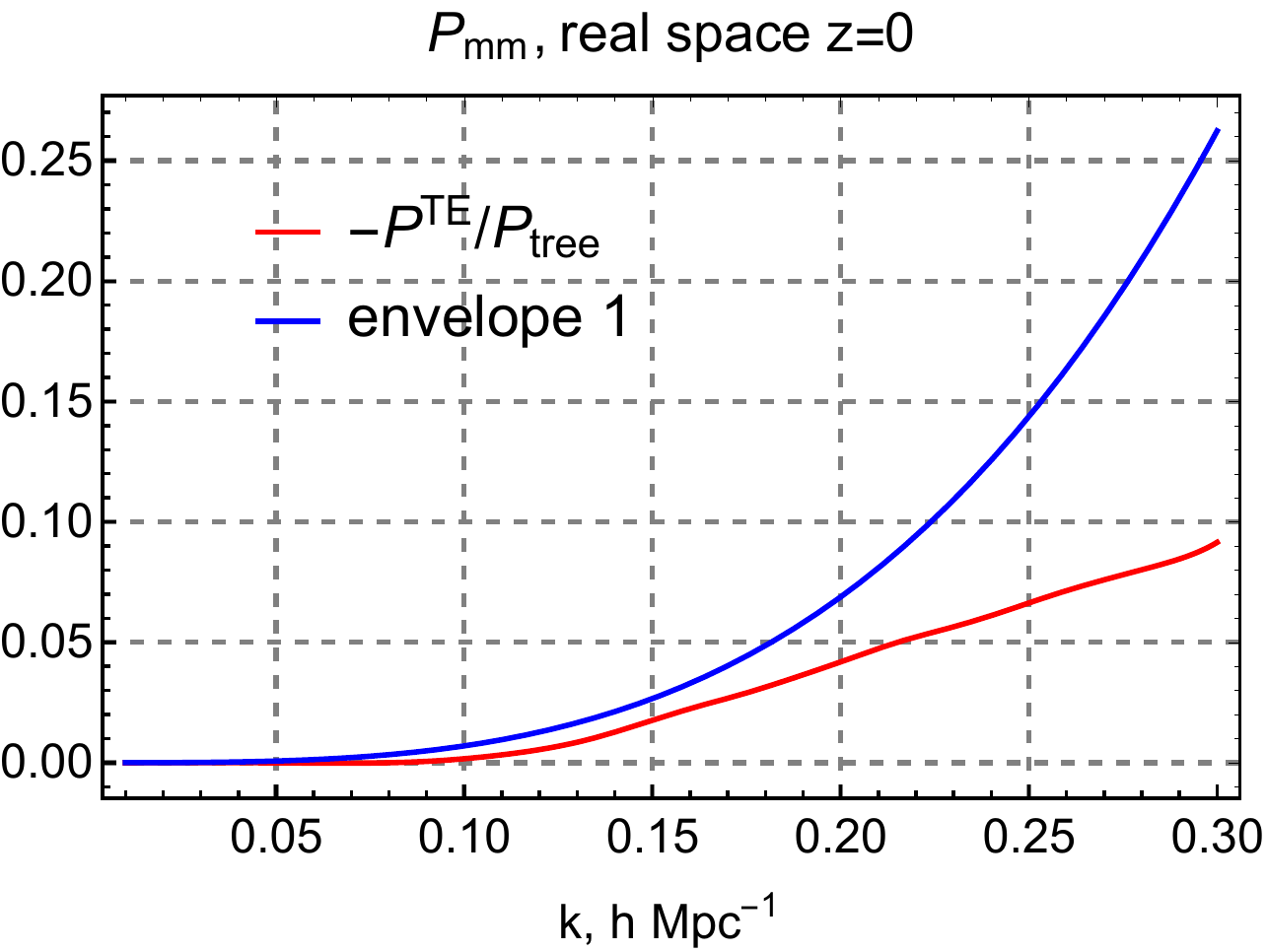}
		\includegraphics[width=0.49\textwidth]{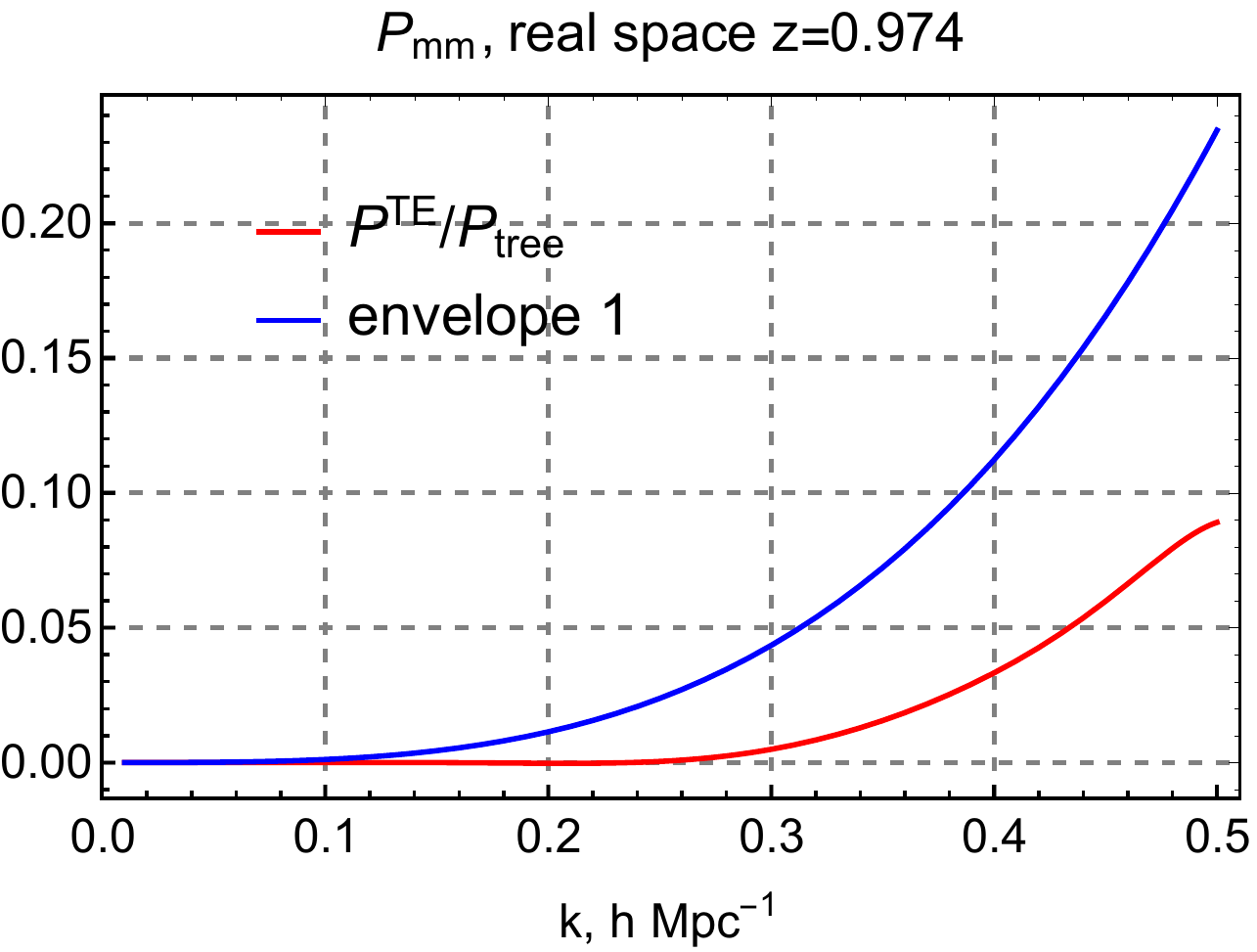}
		\end{center}
     	\caption{\label{fig:Edmrs}
Theoretical error envelopes for dark matter in real space, normalized 
to the tree-level spectra. 
We show the envelope suggested by Ref.~\cite{Baldauf:2016sjb} (envelope 1)
and the envelope $\bar P^{\rm (TE)}$ measured directly from the data. 
	}
\end{figure*}

\subsection{Comparison to perturbation theory}

The key ingredients of the theoretical error approach are the envelope for the 
theoretical error covariance $E(k)$ and the theoretical error mean $\bar P^{\rm (TE)}(k)$. 
By definition the theoretical error is the difference
between the full model describing the data and an analytic approximation to that model.
This suggests that a reasonable choice for the mean should be a residual between some simulation result (or the data) and the approximate model fitted to some sufficiently low $\kmax^{\rm fid.}$. The theoretical error covariance
is then used for $k>\kmax^{\rm fid.}$. 

As a second step, we must define the relationship between $E$ and $\bar P^{\rm (TE)}$. 
The toy model of the theoretical error entirely given by a higher-order counterterm suggests that 
the mean and the envelope should have the same order of magnitude. We impose an even stronger relation 
\be 
\label{eq:Ewe}
E(k,z)=\bar P^{\rm (TE)}(k,z)\,,
\ee
such that the theoretical error likelihood is characterized by a single shape.
The original work~\cite{Baldauf:2016sjb} assumed the zero mean and the following envelope:
\be
\label{eq:E1dm}
\text{envelope 1:}\quad E(k.z)=D^6(z)\left(\frac{k}{0.45~\hMpc} \right)^{3.3}P_{\rm lin}(k,z=0)\,,
\ee
where $D(z)$ is the linear growth factor.
This envelope is a smooth fit to the two-loop dark matter power spectrum computed in standard perturbation theory (SPT)~\cite{Bernardeau:2001qr}, after subtracting the leading UV part. 
It should be stressed that this two-loop SPT correction still has a strong residual unphysical sensitivity to the ultraviolet 
modes at the sub-leading order~\cite{Blas:2013bpa,Blas:2013aba,Foreman:2015lca}, and hence the actual effective field theory result is 
smaller that the SPT prediction. 
Nevertheless, we expect Eq.~\eqref{eq:E1dm} to be order-of-magnitude accurate.
In Fig.~\ref{fig:Edmrs} we compare the prediction of Eq.~\eqref{eq:E1dm}
to the envelope that we extracted from the data. 
We used $\kmax^{\rm fid.}=0.1~\hMpc$ for $z=0$ and $\kmax^{\rm fid.}=0.26~\hMpc$ for $z=0.974$.
Indeed, as expected, we can see that Eq.~\eqref{eq:E1dm} 
and our envelope~\eqref{eq:Ewe} agree by order of magnitude, but Eq.~\eqref{eq:E1dm} 
systematically 
overestimates the difference between the data and the prediction of one-loop perturbation theory by a factor of $\sim 3$.

\begin{figure*}[ht!]
	\begin{center}
		\includegraphics[width=1\textwidth]{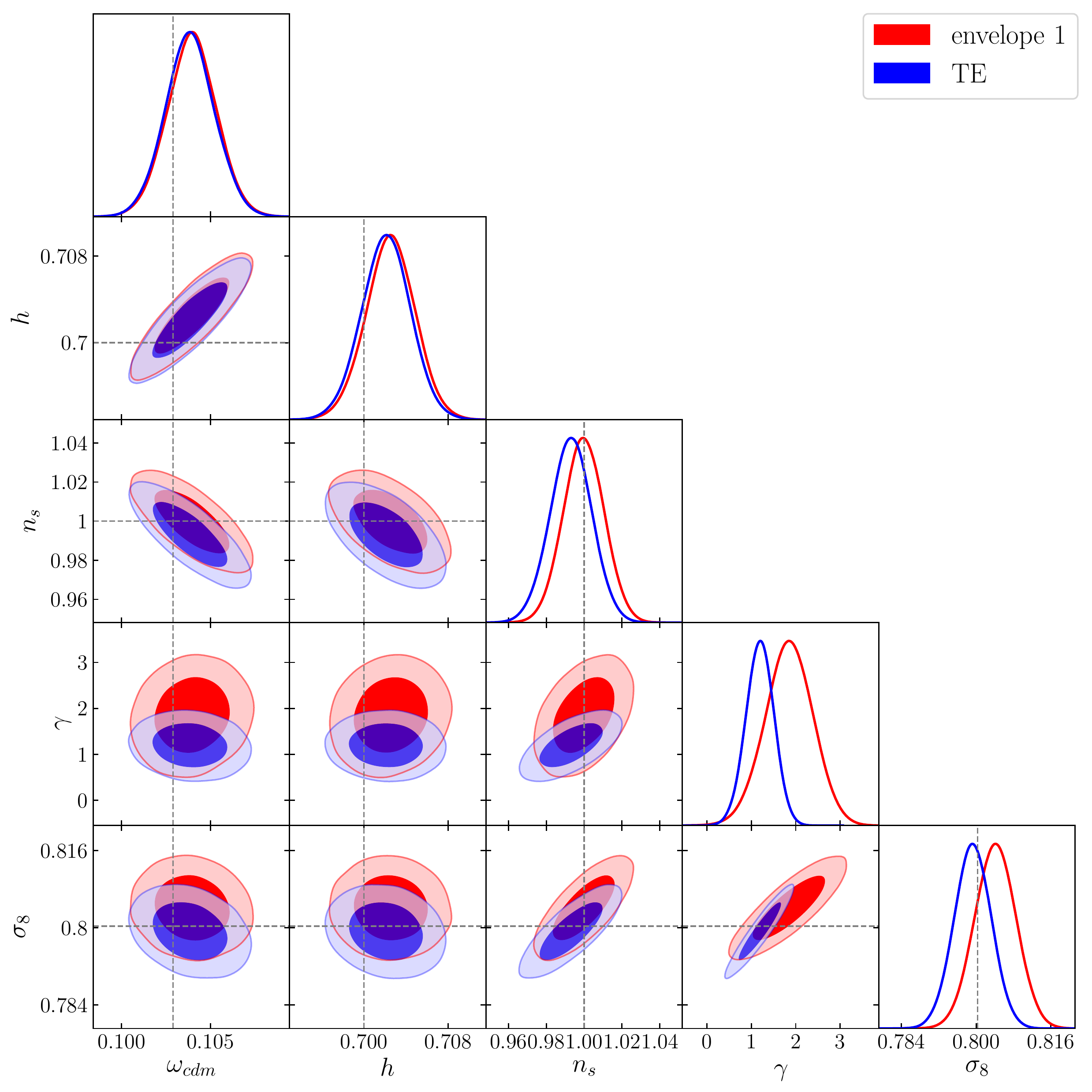}
		\end{center}
     	\caption{\label{fig:triang_old}
Triangle plot with cosmological constraints from the real space dark matter
power spectrum at $z=0$ for two choices of the theoretical error likelihood.
	}
\end{figure*}

\begin{table}[h!]
	\begin{center}
		\begin{tabular}{|c|c|c|c|}  \hline 
			\small{par.} & fid. & envelope 1 & TE \\  \hline\hline
			$\omega_{cdm}$ & $0.1029$  & $0.1040^{+1.4\cdot 10^{-3}}_{-1.4\cdot 10^{-3}}$   &
			$0.1039^{+1.4\cdot 10^{-3}}_{-1.5\cdot 10^{-3}}$ \\ \hline
			$h$ & $0.7$& $0.7025^{+2.4\cdot 10^{-3}}_{-2.3\cdot 10^{-3}}$ &
			$0.7021^{+2.4\cdot 10^{-3}}_{-2.4\cdot 10^{-3}}$ \\ \hline
			$n_s$ & $1$ & $0.9998^{+0.011}_{-0.011}$&
			$0.9930^{+0.011}_{-0.011}$ \\ \hline
			$A$ & $1$ & $0.9950^{+0.015}_{-0.016}$ &
			$0.9897^{+0.016}_{-0.016}$ \\ \hline
			$\Omega_m$ & $0.25$ & $0.2504^{+1.7\cdot10^{-3}}_{-1.7\cdot10^{-3}}$ &
			$0.2504^{+1.6\cdot10^{-3}}_{-1.7\cdot10^{-3}}$ \\ \hline
			$\sigma_8$  & $0.8003$ & $ 0.8041^{+4.5\cdot 10^{-3}}_{-4.5\cdot 10^{-3}}$ &
			$0.7992^{+4.1\cdot 10^{-3}}_{-4.1\cdot 10^{-3}}$ \\ \hline\hline
			$\gamma$ & -- & $1.85^{+0.56}_{-0.53}$ &
			$1.19^{+0.33}_{-0.32}$ \\ \hline
		\end{tabular}
		\caption{
			The marginalized 1d intervals for the cosmological parameters 
			estimated from the Las Damas real space dark matter power spectra at $z=0$ for two different theoretical error prescriptions. 
			The table contains fitted parameters (first column), fiducial values used 
			in simulations (second column), the results for the envelope 1 template with $\bar P^{\rm (TE)}=0$ (third columns)
			and that for the baseline theoretical error analysis TE (fourth column). 
			$\gamma$ is quoted in units $[\Mpch]^2$.
		}
		\label{tab:cos_dm_rs_old}
	\end{center}
\end{table} 

It is instructive to compare the two different theoretical error prescriptions
at the level of the cosmological constraints. To that end we have run the cosmological analysis of the
real space dark matter power spectrum at $z=0$ with $\bar P^{\rm (TE)}=0$ and the envelope~Eq.~\eqref{eq:E1dm}.
The resulting 2d posteriors are shown in Fig.~\ref{fig:triang_old} whereas 1d marginalized constraints are listed in Tab. \ref{tab:cos_dm_rs_old}. One can see that only the measurements of $\sigma_8$
and the effective sound speed
are significantly affected. This shows that the use of the perturbation theory-inspired templates can 
overestimate the actual errorbars on the amplitude parameters. However, the shape and distance parameters
$\omega_{cdm},n_s,h$
are expected to be less affected by this choice. 
A similar situation was found in Ref.~\cite{Philcox:2020vvt}, which 
showed that the BAO measurements using the theoretical error covariance
do not depend on exact shape of the 
theoretical error envelope.

The situation becomes more complicated for redshift space multipoles,
where the complete two-loop calculation has not yet been done.
Thus, even the perturbation theory estimates can be very uncertain. 
One can consider two possible estimates,
\be
\label{eq:E2dm}
\begin{split}
&\text{envelope 1:}\quad E_\ell(k,z)=D^6(z)\left(\frac{k}{0.45~\hMpc} \right)^{3.3}P_{\ell,~{\rm tree}}(k,z=0)\,,\\
&\text{envelope 2:}\quad E_\ell(k,z)=\frac{P_{\ell,~\rm 1-loop}^2(k,z)}{P_{\ell,~{\rm tree}}(k,z)}\,.
\end{split}
\ee
The envelope 2 is based on the there is a disconnected
two-loop diagram which involves a product of two one-loop diagrams, but without
an extra propagator $P_{\ell,~{\rm tree}}$.

\begin{figure*}[ht!]
	\begin{center}
		\includegraphics[width=0.49\textwidth]{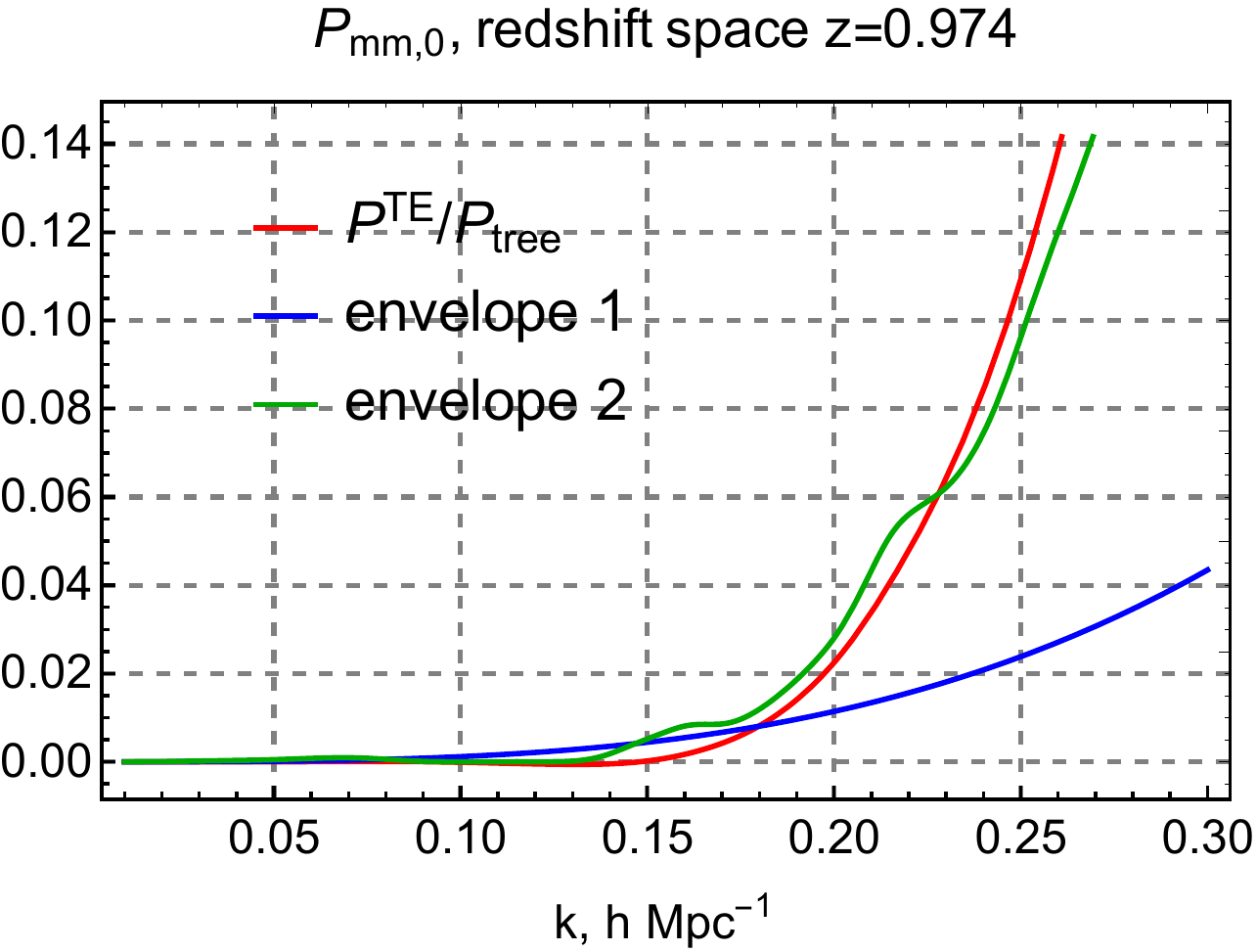}
		\includegraphics[width=0.49\textwidth]{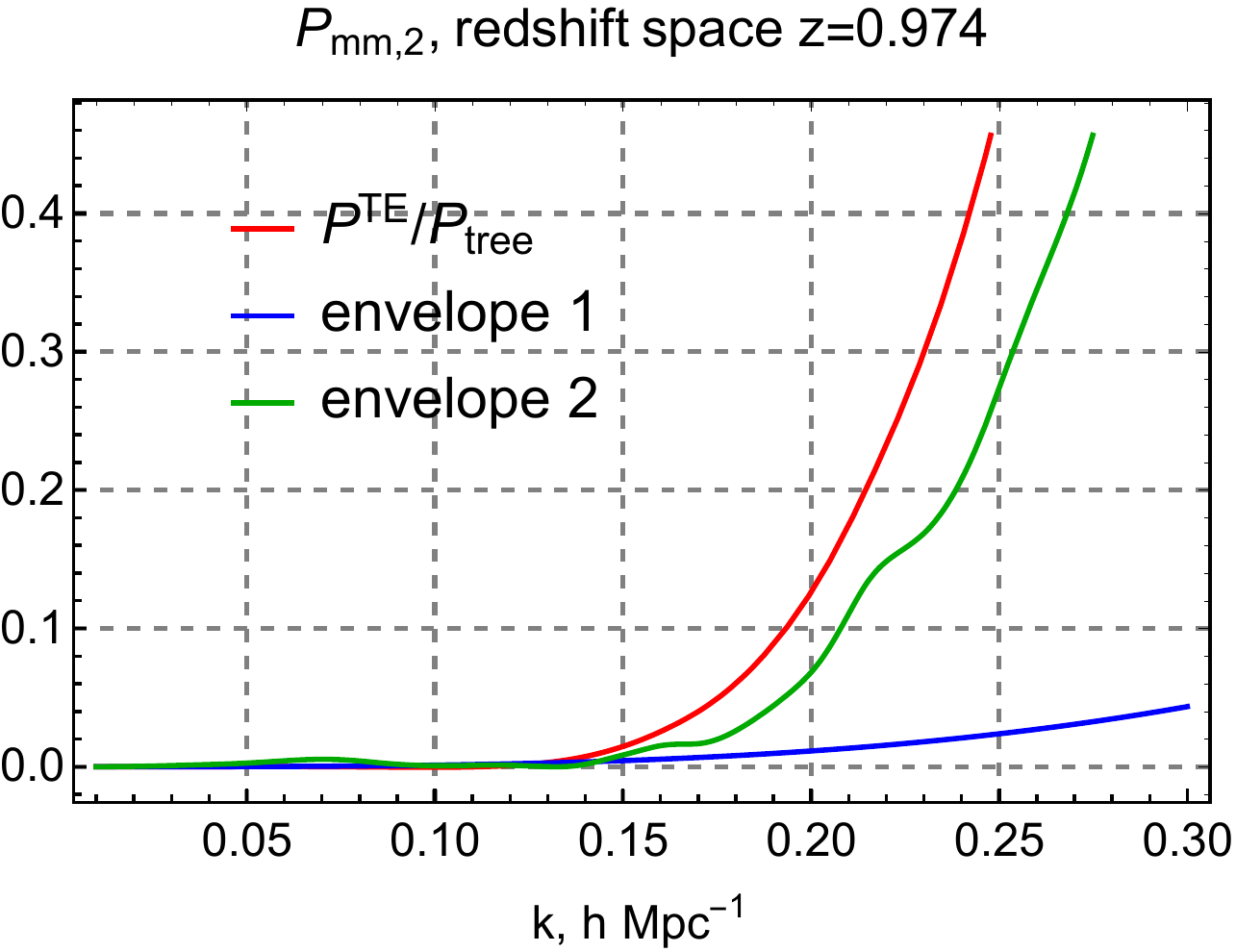}
		\end{center}
     	\caption{\label{fig:Edmzs}
Theoretical error envelopes for the monopole (left panel) and quadrupole 
(right panel) moments of the power spectrum of dark matter in redshift space, normalized to the 
tree-level spectra.
See Eq.~\eqref{eq:E2dm} for the description of the estimates.
	}
\end{figure*}
\begin{figure*}[ht!]
	\begin{center}
		\includegraphics[width=0.49\textwidth]{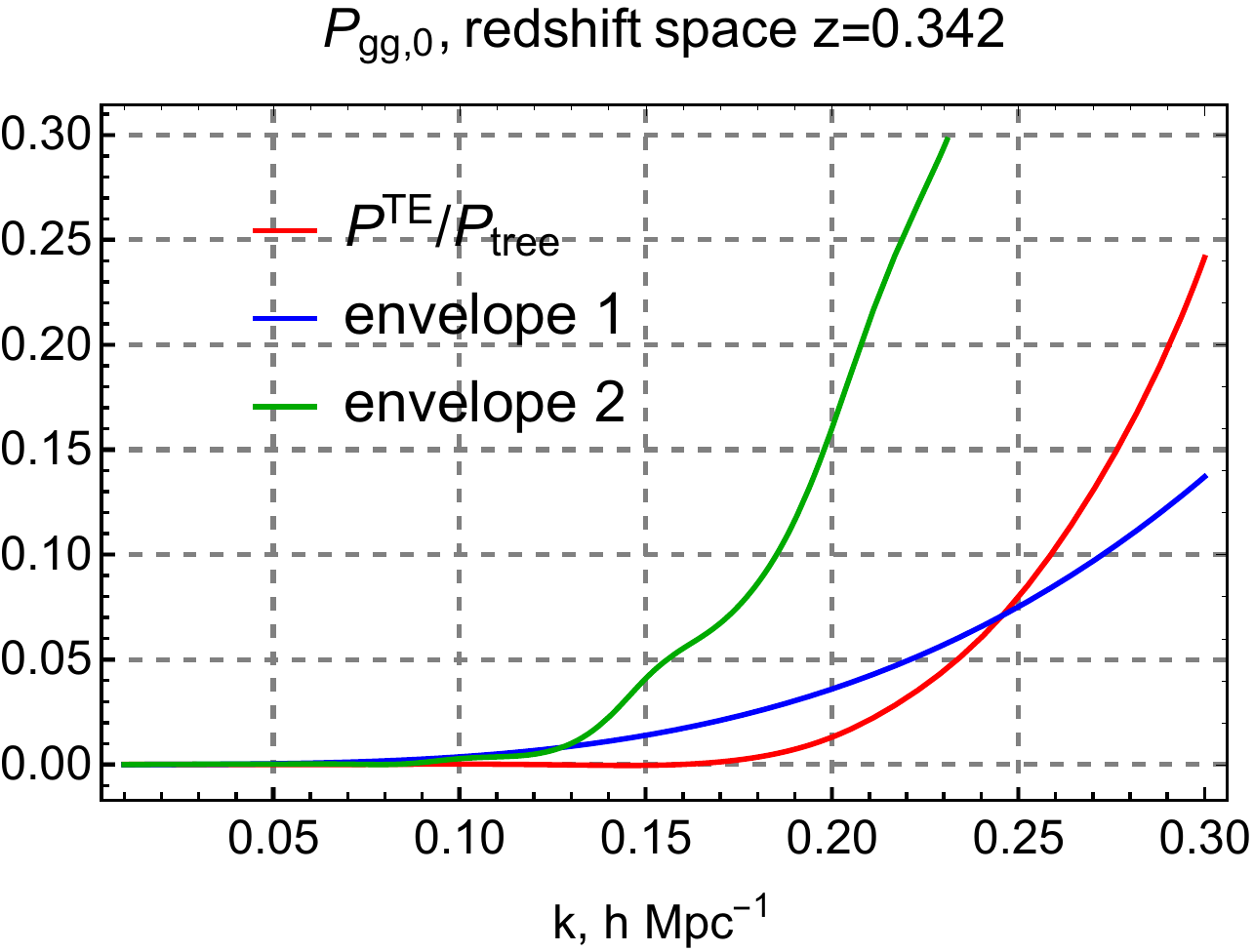}
		\includegraphics[width=0.49\textwidth]{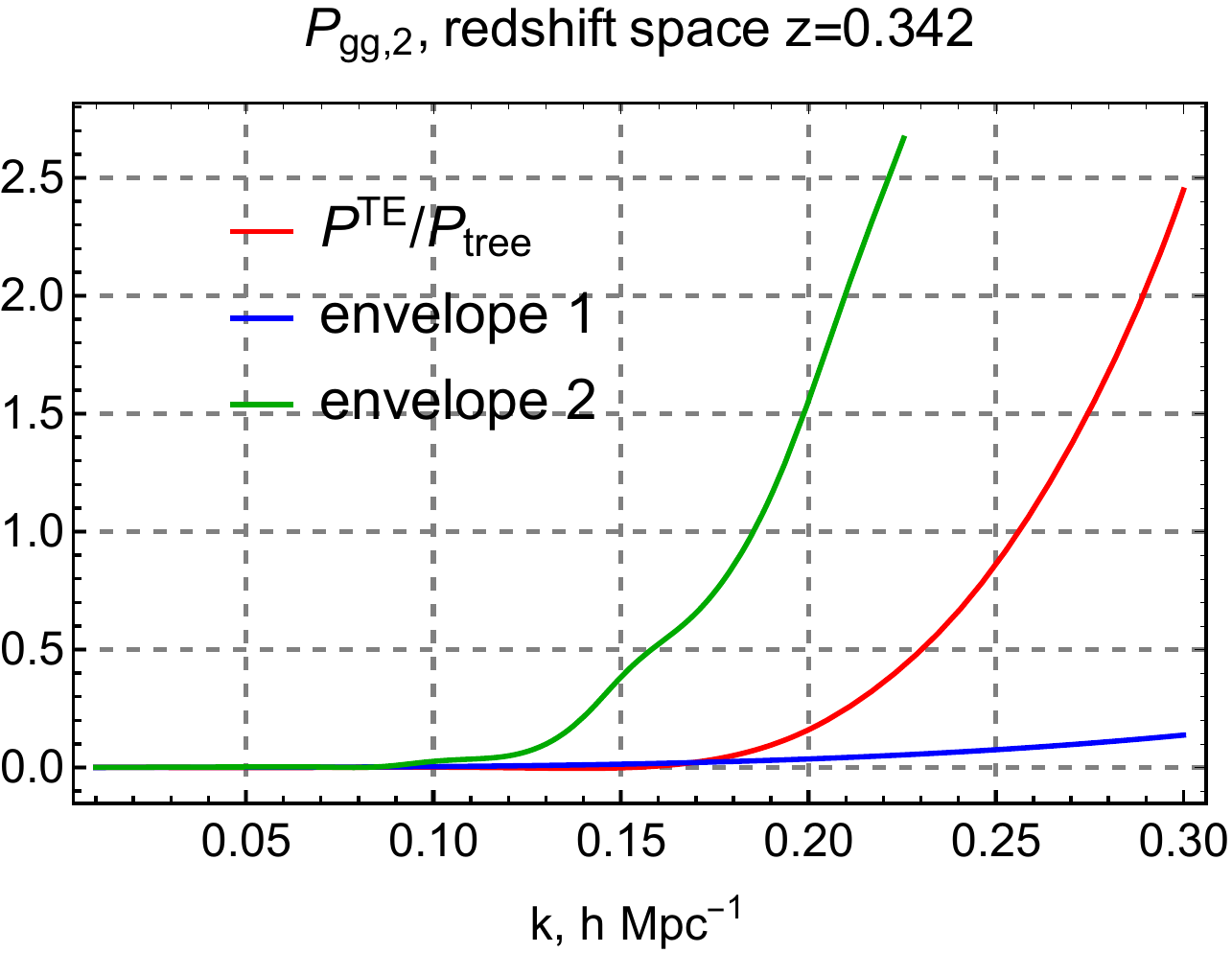}
		\end{center}
     	\caption{\label{fig:Edgalzs}
Same as Fig.~\ref{fig:Edmzs} but for galaxies in redshift space.
	}
\end{figure*}

Comparing these two estimates with our baseline choice (at $\kmax^{\rm fid.}=0.12~\hMpc$) in Fig.~\ref{fig:Edmzs},
one sees that `envelope 2' describes the actual difference between the 
1-loop PT model and the data surprisingly well. On the contrary, `envelope 1' underestimates the theoretical error
quite significantly, by more than on order of magnitude. 

Finally, Fig.~\ref{fig:Edgalzs} shows the envelopes for the redshift space galaxies. Our envelope 
is computed using $\kmax^{\rm fid.}=0.16~\hMpc$.
We can see that 
`envelope 1' agrees well our theoretical error for the monopole, while `envelope 2' 
largely overestimates it on short scales. In contrast, `envelope 1' significantly underestimates
the theoretical error for the quadrupole, whereas `envelope 1' still overestimates it;
our envelope lies in between these two perturbation theory estimates.
All in all, we conclude that the redshift space theoretical error envelopes 
that we use in this paper roughly agree with the perturbation
theory estimates, but the latter are quite uncertain.

\begin{figure*}[ht!]
	\begin{center}
		\includegraphics[width=0.49\textwidth]{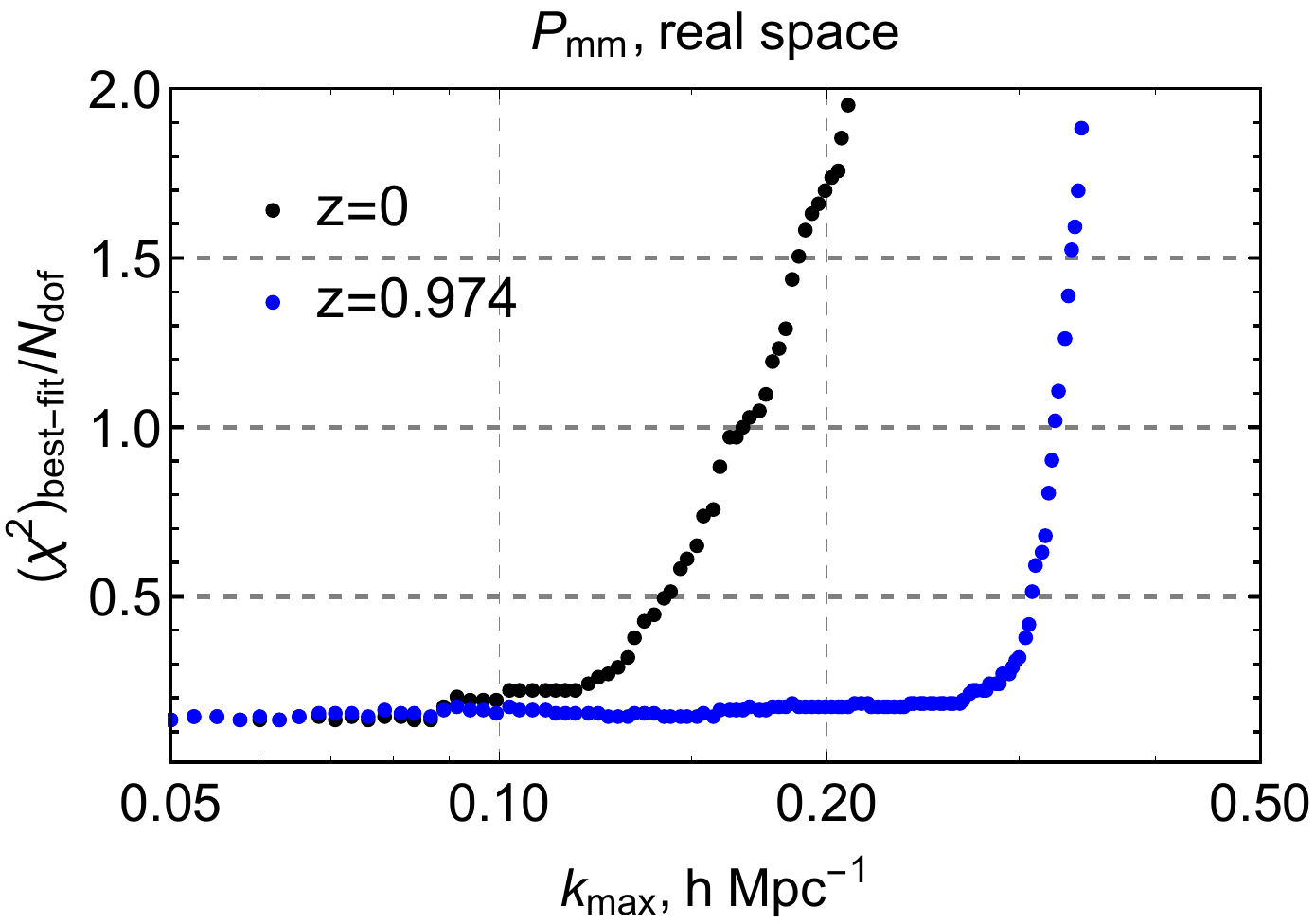}
		\includegraphics[width=0.49\textwidth]{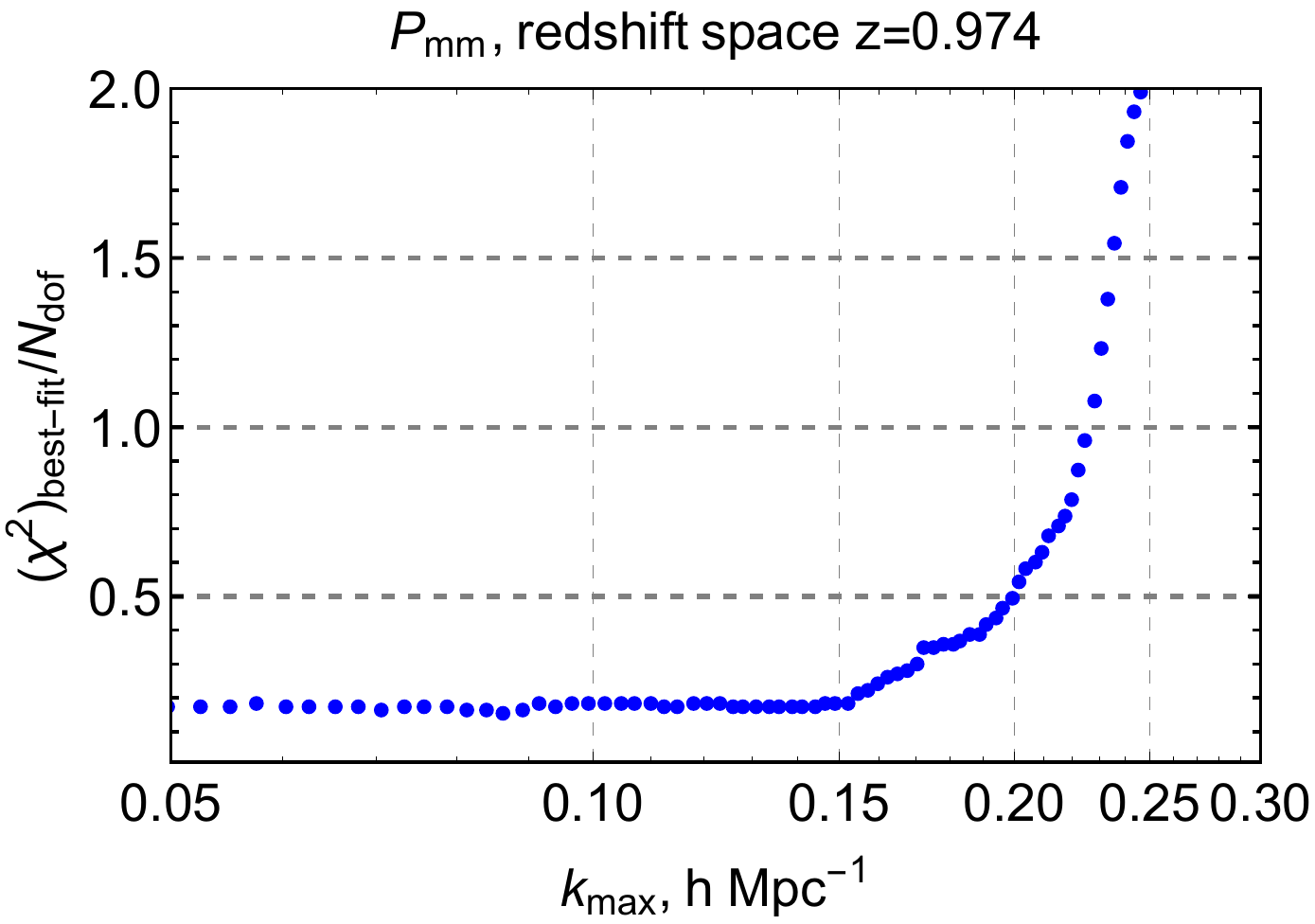}
		\includegraphics[width=0.49\textwidth]{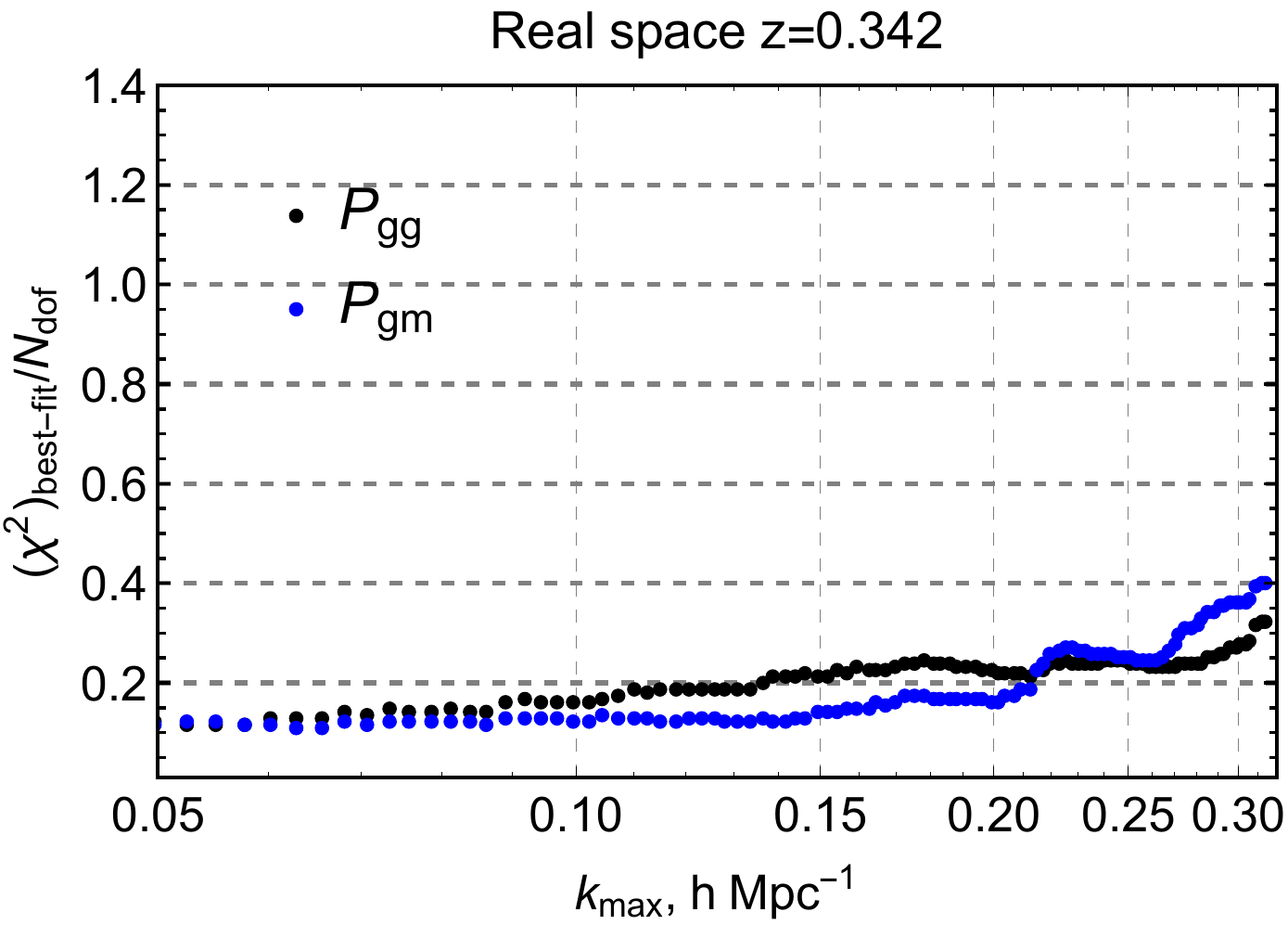}
		\includegraphics[width=0.49\textwidth]{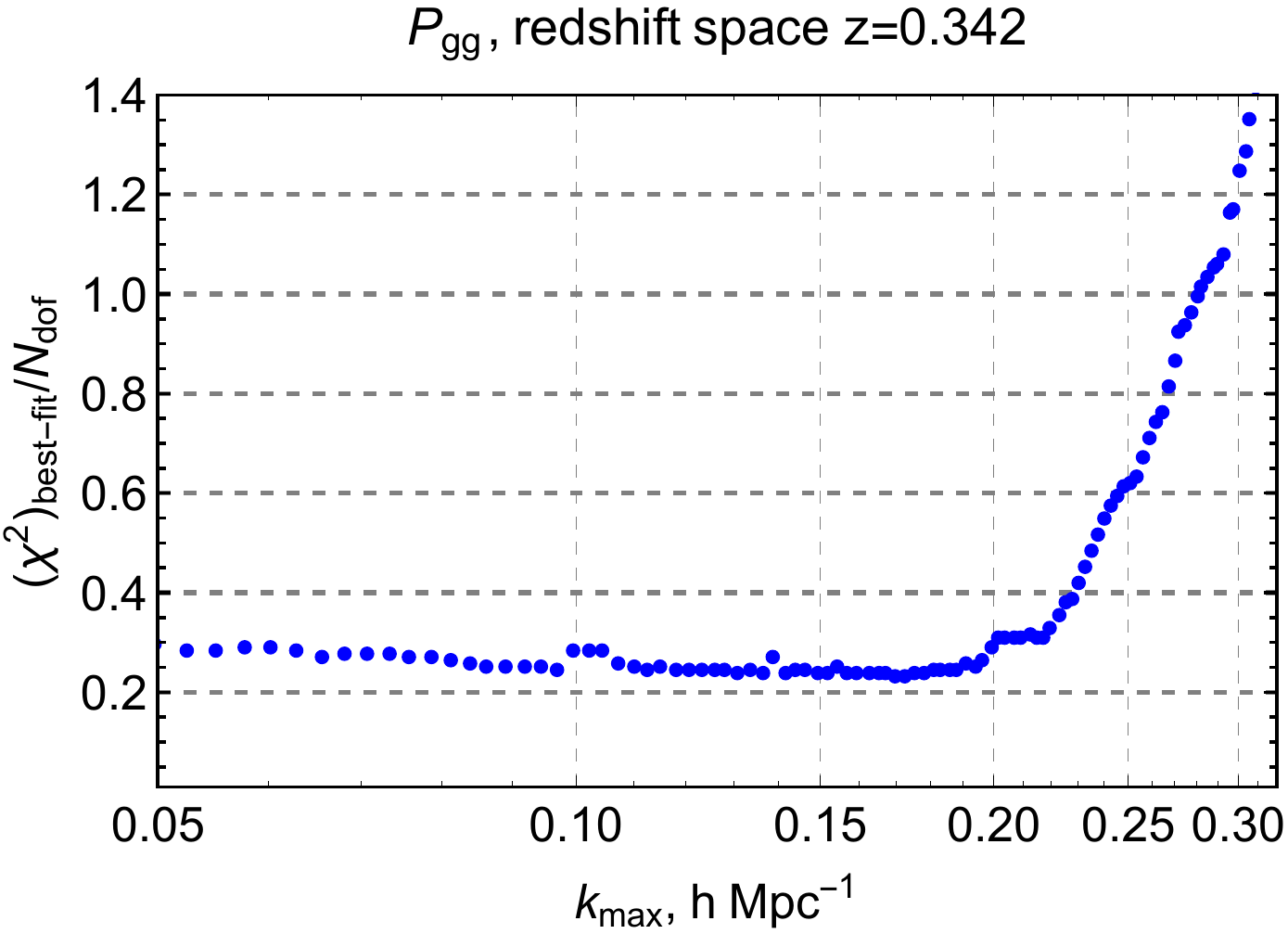}
		\end{center}
     	\caption{
     	\label{fig:chisq_dm_rs}
     	Best-fit reduced $\chi^2/N_{\rm dof}$
     	as a function of $\kmax$ for various likelihoods considered in this paper. 
	}
\end{figure*}

\subsection{Choice of $\kmax^{\rm fid.}$}

In  this section we argue our choice of $\kmax^{\rm fid.}$ in the theoretical error analyses.
For that, we fix the cosmological
parameters to their fiducial 
values and find the best-fitting
values of nuisance parameters 
varying $\kmax$. Then, we extract 
the best-fit reduced 
statistic $\chi^2/N_{\rm dof}$ ($N_{\rm dof}=N_{\rm bins}-N_{\rm params}$)
for each $\kmax$. 
The results for the real space dark matter are shown
in the upper left panel of Fig.~\ref{fig:chisq_dm_rs}.
Note that the typical values of $\chi^2/N_{\rm dof}$ are around $0.2$, which is a result of using 
the reduced volume
$V=100~(\text{Gpc}/h)^3$
in the covariance instead of the 
true cumulative volume of the LasDamas Oriana simulation
$V=553~(\text{Gpc}/h)^3$.
We see that the $\chi^2$ profile 
blows up at $\kmax=0.14~\hMpc$
for $z=0$ and at 
$\kmax=0.30~\hMpc$ at $z=0.974$.
This is a indication that the one-loop 
perturbation theory model 
becomes invalid at these scales. 
Thus, we choose smaller 
fiducial cuts $\kmax^{\rm fid.}$,
where the $\chi^2$ profile is still flat: $\kmax^{\rm fid.}=0.10~\hMpc$
for $z=0$ and 
$\kmax^{\rm fid.}=0.26~\hMpc$ at $z=0.974$. 
Operationally, it is suggestive to use 
\be 
\kmax^{\rm fid.}=k_{\rm 2-loop}-\Delta k'\,,\quad 
\Delta k'=0.04~\hMpc\,.
\ee 

The fiducial $\kmax$ for other cases 
are chosen in a completely similar fashion.
The results for dark matter in redshfit space are shown in the upper right panel of Fig.~\ref{fig:chisq_dm_rs}. We found that the reduced $\chi^2/N_{\rm dof}$ statistics
remains flat up to $\kmax=0.16~\hMpc$. However, in order to be conservative, we choose $\kmax^{\rm fid.}=0.12~\hMpc$ in our analysis. 

The results for the real space galaxies are shown in the lower left panel of Fig.~\ref{fig:chisq_dm_rs}. In this case, the picture 
is not so obvious due to the large shot noise contribution in the covariance, which can be larger than the theoretical error covariance.
For this reason, 
one-loop perturbation theory provides accurate description of galaxy power spectrum up to nonlinear scale $\kmax\approx 0.3~\hMpc$. Since we do
not see any sign of the bias up to $\kmax=0.3\hMpc$, we conclude 
that the theoretical error is negligibly 
smaller than the statistical covariance dominated by shot noise.
The situation is somewhat different for galaxy-matter cross spectrum for which the 
$\chi^2$ profile shows some
scale-dependence  for
$\kmax>0.22~\hMpc$. In this case, we choose $\kmax^{\rm fid.}=0.18~\hMpc$ in our theoretical error analysis of $P_{\rm gm}$. 

Finally, we display the bestfit $\chi^2/N_{\rm dof}$ profile for galaxies in redshift space in the lower right panel of Fig.~\ref{fig:chisq_dm_rs}. One can see that the $\chi^2$ profile blows up at $\kmax>0.2~\hMpc$. Given this reason, we choose $\kmax^{\rm fid.}=0.16~\hMpc$. 

It should be stressed that using the $\chi^2(\kmax)$ profile
is inappropriate for defining 
the baseline cut of the $\kmax$-analysis because the fit can be biased at $\kmax$ lower than $k_{\rm 2-loop}$ when the cosmological parameters are varied. Indeed, we see that $k_{\rm 2-loop}$ is typically larger 
than the baseline cuts $\kmax$
in our main analyses.
Hence, using the $\chi^2$ profile is appropriate only for the theoretical error and not to 
fix the baseline $\kmax$.

\section{Supplementary material on the fingers-of-God modeling}
\label{app:fog}

We show the 1d marginalized
limits on the parameters of different FoG models from the redshift-space spectrum analyses for $z=0$ in 
Table~\ref{tab:Sigma_z0}
and for $z=0.974$
in Table~\ref{tab:Sigma_z1}.

\begin{figure*}[ht!]
	\begin{center}
		\includegraphics[width=1\textwidth]{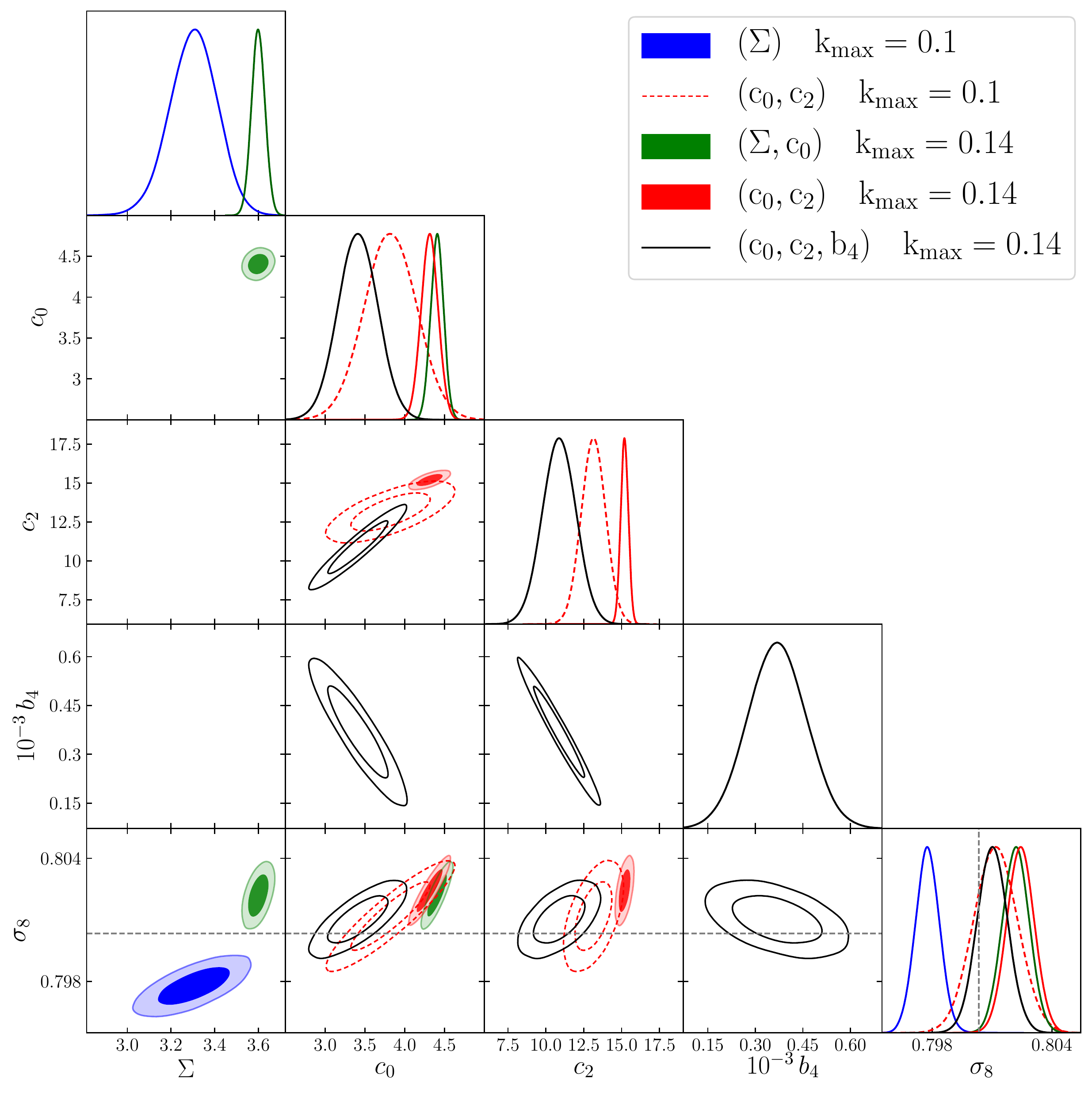}
		\end{center}
     	\caption{\label{fig:sigma2}
Same as Fig.~\ref{fig:rsdSigma_z1}, but with the additional ``EFT+FoG'' model~\eqref{eq:fog2}, characterized by the parameters $(c_0,\Sigma)$.
	}
\end{figure*}

Fig.~\ref{fig:sigma2}
displays the parameter 
constraints for different FoG models including ``EFT+FoG'',
which are extracted from the
matter redshift space
power spectrum at $z=0.974$.

\begin{table}[h!]
	\begin{center}
		\begin{tabular}{|c|c|c|c|c|}  \hline 
			\small{par.} & fid. & \multicolumn{3}{|c|}{$z=0$} \\  \hline
			&  & $(\Sigma)$ & $(c_0,c_2)$  & $(c_0,c_2,b_4)$ \\  \hline\hline
			$A$ & $1$ & 
			$0.9873^{+1.5\cdot10^{-3}}_{-1.5\cdot10^{-3}}$&
			$1.0063^{+2.9\cdot10^{-3}}_{-2.9\cdot10^{-3}}$ &
			$1.0031^{+3.3\cdot10^{-3}}_{-3.3\cdot10^{-3}}$ \\ \hline
			$\sigma_8$  & $0.8003$ & 
			$0.7952^{+6.0\cdot 10^{-4}}_{-6.0\cdot 10^{-4}}$ &
			$0.8028^{+1.2\cdot 10^{-3}}_{-1.2\cdot 10^{-3}}$ &
			$0.8015^{+1.3\cdot 10^{-3}}_{-1.3\cdot 10^{-3}}$ \\ \hline\hline
			$\Sigma$ & -- & 
			$7.75^{+0.15}_{-0.15}$ &
			-- &
			-- \\ \hline
			$c_0$ & -- & 
			-- &
			$5.30^{+0.28}_{-0.27}$ &
			$4.19^{+0.60}_{-0.60}$ \\ \hline
			$c_2$ & -- & 
			-- &
			$25.71^{+0.94}_{-0.95}$ &
			$17.12^{+4.30}_{-4.21}$ \\ \hline
			$10^{-3}b_4$ & -- & 
			-- &
			-- &
			$12.63^{+5.99}_{-6.14}$ \\ \hline
		\end{tabular}
		\caption{
			The marginalized 1d intervals for the amplitude and  nuisance parameters 
			estimated from the monopole and quadrupole moments of the Las Damas dark matter redshift space power spectrum
			at $z=0$ for fixed fiducial cosmology. 
			We show the fitted parameters (first column), fiducial values used 
			in simulations (second column), and the resulting parameter constraints in different models: $(\Sigma)$ (third column) , $(c_0,c_2)$ (fourth column) and $(c_0,c_2,b_4)$ (fifth column) at $k_{\max}=0.1\,h$/Mpc.	$c_0,c_2,b_4$ are quoted in units $[\Mpch]^2,~[\Mpch]^2,~[\Mpch]^4$, respectively.
		}
		\label{tab:Sigma_z0}
	\end{center}
\end{table} 

\begin{table}[h!]
	\begin{center}
		\begin{tabular}{|c|c|c|c|c|c|}  \hline 
			\small{par.}  & \multicolumn{5}{|c|}{$z=0.974$} \\  \hline
			              & \multicolumn{2}{|c|}{$\kmax=0.1\,h$/Mpc} & \multicolumn{3}{|c|}{$\kmax=0.14\,h$/Mpc} \\  \hline
			&  $(\Sigma)$ & $(c_0,c_2)$  & $(\Sigma,c_0)$ & $(c_0,c_2)$ & $(c_0,c_2,b_4)$ \\  \hline\hline
			$A$ & 
			$\!\!0.9936^{+1.5\cdot10^{-3}}_{-1.5\cdot10^{-3}}\!\!$ &
			$\!\!1.0021^{+2.8\cdot10^{-3}}_{-2.8\cdot10^{-3}}\!\!$ &
			$\!\!1.0046^{+1.7\cdot10^{-3}}_{-1.7\cdot10^{-3}}\!\!$ &
			$\!\!1.0052^{+1.8\cdot10^{-3}}_{-1.8\cdot10^{-3}}\!\!$ &
			$\!\!1.0017^{+1.9\cdot 10^{-3}}_{-2.0\cdot 10^{-3}}\!\!$\\ \hline
			$\sigma_8$  & 
			$\!\!0.7977^{+6.2\cdot 10^{-4}}_{-6.1\cdot 10^{-4}}\!\!$ &
			$\!\!0.8011^{+1.1\cdot 10^{-3}}_{-1.1\cdot 10^{-3}}\!\!$ &
			$\!\!0.8021^{+6.9\cdot 10^{-4}}_{-6.9\cdot 10^{-4}}\!\!$ &
			$\!\!0.8024^{+7.0\cdot 10^{-4}}_{-7.0\cdot 10^{-4}}\!\!$ &
			$\!\!0.8010^{+7.8\cdot 10^{-4}}_{-7.9\cdot 10^{-4}}\!\!$  \\  \hline\hline
			$\Sigma$ & 
			$3.30^{+0.11}_{-0.11}$ &
			-- &
			$3.60^{+0.03}_{-0.03}$ &
			-- &
			-- \\ \hline
			$c_0$ & 
			-- &
			$3.82^{+0.34}_{-0.34}$ &
			$0.94^{+0.08}_{-0.08}$ &
			$4.31^{+0.11}_{-0.11}$ &
			$3.41^{+0.26}_{-0.26}$\\ \hline
			$c_2$ & 
			-- &
			$13.14^{+0.83}_{-0.82}$ &
			-- &
			$15.19^{+0.30}_{-0.30}$ &
			$10.88^{+1.15}_{-1.13}$\\ \hline
			$10^{-3}b_4$ & 
			-- &
			-- &
			-- &
			-- &
			$0.37^{+0.09}_{-0.10}$\\ \hline
		\end{tabular}
		\caption{
			The marginalized 1d intervals for the amplitude and  nuisance parameters 
			estimated from the monopole and quadrupole moments of the Las Damas dark matter redshift space power spectrum
			at $z=0.974$ for fixed fiducial cosmology. 
			We show the fitted parameters (first column), the resulting parameter constraints in different models: $(\Sigma)$ (second column), $(c_0,c_2)$ (third column) at $k_{\max}=0.1\,h$/Mpc and $(\Sigma,c_0)$ (fourth column), $(c_0,c_2)$ (fifth column), $(c_0,c_2,b_4)$ (sixth column) at $k_{\max}=0.14\,h$/Mpc. $c_0,c_2,b_4$ are quoted in units $[\Mpch]^2,~[\Mpch]^2,~[\Mpch]^4$, respectively.
		}
		\label{tab:Sigma_z1}
	\end{center}
\end{table} 

\section{Dependence on fiducial cosmology}
\label{app:fid}

In our algorithm, the theory prediction for the TE mean has been calculated using 
the {\it true} cosmology of mock catalogs. As the actual cosmological parameters are {\it a prior} unknown, we need to check whether our results depend on the choice of the fiducial cosmology. 
To estimate the corresponding uncertainty we repeat the TE analysis with a different fiducial cosmology.
To that end we use a  
set of cosmological 
parameters from a
randomly chosen 
step from MCMC
chains of the baseline $\kmax$-analyses
for dark matter
and galaxies in real and redshift spaces.
In each of these cases, we require that this new parameter set
deviates noticeably from the true mock cosmology, but 
still 
stays within the \mbox{$95\%$ CL}. Then, for this chosen cosmology we calculate the TE mean following the algorithm described in the main text. We will fit the real space dark matter ($z=0$), redshift space dark matter ($z=0.974$) and redshift space galaxies ($z=0.342$) mock data and compare the resulting posterior distribution with the results of Sec. \ref{sec:dm_rs}, \ref{sec:dm_rsd} and \ref{sec:gal_rsd}.

\begin{figure}[ht!]
	\begin{center}
		\includegraphics[width=1\textwidth]{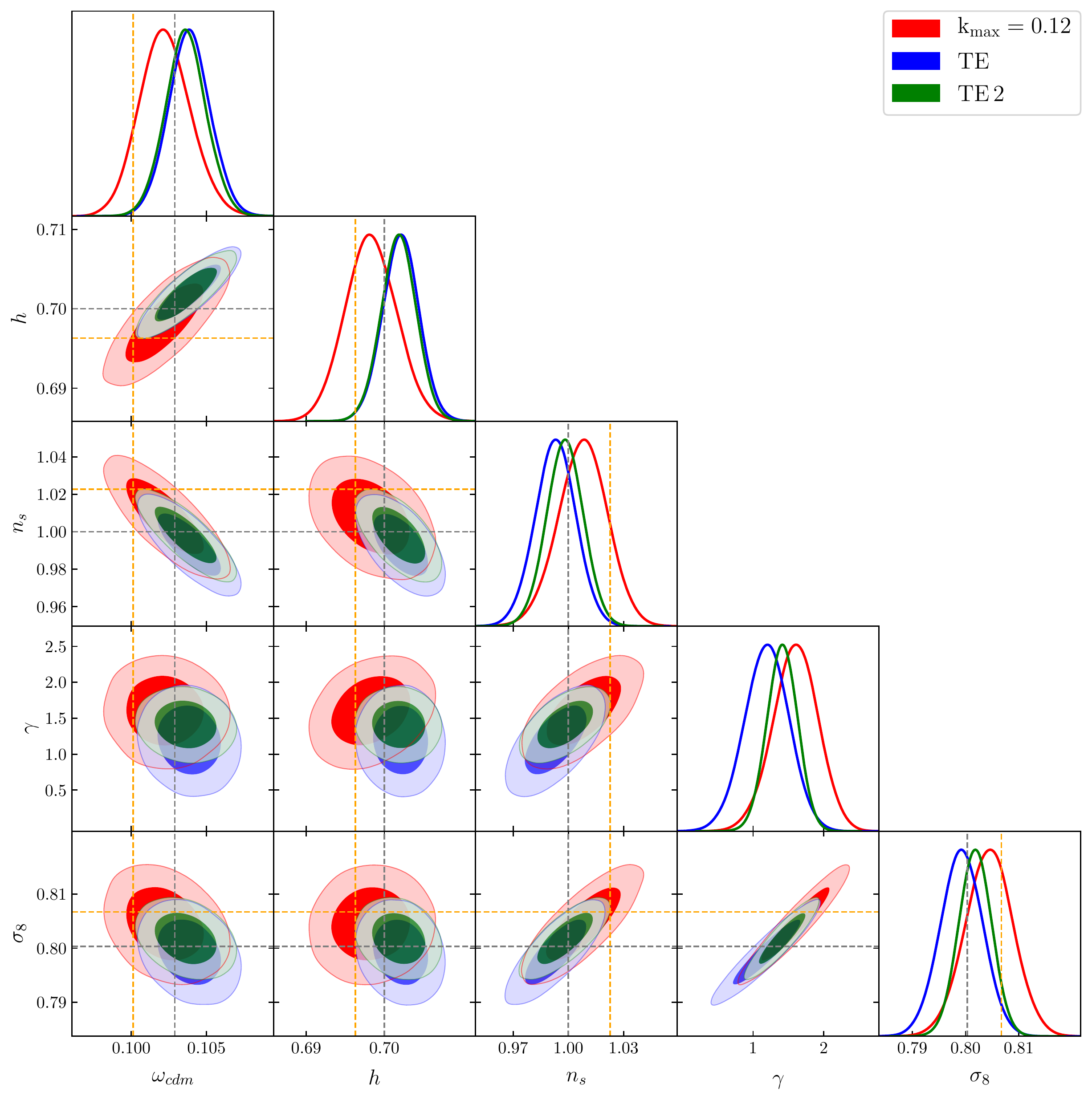}
		\end{center}
     	\caption{\label{fig:dm_rs_z0_fid}
Triangle plot for the cosmological and nuisance parameters measured from the real space dark matter
power spectrum of the Las Damas simulations at $z=0$ for the $\kmax$-analysis and two choices of the fiducial cosmology in the TE approach. The dashed orange lines mark the fiducial cosmological parameters used in the $\rm TE\,2$ analysis, whereas the grey dashed lines mark the 
fiducial cosmology used 
in the baseline analysis.
It coincides with the true
mock cosmology.
	}
\end{figure}

Let us begin with the real space dark matter ($z=0$). The marginalized 1d parameter constraints are listed in Tab. \ref{tab:cos_dm_real_fid} (fifth column).
\begin{table}[h!]
	\begin{center}
		\begin{tabular}{|c|c|c|c|c|}  \hline 
			\small{par.} & fid. & \multicolumn{3}{|c|}{$z=0$}\\  \hline
			&  & $k_{\text{max}}=0.12$ & TE & TE 2 \\  \hline\hline
			$\omega_{cdm}$ & $0.1029$  & $0.1023^{+1.6\cdot 10^{-3}}_{-1.8\cdot 10^{-3}}$   &
			$0.1039^{+1.4\cdot 10^{-3}}_{-1.5\cdot 10^{-3}}$ &
			$0.1036^{+1.4\cdot 10^{-3}}_{-1.4\cdot 10^{-3}}$ \\ \hline
			$h$ & $0.7$& $0.6983^{+3.3\cdot 10^{-3}}_{-3.4\cdot 10^{-3}}$ &
			$0.7021^{+2.4\cdot 10^{-3}}_{-2.4\cdot 10^{-3}}$ &
			$0.7018^{+2.3\cdot 10^{-3}}_{-2.3\cdot 10^{-3}}$ \\ \hline
			$n_s$ & $1$ & $1.008^{+0.014}_{-0.013}$&
			$0.9930^{+0.011}_{-0.011}$ &
			$0.9981^{+0.010}_{-0.010}$ \\ \hline
			$A$ & $1$ & $1.014^{+0.020}_{-0.020}$ &
			$0.9897^{+0.016}_{-0.016}$ &
			$0.9952^{+0.015}_{-0.015}$ \\ \hline
			$\Omega_m$ & $0.25$ & $0.2499^{+2.2\cdot10^{-3}}_{-2.3\cdot10^{-3}}$ &
			$0.2504^{+1.6\cdot10^{-3}}_{-1.7\cdot10^{-3}}$ &
			$0.2502^{+1.6\cdot10^{-3}}_{-1.7\cdot10^{-3}}$ \\ \hline
			$\sigma_8$  & $0.8003$ & $ 0.8044^{+4.6\cdot 10^{-3}}_{-4.5\cdot 10^{-3}}$ &
			$0.7992^{+4.1\cdot 10^{-3}}_{-4.1\cdot 10^{-3}}$ &
			$0.8018^{+3.2\cdot 10^{-4}}_{-3.2\cdot 10^{-4}}$ \\ \hline
			$\gamma$ & -- & $1.60^{+0.33}_{-0.31}$ &
			$1.19^{+0.33}_{-0.32}$ &
			$1.41^{+0.22}_{-0.22}$ \\ \hline
		\end{tabular}
		\caption{
			The marginalized 1d intervals for the cosmological parameters 
			estimated from the Las Damas real space dark matter power spectra at $z=0$. 
			The table contains fitted parameters (first column), fiducial values used 
			in simulations (second column), the results of 
			the baseline $k_{\text{max}}$ analysis (third column) and outcomes of TE approaches with default (fourth column) and new fiducial cosmology (fifth column) termed $\rm TE\,2$.
			$\gamma$ is quoted in units $[\Mpch]^2$.
		}
		\label{tab:cos_dm_real_fid}
	\end{center}
\end{table}
The 2d posterior distributions are shown in Fig. \ref{fig:dm_rs_z0_fid} (green contours).
We have found that posterior distributions for all parameters except for $\sigma_8$, $n_s$ and $\gamma$ are not altered by the different choice of the fiducial cosmology. However, the error bars on $\sigma_8$, $n_s$ and $\gamma$ decrease by $20\%$, $10\%$ and $30\%$, respectively, compared to the baseline TE analysis. These changes can be readily understood. First, $\sigma_8$ and $n_s$ control the amplitude of the theoretical error envelope $E$. It implies that
the uncertainty in the 
choice of fiducial cosmology
for $E$ propagates into uncertainty for $\sigma_8$ and $n_s$ measurements. Second, $\sigma_8$ and $\gamma$ are strongly degenerate, and hence the aforementioned 
uncertainty propagates into the $\gamma$ constraint
and worsens it as well.

\begin{figure}[ht!]
	\begin{center}
		\includegraphics[width=1\textwidth]{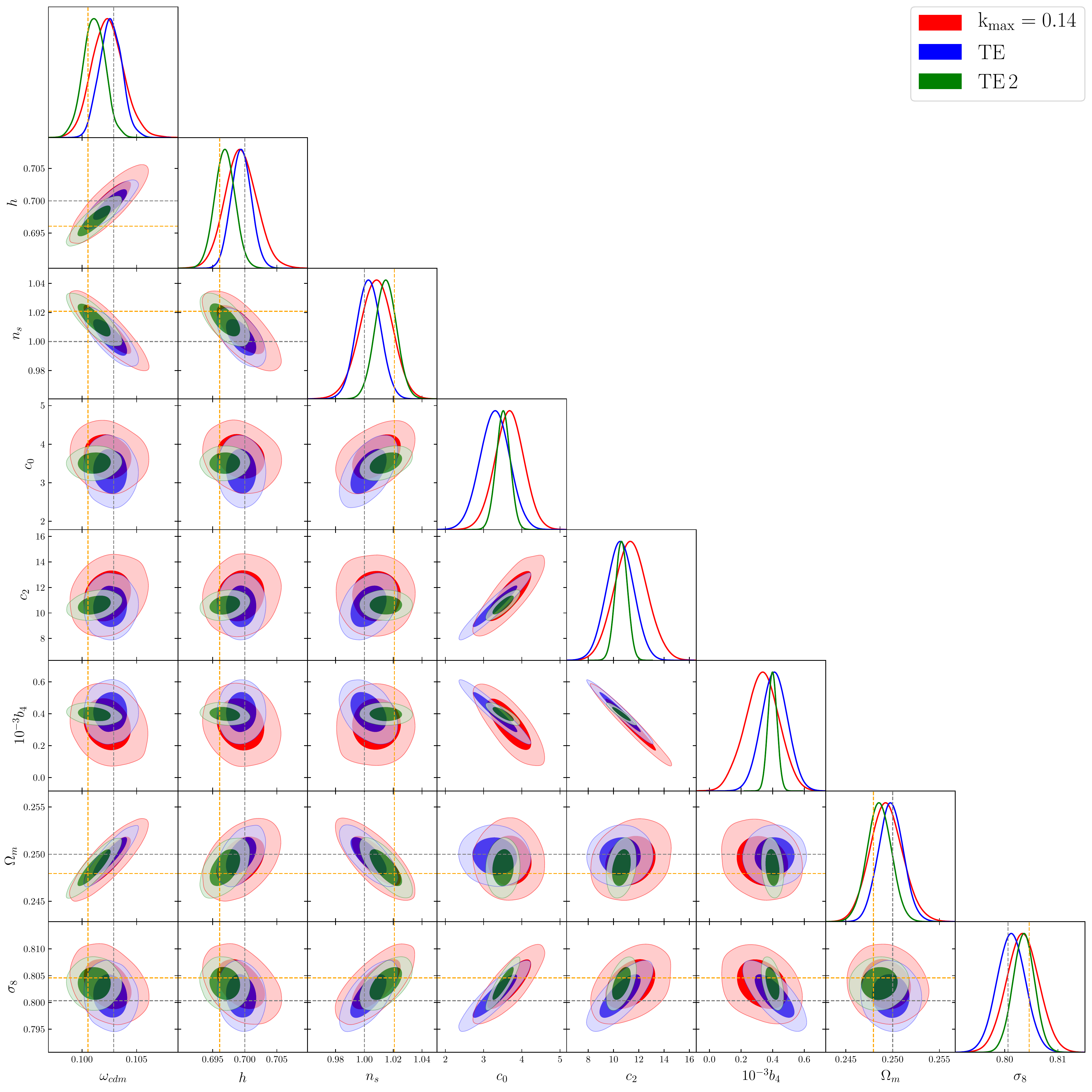}
		\end{center}
     	\caption{\label{fig:dm_rsd_z1_fid}
Triangle plot for the cosmological and nuisance parameters 
		measured from the redshift space dark matter power spectrum
		of the LasDamas simulations at $z=0.974$ for the $\kmax$-analysis and two choices of the fiducial cosmology in the TE approach.
			$c_0,c_2,b_4$ are quoted in units $[\Mpch]^2,~[\Mpch]^2,~[\Mpch]^4$, respectively. Crossing of the dashed orange lines outlines the fiducial cosmological parameters used in the $\rm TE\,2$ analysis.
	}
\end{figure}

Now let us consider
the case of redshift space dark matter ($z=0.974$). The marginalized 1d parameter constraints are listed in Tab. \ref{tab:cos_dm_rsd0_fid} (fifth column).
\begin{table}[h!]
	\begin{center}
		\begin{tabular}{|c|c|c|c|c|}  \hline 
			\small{par.} & fid. & \multicolumn{3}{|c|}{$z=0.974$} \\  \hline
			&  & $k_{\text{max}}=0.14\,h$/Mpc & TE & TE 2 \\  \hline\hline
			$\omega_{cdm}$ & $0.1029$  & $0.1023^{+1.4\cdot 10^{-3}}_{-1.5\cdot 10^{-3}}$ &
			$0.1026^{+1.1\cdot 10^{-3}}_{-1.0\cdot 10^{-3}}$ &
			$0.1011^{+1.1\cdot 10^{-3}}_{-1.0\cdot 10^{-3}}$  \\ \hline
			$h$ & $0.7$& $0.6993^{+2.4\cdot 10^{-3}}_{-2.6\cdot 10^{-3}}$ &
			$0.6995^{+1.6\cdot 10^{-3}}_{-1.6\cdot 10^{-3}}$ &
			$0.6969^{+1.6\cdot 10^{-3}}_{-1.6\cdot 10^{-3}}$ \\ \hline
			$n_s$ & $1$ & $1.0083^{+0.012}_{-0.011}$&
			$1.0028^{+8.5\cdot 10^{-3}}_{-8.5\cdot 10^{-3}}$ &
			$1.0149^{+7.6\cdot 10^{-3}}_{-7.7\cdot 10^{-3}}$ \\ \hline
			$A$ & $1$ & $1.0096^{+0.016}_{-0.016}$&
			$1.0043^{+0.010}_{-0.011}$ &
			$1.0221^{+0.010}_{-0.011}$ \\ \hline
			$\Omega_m$ & $0.25$ & $0.2493^{+1.7\cdot10^{-3}}_{-1.8\cdot10^{-3}}$ &
			$0.2498^{+1.4\cdot10^{-3}}_{-1.3\cdot10^{-3}}$ &
			$0.2486^{+1.3\cdot10^{-3}}_{-1.3\cdot10^{-3}}$ \\ \hline
			$\sigma_8$  & $0.8003$ & $0.8033^{+3.1\cdot10^{-3}}_{-3.1\cdot10^{-3}}$ &
			$0.8012^{+2.8\cdot 10^{-3}}_{-2.8\cdot 10^{-3}}$ &
			$0.8036^{+2.1\cdot 10^{-3}}_{-2.1\cdot 10^{-3}}$ \\ \hline\hline
			$c_0$ & -- & $3.66^{+0.39}_{-0.37}$ &
			$3.29^{+0.40}_{-0.39}$ &
			$3.51^{+0.19}_{-0.19}$ \\ \hline
			$c_2$ & -- & $11.26^{+1.32}_{-1.25}$ &
			$10.52^{+1.11}_{-1.10}$ &
			$10.62^{+0.50}_{-0.50}$ \\ \hline
			$10^{-3}b_4$ & -- & $0.34^{+0.10}_{-0.11}$ &
			$0.41^{+0.08}_{-0.08}$ &
			$0.40^{+0.03}_{-0.03}$ \\ \hline
		\end{tabular}
		\caption{
			The marginalized 1d intervals for the cosmological and nuisance parameters 
			estimated from the monopole and quadrupole moments of the Las Damas dark matter redshift space power spectrum
			at $z=0.974$. 
			We show the fitted parameters (first column), fiducial values used 
			in simulations (second column), and the resulting parameter constraints for the baseline $k_{\max}$ analysis (third column) and the theoretical error approach with default (fourth column) and new fiducial cosmology (fifth column) termed $\rm TE\,2$.
			$c_0,c_2,b_4$ are quoted in units $[\Mpch]^2,~[\Mpch]^2,~[\Mpch]^4$, respectively.
		}
		\label{tab:cos_dm_rsd0_fid}
	\end{center}
\end{table} 
The 2d posterior distributions are shown in Fig. \ref{fig:dm_rsd_z1_fid} (green contours).
Just like in real space, changing the fiducial cosmology does not impact uncertainties on $\omega_{cdm}$ and $h$. However, the error bars on $\sigma_8$ and $n_s$ decrease by $25\%$ and $10\%$. The impact on the counterterms is even stronger. Namely, the constraints on $c_0$, $c_2$, $b_4$ become twice weaker compared to those which we 
have obtained in the baseline TE analysis. We have also found that the posterior distributions are shifted towards the new fiducial cosmology. 
Nevertheless, the new 2d posteriors enclose the true mock parameters within 95$\%$ CL as shown in Fig. \ref{fig:dm_rsd_z1_fid}.

\begin{figure}[ht!]
	\begin{center}
		\includegraphics[width=1\textwidth]{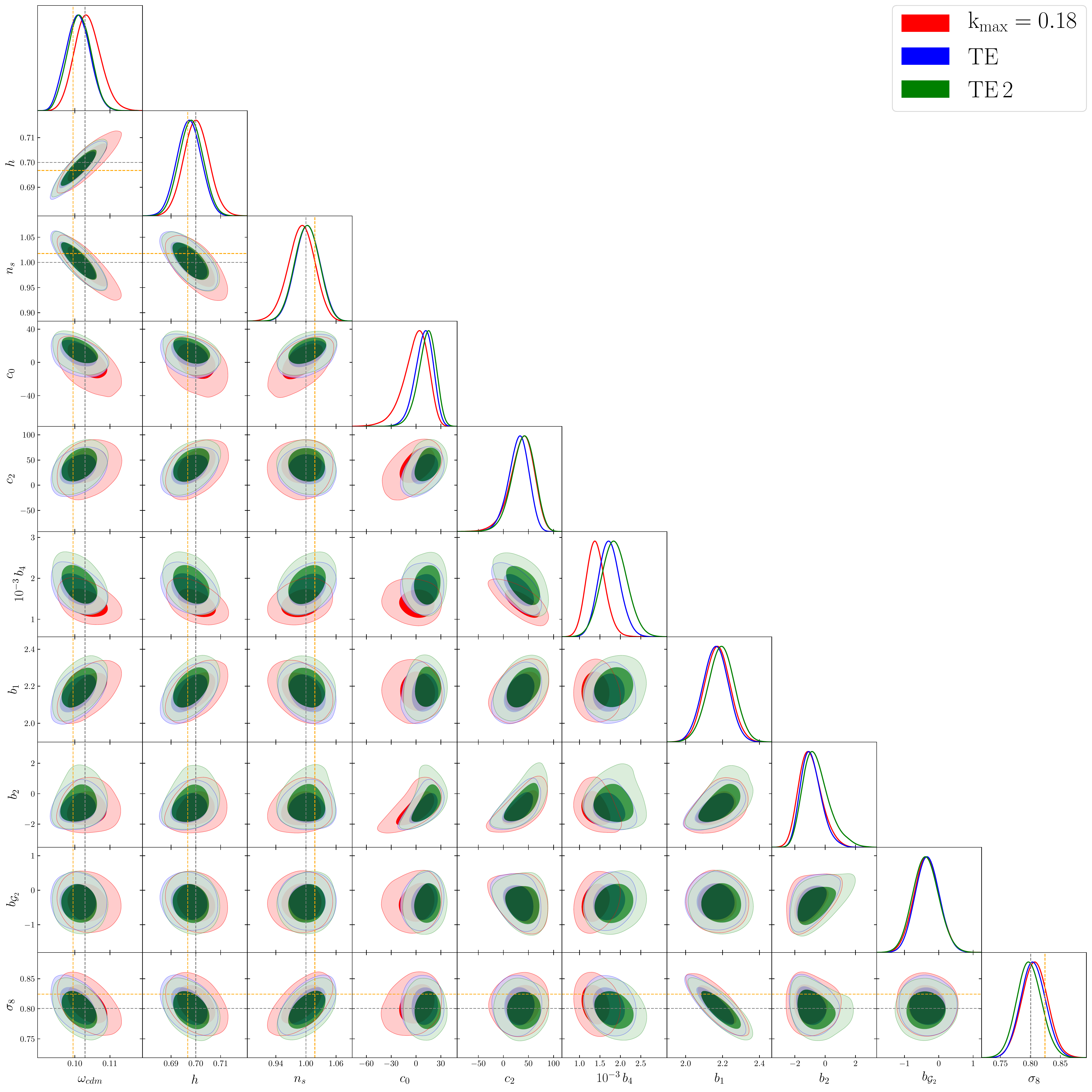}\end{center}
     	\caption{\label{fig:gal_rsd_fid}
		Triangle plot for the cosmological and nuisance parameters 
		measured from the redshift-space galaxy power spectrum
		of the LasDamas simulations at $z=0.342$ for the $\kmax$-analysis and two choices of the fiducial cosmology in the TE approach.
			$c_0,c_2,b_4$ are quoted in units $[\Mpch]^2,~[\Mpch]^2,~[\Mpch]^4$, respectively. Crossing of the dashed orange lines outlines the fiducial cosmological parameters used in the $\rm TE\,2$ analysis.
	}
\end{figure}

Finally, we discuss redshift space galaxies ($z=0.342$). The marginalized 1d parameter constraints are listed in Tab. \ref{tab:gal_rsd2} (fifth column).
\begin{table}[h!]
	\begin{center}
		\begin{tabular}{|c|c|c|c|c|}  \hline 
			\small{par.} & fid. & $k_\maxx=0.18\,h$/Mpc & TE & TE 2 \\  \hline
			\hline
			$\omega_{cdm}$ & $0.1029$  & $0.1037^{+3.5\cdot 10^{-3}}_{-4.1\cdot 10^{-3}}$ &
			$0.1009^{+3.5\cdot 10^{-3}}_{-3.5\cdot 10^{-3}}$ &
			$0.1013^{+3.4\cdot 10^{-3}}_{-3.4\cdot 10^{-3}}$ \\ \hline
			$h$ & $0.7$& $0.7002^{+5.1\cdot 10^{-3}}_{-5.3\cdot 10^{-3}}$ &
			$0.6973^{+5.0\cdot 10^{-3}}_{-5.0\cdot 10^{-3}}$ &
			$0.6981^{+4.9\cdot 10^{-3}}_{-4.9\cdot 10^{-3}}$ \\ \hline
			$n_s$ & $1$ & $0.9909^{+0.028}_{-0.026}$ &
			$1.0031^{+0.025}_{-0.025}$ &
			$1.0021^{+0.025}_{-0.025}$ \\ \hline
			$A$ & $1$ & $1.0169^{+0.073}_{-0.079}$& 
			$1.0571^{+0.069}_{-0.081}$ &
			$1.0174^{+0.066}_{-0.079}$ \\ \hline
			$\Omega_m$ & $0.25$ & $0.2513^{+4.6\cdot10^{-3}}_{-5.2\cdot10^{-3}}$ &
			$0.2479^{+4.3\cdot10^{-3}}_{-4.7\cdot10^{-3}}$ &
			$0.2480^{+4.1\cdot10^{-3}}_{-4.4\cdot10^{-3}}$ \\ \hline
			$\sigma_8$  & $0.8003$ & $ 0.8069^{+0.022}_{-0.022}$ &
			$0.8044^{+0.021}_{-0.021}$  &
			$0.7980^{+0.021}_{-0.020}$  \\  \hline\hline
			$c_0$ & -- & $0.77^{+17.93}_{-11.48}$ & 
			$9.40^{+12.45}_{-9.68}$ &
			$12.88^{+12.56}_{-9.41}$ \\ \hline
			$c_2$ & -- & $36.22^{+28.01}_{-20.61}$ & 
			$29.61^{+23.11}_{-18.30}$ &
			$38.76^{+25.64}_{-20.78}$ \\ \hline
			$10^{-3}b_4$ & -- & $1.40^{+0.21}_{-0.26}$ & 
			$1.73^{+0.25}_{-0.28}$ &
			$1.85^{+0.31}_{-0.34}$ \\ \hline
			$b_1$ & $-$ & $2.17^{+0.07}_{-0.07}$ & 
			$2.16^{+0.07}_{-0.07}$ &
			$2.19^{+0.07}_{-0.07}$ \\ \hline
			$b_2$ & -- & $-0.96^{+0.64}_{-0.87}$ & 
			$-0.92^{+0.57}_{-0.78}$ &
			$-0.57^{+0.65}_{-1.03}$ \\ \hline
			$b_{\mathcal{G}_2}$ & -- & $-0.350^{+0.364}_{-0.388}$ &
			$-0.330^{+0.351}_{-0.368}$ &
			$-0.377^{+0.372}_{-0.395}$ \\ \hline
		\end{tabular}
		\caption{
			The marginalized 1d intervals for the cosmological parameters 
			estimated from the Las Damas redshift space galaxy power spectra at $z=0.342$. 
			The table contains fitted parameters (first column), fiducial values used 
			in simulations (second column), and the results of the baseline $k_\maxx$ analysis (third column) and the outcome of the theoretical error approach with default (fourth columns) and new (fifth columns) fiducial cosmology. 
			$c_0,c_2,b_4$ are quoted in units $[\Mpch]^2,~[\Mpch]^2,~[\Mpch]^4$, respectively.
		}
		\label{tab:gal_rsd2}
	\end{center}
\end{table} 
The 2d posterior distributions are shown in Fig. \ref{fig:gal_rsd_fid} (green contours).
In this case, we do not find any significant difference in posterior distributions of the cosmological parameters due to the change of the fiducial cosmology. 

Our analysis suggests the following  conclusions.  The improvement
that we have found for dark matter in real space and redshift space on $\sigma_8$ and $n_s$ parameters may be artificially caused by 
the specific choice of the fiducial cosmology.
However, in the realistic analysis, the TE approach  yields robust cosmological constraints which do not depend on the choice of the fiducial cosmology. In this case the TE approach does not provide any information gain in comparison with the analysis with the sharp momentum cut and only serves to optimise the choice of $k_{\rm max}$. Finally, we would like to point out that we have also studied the convergence of the results w.r.t. variations
of $k^{\rm fid.}_{\rm max}$, and 
found statistically indistinguishable  results for 
all considered cases.

\bibliographystyle{JHEP}
\bibliography{short}

\providecommand{\href}[2]{#2}\begingroup\raggedright\begin{thebibliography}{10}

\bibitem{Ivanov:2019pdj}
M.~M. Ivanov, M.~Simonovi\'c and M.~Zaldarriaga, \emph{{Cosmological Parameters
  from the BOSS Galaxy Power Spectrum}},
  \href{https://doi.org/10.1088/1475-7516/2020/05/042}{\emph{JCAP} {\bfseries
  05} (2020) 042} [\href{https://arxiv.org/abs/1909.05277}{{\ttfamily
  1909.05277}}].

\bibitem{Colas:2019ret}
T.~Colas, G.~D'amico, L.~Senatore, P.~Zhang and F.~Beutler, \emph{{Efficient
  Cosmological Analysis of the SDSS/BOSS data from the Effective Field Theory
  of Large-Scale Structure}},
  \href{https://doi.org/10.1088/1475-7516/2020/06/001}{\emph{JCAP} {\bfseries
  06} (2020) 001} [\href{https://arxiv.org/abs/1909.07951}{{\ttfamily
  1909.07951}}].

\bibitem{Troster:2019ean}
T.~Tr\"oster et~al., \emph{{Cosmology from large-scale structure: Constraining
  $\Lambda$CDM with BOSS}},
  \href{https://doi.org/10.1051/0004-6361/201936772}{\emph{Astron. Astrophys.}
  {\bfseries 633} (2020) L10}
  [\href{https://arxiv.org/abs/1909.11006}{{\ttfamily 1909.11006}}].

\bibitem{Ivanov:2019hqk}
M.~M. Ivanov, M.~Simonovi\'c and M.~Zaldarriaga, \emph{{Cosmological Parameters
  and Neutrino Masses from the Final Planck and Full-Shape BOSS Data}},
  \href{https://doi.org/10.1103/PhysRevD.101.083504}{\emph{Phys. Rev. D}
  {\bfseries 101} (2020) 083504}
  [\href{https://arxiv.org/abs/1912.08208}{{\ttfamily 1912.08208}}].

\bibitem{Philcox:2020vvt}
O.~H. Philcox, M.~M. Ivanov, M.~Simonovi\'c and M.~Zaldarriaga,
  \emph{{Combining Full-Shape and BAO Analyses of Galaxy Power Spectra: A 1.6%
  CMB-independent constraint on H0}},
  \href{https://doi.org/10.1088/1475-7516/2020/05/032}{\emph{JCAP} {\bfseries
  05} (2020) 032} [\href{https://arxiv.org/abs/2002.04035}{{\ttfamily
  2002.04035}}].

\bibitem{Philcox:2020xbv}
O.~H. Philcox, B.~D. Sherwin, G.~S. Farren and E.~J. Baxter, \emph{{Determining
  the Hubble Constant without the Sound Horizon: Measurements from Galaxy
  Surveys}},  \href{https://arxiv.org/abs/2008.08084}{{\ttfamily 2008.08084}}.

\bibitem{Chudaykin:2020ghx}
A.~Chudaykin, K.~Dolgikh and M.~M. Ivanov, \emph{{Constraints on the curvature
  of the Universe and dynamical dark energy from the Full-shape and BAO data}},
   \href{https://arxiv.org/abs/2009.10106}{{\ttfamily 2009.10106}}.

\bibitem{Laureijs:2011gra}
{\scshape EUCLID} collaboration, R.~Laureijs et~al., \emph{{Euclid Definition
  Study Report}},  \href{https://arxiv.org/abs/1110.3193}{{\ttfamily
  1110.3193}}.

\bibitem{Amendola:2016saw}
L.~Amendola et~al., \emph{{Cosmology and fundamental physics with the Euclid
  satellite}}, \href{https://doi.org/10.1007/s41114-017-0010-3}{\emph{Living
  Rev. Rel.} {\bfseries 21} (2018) 2}
  [\href{https://arxiv.org/abs/1606.00180}{{\ttfamily 1606.00180}}].

\bibitem{Aghamousa:2016zmz}
{\scshape DESI} collaboration, A.~Aghamousa et~al., \emph{{The DESI Experiment
  Part I: Science,Targeting, and Survey Design}},
  \href{https://arxiv.org/abs/1611.00036}{{\ttfamily 1611.00036}}.

\bibitem{Chudaykin:2019ock}
A.~Chudaykin and M.~M. Ivanov, \emph{{Measuring neutrino masses with
  large-scale structure: Euclid forecast with controlled theoretical error}},
  \href{https://doi.org/10.1088/1475-7516/2019/11/034}{\emph{JCAP} {\bfseries
  1911} (2019) 034} [\href{https://arxiv.org/abs/1907.06666}{{\ttfamily
  1907.06666}}].

\bibitem{Audren:2012vy}
B.~Audren, J.~Lesgourgues, S.~Bird, M.~G. Haehnelt and M.~Viel, \emph{{Neutrino
  masses and cosmological parameters from a Euclid-like survey: Markov Chain
  Monte Carlo forecasts including theoretical errors}},
  \href{https://doi.org/10.1088/1475-7516/2013/01/026}{\emph{JCAP} {\bfseries
  1301} (2013) 026} [\href{https://arxiv.org/abs/1210.2194}{{\ttfamily
  1210.2194}}].

\bibitem{Yankelevich:2018uaz}
V.~Yankelevich and C.~Porciani, \emph{{Cosmological information in the
  redshift-space bispectrum}},
  \href{https://doi.org/10.1093/mnras/sty3143}{\emph{Mon. Not. Roy. Astron.
  Soc.} {\bfseries 483} (2019) 2078}
  [\href{https://arxiv.org/abs/1807.07076}{{\ttfamily 1807.07076}}].

\bibitem{Ivanov:2020ril}
M.~M. Ivanov, E.~McDonough, J.~C. Hill, M.~Simonovi\'c, M.~W. Toomey,
  S.~Alexander et~al., \emph{{Constraining Early Dark Energy with Large-Scale
  Structure}},  \href{https://arxiv.org/abs/2006.11235}{{\ttfamily
  2006.11235}}.

\bibitem{Scoccimarro:1999kp}
R.~Scoccimarro, M.~Zaldarriaga and L.~Hui, \emph{{Power spectrum correlations
  induced by nonlinear clustering}},
  \href{https://doi.org/10.1086/308059}{\emph{Astrophys. J.} {\bfseries 527}
  (1999) 1} [\href{https://arxiv.org/abs/astro-ph/9901099}{{\ttfamily
  astro-ph/9901099}}].

\bibitem{Wadekar:2019rdu}
D.~Wadekar and R.~Scoccimarro, \emph{{The Galaxy Power Spectrum Multipoles
  Covariance in Perturbation Theory}},
  \href{https://arxiv.org/abs/1910.02914}{{\ttfamily 1910.02914}}.

\bibitem{Yamamoto:2005dz}
K.~Yamamoto, M.~Nakamichi, A.~Kamino, B.~A. Bassett and H.~Nishioka, \emph{{A
  Measurement of the quadrupole power spectrum in the clustering of the 2dF QSO
  Survey}}, \href{https://doi.org/10.1093/pasj/58.1.93}{\emph{Publ. Astron.
  Soc. Jap.} {\bfseries 58} (2006) 93}
  [\href{https://arxiv.org/abs/astro-ph/0505115}{{\ttfamily
  astro-ph/0505115}}].

\bibitem{Grieb:2015bia}
J.~N. Grieb, A.~G. Sánchez, S.~Salazar-Albornoz and C.~Dalla~Vecchia,
  \emph{{Gaussian covariance matrices for anisotropic galaxy clustering
  measurements}}, \href{https://doi.org/10.1093/mnras/stw065}{\emph{Mon. Not.
  Roy. Astron. Soc.} {\bfseries 457} (2016) 1577}
  [\href{https://arxiv.org/abs/1509.04293}{{\ttfamily 1509.04293}}].

\bibitem{Blake:2018tou}
C.~Blake, P.~Carter and J.~Koda, \emph{{Power spectrum multipoles on the curved
  sky: an application to the 6-degree Field Galaxy Survey}},
  \href{https://doi.org/10.1093/mnras/sty1814}{\emph{Mon. Not. Roy. Astron.
  Soc.} {\bfseries 479} (2018) 5168}
  [\href{https://arxiv.org/abs/1801.04969}{{\ttfamily 1801.04969}}].

\bibitem{Li:2018scc}
Y.~Li, S.~Singh, B.~Yu, Y.~Feng and U.~Seljak, \emph{{Disconnected Covariance
  of 2-point Functions in Large-Scale Structure}},
  \href{https://doi.org/10.1088/1475-7516/2019/01/016}{\emph{JCAP} {\bfseries
  01} (2019) 016} [\href{https://arxiv.org/abs/1811.05714}{{\ttfamily
  1811.05714}}].

\bibitem{Kitaura:2015uqa}
F.-S. Kitaura et~al., \emph{{The clustering of galaxies in the SDSS-III Baryon
  Oscillation Spectroscopic Survey: mock galaxy catalogues for the BOSS Final
  Data Release}}, \href{https://doi.org/10.1093/mnras/stv2826}{\emph{Mon. Not.
  Roy. Astron. Soc.} {\bfseries 456} (2016) 4156}
  [\href{https://arxiv.org/abs/1509.06400}{{\ttfamily 1509.06400}}].

\bibitem{Zhao:2020bib}
C.~Zhao et~al., \emph{{The Completed SDSS-IV extended Baryon Oscillation
  Spectroscopic Survey: one thousand multi-tracer mock catalogues with redshift
  evolution and systematics for galaxies and quasars of the final data
  release}},  \href{https://arxiv.org/abs/2007.08997}{{\ttfamily 2007.08997}}.

\bibitem{Lin:2020nef}
S.~Lin et~al., \emph{{The Completed SDSS-IV Extended Baryon Oscillation
  Spectroscopic Survey: GLAM-QPM mock galaxy catalogs for the Emission Line
  Galaxy Sample}},  \href{https://arxiv.org/abs/2007.08996}{{\ttfamily
  2007.08996}}.

\bibitem{Hartlap:2006kj}
J.~Hartlap, P.~Simon and P.~Schneider, \emph{{Why your model parameter
  confidences might be too optimistic: Unbiased estimation of the inverse
  covariance matrix}},
  \href{https://doi.org/10.1051/0004-6361:20066170}{\emph{Astron. Astrophys.}
  {\bfseries 464} (2007) 399}
  [\href{https://arxiv.org/abs/astro-ph/0608064}{{\ttfamily
  astro-ph/0608064}}].

\bibitem{Percival:2013sga}
W.~J. Percival et~al., \emph{{The Clustering of Galaxies in the SDSS-III Baryon
  Oscillation Spectroscopic Survey: Including covariance matrix errors}},
  \href{https://doi.org/10.1093/mnras/stu112}{\emph{Mon. Not. Roy. Astron.
  Soc.} {\bfseries 439} (2014) 2531}
  [\href{https://arxiv.org/abs/1312.4841}{{\ttfamily 1312.4841}}].

\bibitem{Sellentin:2015waz}
E.~Sellentin and A.~F. Heavens, \emph{{Parameter inference with estimated
  covariance matrices}},
  \href{https://doi.org/10.1093/mnrasl/slv190}{\emph{Mon. Not. Roy. Astron.
  Soc.} {\bfseries 456} (2016) L132}
  [\href{https://arxiv.org/abs/1511.05969}{{\ttfamily 1511.05969}}].

\bibitem{Philcox:2020zyp}
O.~H. Philcox, M.~M. Ivanov, M.~Zaldarriaga, M.~Simonovic and M.~Schmittfull,
  \emph{{Fewer Mocks and Less Noise: Reducing the Dimensionality of
  Cosmological Observables with Subspace Projections}},
  \href{https://arxiv.org/abs/2009.03311}{{\ttfamily 2009.03311}}.

\bibitem{Baldauf:2016sjb}
T.~Baldauf, M.~Mirbabayi, M.~Simonović and M.~Zaldarriaga, \emph{{LSS
  constraints with controlled theoretical uncertainties}},
  \href{https://arxiv.org/abs/1602.00674}{{\ttfamily 1602.00674}}.

\bibitem{Steele:2020tak}
T.~Steele and T.~Baldauf, \emph{{Precise Calibration of the One-Loop Bispectrum
  in the Effective Field Theory of Large Scale Structure}},
  \href{https://arxiv.org/abs/2009.01200}{{\ttfamily 2009.01200}}.

\bibitem{DAmico:2019fhj}
G.~D'Amico, J.~Gleyzes, N.~Kokron, D.~Markovic, L.~Senatore, P.~Zhang et~al.,
  \emph{{The Cosmological Analysis of the SDSS/BOSS data from the Effective
  Field Theory of Large-Scale Structure}},
  \href{https://arxiv.org/abs/1909.05271}{{\ttfamily 1909.05271}}.

\bibitem{Nishimichi:2020tvu}
T.~Nishimichi, G.~D'Amico, M.~M. Ivanov, L.~Senatore, M.~Simonovic, M.~Takada
  et~al., \emph{{Blinded challenge for precision cosmology with large-scale
  structure: results from effective field theory for the redshift-space galaxy
  power spectrum}},  \href{https://arxiv.org/abs/2003.08277}{{\ttfamily
  2003.08277}}.

\bibitem{Rossi:2020wxx}
G.~Rossi et~al., \emph{{The Completed SDSS-IV Extended Baryon Oscillation
  Spectroscopic Survey: N-body Mock Challenge for Galaxy Clustering
  Measurements}},  \href{https://arxiv.org/abs/2007.09002}{{\ttfamily
  2007.09002}}.

\bibitem{Alam:2020jvh}
S.~Alam et~al., \emph{{The Completed SDSS-IV extended Baryon Oscillation
  Spectroscopic Survey: N-body Mock Challenge for the eBOSS Emission Line
  Galaxy Sample}},  \href{https://arxiv.org/abs/2007.09004}{{\ttfamily
  2007.09004}}.

\bibitem{DAmico:2020kxu}
G.~D'Amico, L.~Senatore and P.~Zhang, \emph{{Limits on $w$CDM from the EFTofLSS
  with the PyBird code}},  \href{https://arxiv.org/abs/2003.07956}{{\ttfamily
  2003.07956}}.

\bibitem{Senatore:2014via}
L.~Senatore and M.~Zaldarriaga, \emph{{The IR-resummed Effective Field Theory
  of Large Scale Structures}},
  \href{https://doi.org/10.1088/1475-7516/2015/02/013}{\emph{JCAP} {\bfseries
  1502} (2015) 013} [\href{https://arxiv.org/abs/1404.5954}{{\ttfamily
  1404.5954}}].

\bibitem{Baldauf:2015xfa}
T.~Baldauf, M.~Mirbabayi, M.~Simonović and M.~Zaldarriaga, \emph{{Equivalence
  Principle and the Baryon Acoustic Peak}},
  \href{https://doi.org/10.1103/PhysRevD.92.043514}{\emph{Phys. Rev.}
  {\bfseries D92} (2015) 043514}
  [\href{https://arxiv.org/abs/1504.04366}{{\ttfamily 1504.04366}}].

\bibitem{Vlah:2015zda}
Z.~Vlah, U.~Seljak, M.~Y. Chu and Y.~Feng, \emph{{Perturbation theory,
  effective field theory, and oscillations in the power spectrum}},
  \href{https://doi.org/10.1088/1475-7516/2016/03/057}{\emph{JCAP} {\bfseries
  1603} (2016) 057} [\href{https://arxiv.org/abs/1509.02120}{{\ttfamily
  1509.02120}}].

\bibitem{Blas:2016sfa}
D.~Blas, M.~Garny, M.~M. Ivanov and S.~Sibiryakov, \emph{{Time-Sliced
  Perturbation Theory II: Baryon Acoustic Oscillations and Infrared
  Resummation}},
  \href{https://doi.org/10.1088/1475-7516/2016/07/028}{\emph{JCAP} {\bfseries
  1607} (2016) 028} [\href{https://arxiv.org/abs/1605.02149}{{\ttfamily
  1605.02149}}].

\bibitem{Senatore:2017pbn}
L.~Senatore and G.~Trevisan, \emph{{On the IR-Resummation in the EFTofLSS}},
  \href{https://doi.org/10.1088/1475-7516/2018/05/019}{\emph{JCAP} {\bfseries
  1805} (2018) 019} [\href{https://arxiv.org/abs/1710.02178}{{\ttfamily
  1710.02178}}].

\bibitem{Ivanov:2018gjr}
M.~M. Ivanov and S.~Sibiryakov, \emph{{Infrared Resummation for Biased Tracers
  in Redshift Space}},
  \href{https://doi.org/10.1088/1475-7516/2018/07/053}{\emph{JCAP} {\bfseries
  1807} (2018) 053} [\href{https://arxiv.org/abs/1804.05080}{{\ttfamily
  1804.05080}}].

\bibitem{Chen:2020fxs}
S.-F. Chen, Z.~Vlah and M.~White, \emph{{Consistent Modeling of Velocity
  Statistics and Redshift-Space Distortions in One-Loop Perturbation Theory}},
  \href{https://doi.org/10.1088/1475-7516/2020/07/062}{\emph{JCAP} {\bfseries
  07} (2020) 062} [\href{https://arxiv.org/abs/2005.00523}{{\ttfamily
  2005.00523}}].

\bibitem{Vasudevan:2019ewf}
A.~Vasudevan, M.~M. Ivanov, S.~Sibiryakov and J.~Lesgourgues,
  \emph{{Time-sliced perturbation theory with primordial non-Gaussianity and
  effects of large bulk flows on inflationary oscillating features}},
  \href{https://doi.org/10.1088/1475-7516/2019/09/037}{\emph{JCAP} {\bfseries
  09} (2019) 037} [\href{https://arxiv.org/abs/1906.08697}{{\ttfamily
  1906.08697}}].

\bibitem{Chen:2020ckc}
S.-F. Chen, Z.~Vlah and M.~White, \emph{{Modeling features in the
  redshift-space halo power spectrum with perturbation theory}},
  \href{https://arxiv.org/abs/2007.00704}{{\ttfamily 2007.00704}}.

\bibitem{Chudaykin:2020aoj}
A.~Chudaykin, M.~M. Ivanov and M.~Simonovi\'c, \emph{{CLASS-PT: non-linear
  perturbation theory extension of the Boltzmann code CLASS}},
  \href{https://arxiv.org/abs/2004.10607}{{\ttfamily 2004.10607}}.

\bibitem{Tegmark:1997rp}
M.~Tegmark, \emph{{Measuring cosmological parameters with galaxy surveys}},
  \href{https://doi.org/10.1103/PhysRevLett.79.3806}{\emph{Phys. Rev. Lett.}
  {\bfseries 79} (1997) 3806}
  [\href{https://arxiv.org/abs/astro-ph/9706198}{{\ttfamily
  astro-ph/9706198}}].

\bibitem{Tegmark:1996qt}
M.~Tegmark, \emph{{How to measure CMB power spectra without losing
  information}}, \href{https://doi.org/10.1103/PhysRevD.55.5895}{\emph{Phys.
  Rev. D} {\bfseries 55} (1997) 5895}
  [\href{https://arxiv.org/abs/astro-ph/9611174}{{\ttfamily
  astro-ph/9611174}}].

\bibitem{Wadekar:2020rdu}
D.~Wadekar, M.~M. Ivanov and R.~Scoccimarro, \emph{{Cosmological constraints
  from BOSS with analytic covariance matrices}},
  \href{https://arxiv.org/abs/2009.00622}{{\ttfamily 2009.00622}}.

\bibitem{Blas:2015qsi}
D.~Blas, M.~Garny, M.~M. Ivanov and S.~Sibiryakov, \emph{{Time-Sliced
  Perturbation Theory for Large Scale Structure I: General Formalism}},
  \href{https://doi.org/10.1088/1475-7516/2016/07/052}{\emph{JCAP} {\bfseries
  1607} (2016) 052} [\href{https://arxiv.org/abs/1512.05807}{{\ttfamily
  1512.05807}}].

\bibitem{Baumann:2010tm}
D.~Baumann, A.~Nicolis, L.~Senatore and M.~Zaldarriaga, \emph{{Cosmological
  Non-Linearities as an Effective Fluid}},
  \href{https://doi.org/10.1088/1475-7516/2012/07/051}{\emph{JCAP} {\bfseries
  1207} (2012) 051} [\href{https://arxiv.org/abs/1004.2488}{{\ttfamily
  1004.2488}}].

\bibitem{Carrasco:2012cv}
J.~J.~M. Carrasco, M.~P. Hertzberg and L.~Senatore, \emph{{The Effective Field
  Theory of Cosmological Large Scale Structures}},
  \href{https://doi.org/10.1007/JHEP09(2012)082}{\emph{JHEP} {\bfseries 09}
  (2012) 082} [\href{https://arxiv.org/abs/1206.2926}{{\ttfamily 1206.2926}}].

\bibitem{Perko:2016puo}
A.~Perko, L.~Senatore, E.~Jennings and R.~H. Wechsler, \emph{{Biased Tracers in
  Redshift Space in the EFT of Large-Scale Structure}},
  \href{https://arxiv.org/abs/1610.09321}{{\ttfamily 1610.09321}}.

\bibitem{2009AAS...21342506M}
C.~{McBride}, A.~{Berlind}, R.~{Scoccimarro}, R.~{Wechsler}, M.~{Busha},
  J.~{Gardner} et~al., \emph{{LasDamas Mock Galaxy Catalogs for SDSS}},  in
  \emph{American Astronomical Society Meeting Abstracts \#213}, vol.~213 of
  \emph{American Astronomical Society Meeting Abstracts}, p.~425.06, Jan, 2009.

\bibitem{Ivanov:2020mfr}
M.~M. Ivanov, Y.~Ali-Ha\"\i{}moud and J.~Lesgourgues, \emph{{H0 tension or T0
  tension?}}, \href{https://doi.org/10.1103/PhysRevD.102.063515}{\emph{Phys.
  Rev. D} {\bfseries 102} (2020) 063515}
  [\href{https://arxiv.org/abs/2005.10656}{{\ttfamily 2005.10656}}].

\bibitem{Alam:2016hwk}
{\scshape BOSS} collaboration, S.~Alam et~al., \emph{{The clustering of
  galaxies in the completed SDSS-III Baryon Oscillation Spectroscopic Survey:
  cosmological analysis of the DR12 galaxy sample}},
  \href{https://doi.org/10.1093/mnras/stx721}{\emph{Mon. Not. Roy. Astron.
  Soc.} {\bfseries 470} (2017) 2617}
  [\href{https://arxiv.org/abs/1607.03155}{{\ttfamily 1607.03155}}].

\bibitem{Simonovic:2017mhp}
M.~Simonović, T.~Baldauf, M.~Zaldarriaga, J.~J. Carrasco and J.~A. Kollmeier,
  \emph{{Cosmological perturbation theory using the FFTLog: formalism and
  connection to QFT loop integrals}},
  \href{https://doi.org/10.1088/1475-7516/2018/04/030}{\emph{JCAP} {\bfseries
  1804} (2018) 030} [\href{https://arxiv.org/abs/1708.08130}{{\ttfamily
  1708.08130}}].

\bibitem{Audren:2012wb}
B.~Audren, J.~Lesgourgues, K.~Benabed and S.~Prunet, \emph{{Conservative
  Constraints on Early Cosmology: an illustration of the Monte Python
  cosmological parameter inference code}},
  \href{https://doi.org/10.1088/1475-7516/2013/02/001}{\emph{JCAP} {\bfseries
  1302} (2013) 001} [\href{https://arxiv.org/abs/1210.7183}{{\ttfamily
  1210.7183}}].

\bibitem{Brinckmann:2018cvx}
T.~Brinckmann and J.~Lesgourgues, \emph{{MontePython 3: boosted MCMC sampler
  and other features}},
  \href{https://doi.org/10.1016/j.dark.2018.100260}{\emph{Phys. Dark Univ.}
  {\bfseries 24} (2019) 100260}
  [\href{https://arxiv.org/abs/1804.07261}{{\ttfamily 1804.07261}}].

\bibitem{Lewis:2019xzd}
A.~Lewis, \emph{{GetDist: a Python package for analysing Monte Carlo samples}},
   \href{https://arxiv.org/abs/1910.13970}{{\ttfamily 1910.13970}}.

\bibitem{Kaiser:1987qv}
N.~Kaiser, \emph{{Clustering in real space and in redshift space}}, {\emph{Mon.
  Not. Roy. Astron. Soc.} {\bfseries 227} (1987) 1}.

\bibitem{Bernardeau:2001qr}
F.~Bernardeau, S.~Colombi, E.~Gaztanaga and R.~Scoccimarro, \emph{{Large scale
  structure of the universe and cosmological perturbation theory}},
  \href{https://doi.org/10.1016/S0370-1573(02)00135-7}{\emph{Phys. Rept.}
  {\bfseries 367} (2002) 1}
  [\href{https://arxiv.org/abs/astro-ph/0112551}{{\ttfamily
  astro-ph/0112551}}].

\bibitem{Baldauf:2015aha}
T.~Baldauf, L.~Mercolli and M.~Zaldarriaga, \emph{{Effective field theory of
  large scale structure at two loops: The apparent scale dependence of the
  speed of sound}},
  \href{https://doi.org/10.1103/PhysRevD.92.123007}{\emph{Phys. Rev.}
  {\bfseries D92} (2015) 123007}
  [\href{https://arxiv.org/abs/1507.02256}{{\ttfamily 1507.02256}}].

\bibitem{Jackson:2008yv}
J.~C. Jackson, \emph{{Fingers of God: A critique of Rees' theory of primoridal
  gravitational radiation}},
  \href{https://doi.org/10.1093/mnras/156.1.1P}{\emph{Mon. Not. Roy. Astron.
  Soc.} {\bfseries 156} (1972) 1P}
  [\href{https://arxiv.org/abs/0810.3908}{{\ttfamily 0810.3908}}].

\bibitem{Senatore:2014vja}
L.~Senatore and M.~Zaldarriaga, \emph{{Redshift Space Distortions in the
  Effective Field Theory of Large Scale Structures}},
  \href{https://arxiv.org/abs/1409.1225}{{\ttfamily 1409.1225}}.

\bibitem{Lewandowski:2015ziq}
M.~Lewandowski, L.~Senatore, F.~Prada, C.~Zhao and C.-H. Chuang, \emph{{EFT of
  large scale structures in redshift space}},
  \href{https://doi.org/10.1103/PhysRevD.97.063526}{\emph{Phys. Rev. D}
  {\bfseries 97} (2018) 063526}
  [\href{https://arxiv.org/abs/1512.06831}{{\ttfamily 1512.06831}}].

\bibitem{Neveux:2020voa}
R.~Neveux et~al., \emph{{The Completed SDSS-IV extended Baryon Oscillation
  Spectroscopic Survey: BAO and RSD measurements from the anisotropic power
  spectrum of the Quasar sample between redshift 0.8 and 2.2}},
  \href{https://arxiv.org/abs/2007.08999}{{\ttfamily 2007.08999}}.

\bibitem{Smith:2020stf}
A.~Smith et~al., \emph{{The Completed SDSS-IV Extended Baryon Oscillation
  Spectroscopic Survey: N-body Mock Challenge for the Quasar Sample}},
  \href{https://arxiv.org/abs/2007.09003}{{\ttfamily 2007.09003}}.

\bibitem{Beutler:2016arn}
{\scshape BOSS} collaboration, F.~Beutler et~al., \emph{{The clustering of
  galaxies in the completed SDSS-III Baryon Oscillation Spectroscopic Survey:
  Anisotropic galaxy clustering in Fourier-space}},
  \href{https://doi.org/10.1093/mnras/stw3298}{\emph{Mon. Not. Roy. Astron.
  Soc.} {\bfseries 466} (2017) 2242}
  [\href{https://arxiv.org/abs/1607.03150}{{\ttfamily 1607.03150}}].

\bibitem{Baldauf:2013hka}
T.~Baldauf, U.~s. Seljak, R.~E. Smith, N.~Hamaus and V.~Desjacques, \emph{{Halo
  stochasticity from exclusion and nonlinear clustering}},
  \href{https://doi.org/10.1103/PhysRevD.88.083507}{\emph{Phys. Rev. D}
  {\bfseries 88} (2013) 083507}
  [\href{https://arxiv.org/abs/1305.2917}{{\ttfamily 1305.2917}}].

\bibitem{Hahn:2016kiy}
C.~Hahn, R.~Scoccimarro, M.~R. Blanton, J.~L. Tinker and S.~A.
  Rodríguez-Torres, \emph{{The effect of fibre collisions on the galaxy power
  spectrum multipoles}}, \href{https://doi.org/10.1093/mnras/stx185}{\emph{Mon.
  Not. Roy. Astron. Soc.} {\bfseries 467} (2017) 1940}
  [\href{https://arxiv.org/abs/1609.01714}{{\ttfamily 1609.01714}}].

\bibitem{Schmittfull:2018yuk}
M.~Schmittfull, M.~Simonović, V.~Assassi and M.~Zaldarriaga, \emph{{Modeling
  Biased Tracers at the Field Level}},
  \href{https://doi.org/10.1103/PhysRevD.100.043514}{\emph{Phys.\ Rev.\ D}
  {\bfseries 100} (2019) 043514}
  [\href{https://arxiv.org/abs/1811.10640}{{\ttfamily 1811.10640}}].

\bibitem{Desjacques:2016bnm}
V.~Desjacques, D.~Jeong and F.~Schmidt, \emph{{Large-Scale Galaxy Bias}},
  \href{https://doi.org/10.1016/j.physrep.2017.12.002}{\emph{Phys. Rept.}
  {\bfseries 733} (2018) 1} [\href{https://arxiv.org/abs/1611.09787}{{\ttfamily
  1611.09787}}].

\bibitem{Abidi:2018eyd}
M.~M. Abidi and T.~Baldauf, \emph{{Cubic Halo Bias in Eulerian and Lagrangian
  Space}}, \href{https://doi.org/10.1088/1475-7516/2018/07/029}{\emph{JCAP}
  {\bfseries 1807} (2018) 029}
  [\href{https://arxiv.org/abs/1802.07622}{{\ttfamily 1802.07622}}].

\bibitem{deMattia:2020fkb}
A.~de~Mattia et~al., \emph{{The Completed SDSS-IV extended Baryon Oscillation
  Spectroscopic Survey: measurement of the BAO and growth rate of structure of
  the emission line galaxy sample from the anisotropic power spectrum between
  redshift 0.6 and 1.1}},  \href{https://arxiv.org/abs/2007.09008}{{\ttfamily
  2007.09008}}.

\bibitem{Pandey:2020zyr}
{\scshape Dark Energy Survey} collaboration, S.~Pandey et~al.,
  \emph{{Perturbation theory for modeling galaxy bias: validation with
  simulations of the Dark Energy Survey}},
  \href{https://arxiv.org/abs/2008.05991}{{\ttfamily 2008.05991}}.

\bibitem{Blas:2013bpa}
D.~Blas, M.~Garny and T.~Konstandin, \emph{{On the non-linear scale of
  cosmological perturbation theory}},
  \href{https://doi.org/10.1088/1475-7516/2013/09/024}{\emph{JCAP} {\bfseries
  09} (2013) 024} [\href{https://arxiv.org/abs/1304.1546}{{\ttfamily
  1304.1546}}].

\bibitem{Blas:2013aba}
D.~Blas, M.~Garny and T.~Konstandin, \emph{{Cosmological perturbation theory at
  three-loop order}},
  \href{https://doi.org/10.1088/1475-7516/2014/01/010}{\emph{JCAP} {\bfseries
  1401} (2014) 010} [\href{https://arxiv.org/abs/1309.3308}{{\ttfamily
  1309.3308}}].

\bibitem{Foreman:2015lca}
S.~Foreman, H.~Perrier and L.~Senatore, \emph{{Precision Comparison of the
  Power Spectrum in the EFTofLSS with Simulations}},
  \href{https://doi.org/10.1088/1475-7516/2016/05/027}{\emph{JCAP} {\bfseries
  05} (2016) 027} [\href{https://arxiv.org/abs/1507.05326}{{\ttfamily
  1507.05326}}].

\end{thebibliography}\endgroup

\end{document}